\documentclass[twocolumn]{aastex63}
\usepackage{apjfonts}
\usepackage{aas_macros}

\usepackage{graphicx}	%
\usepackage{amsmath}	%
\usepackage{amssymb}	%
\usepackage{multirow}
\usepackage{bm}		%
\usepackage{hyperref}
\usepackage{ulem} %
\usepackage{algorithm}
\usepackage{algpseudocode}

\newcommand{\RNum}[1]{\uppercase\expandafter{\romannumeral #1\relax}}

\newcommand{\vect}[1]{\boldsymbol{#1}}
\newcommand{\thetat}{\vect{\theta}}

\newcommand{\redmagic}{\rm redMaGiC}
\newcommand{\Tbox}{\rm Chinchilla-T1}

\newcommand{\Mr}{M_{\rm{r}}}

\newcommand{\Msun}{\rm{M}_\odot}

\newcommand{\kpc}{\mbox{kpc}}

\newcommand{\Omegam}{\Omega_{{\rm m}}}
\newcommand{\Omegab}{\Omega_{{\rm b}}}

\newcommand{\redmapper}{{\rm redMaPPer}}

\defcitealias{DES_cluster_cosmology}{DES2020}
\defcitealias{DESY1KP}{DESY1KP}

\usepackage[T1]{fontenc}
\usepackage{ae,aecompl}

\usepackage{eso-pic}%

\usepackage{txfonts}

\shortauthors{To et al}
\shorttitle{Cardinal}
\begin{document}

\title{Buzzard to Cardinal: Improved Mock Catalogs for Large Galaxy Surveys}
\correspondingauthor{Chun-Hao To}
\email{to.87@osu.edu}
\author[0000-0001-7836-2261]{Chun-Hao To}
\affiliation{Center for Cosmology and AstroParticle Physics (CCAPP), Ohio State University, Columbus, OH 43210, USA}
\affiliation{Department of Physics, Ohio State University, Columbus, OH 43210, USA}
\affiliation{Department of Astronomy, Ohio State University, Columbus, OH 43210, USA}
\author[0000-0002-0728-0960]{Joseph DeRose}
\affiliation{Physics Division, Lawrence Berkeley National Laboratory, Berkeley, CA, USA}
\author[0000-0003-2229-011X]{Risa H. Wechsler}
\affiliation{Kavli Institute for Particle Astrophysics and Cosmology}
\affiliation{Physics Department, Stanford University, Stanford, CA, 94305}
\affiliation{SLAC National Accelerator Laboratory, Menlo Park, CA, 94025}
\author[0000-0001-9376-3135]{Eli Rykoff}
\affiliation{Kavli Institute for Particle Astrophysics and Cosmology}
\affiliation{SLAC National Accelerator Laboratory, Menlo Park, CA, 94025}
\author[0000-0002-7904-1707]{Hao-Yi Wu}
\affiliation{Department of Physics, Boise State University, Boise, ID 83725, USA}

\author[0000-0002-0298-4432]{Susmita Adhikari}
\affiliation{Indian Institute of Science Education and Research Pune: Pune, Maharashtra, IN}

\author[0000-0001-8356-2014]{Elisabeth Krause}
\affiliation{Department of Astronomy/Steward Observatory, University of Arizona, 933 North Cherry Avenue, Tucson, AZ 85721-0065}
\author[0000-0002-1666-6275]{Eduardo Rozo}
\affiliation{Department of Physics, University of Arizona, Tucson, AZ 85721, USA}

\author[0000-0001-7775-7261]{David H. Weinberg}
\affiliation{Center for Cosmology and AstroParticle Physics (CCAPP), Ohio State University, Columbus, OH 43210, USA}
\affiliation{Department of Astronomy, Ohio State University, Columbus, OH 43210, USA}

\begin{abstract}
We present the Cardinal mock galaxy catalogs, a new version of the Buzzard simulation that has been updated to support ongoing and future cosmological surveys, including DES, DESI, and LSST. These catalogs are based on a one-quarter sky simulation populated with galaxies out to a redshift of $z=2.35$ to a depth of $m_{\rm{r}}=27$. Compared to the Buzzard mocks, the Cardinal mocks include an updated subhalo abundance matching (SHAM) model that considers orphan galaxies and includes mass-dependent scatter between galaxy luminosity and halo properties. This model can simultaneously fit galaxy clustering and group--galaxy cross-correlations measured in three different luminosity threshold samples. The Cardinal mocks also feature a new color assignment model that can simultaneously fit color-dependent galaxy clustering in three different luminosity bins. We have developed an algorithm that uses photometric data to improve the color assignment model further and have also developed a novel method to improve small-scale lensing below the ray-tracing resolution. These improvements enable the Cardinal mocks to accurately reproduce the abundance of galaxy clusters and the properties of lens galaxies in the Dark Energy Survey data. As such, these simulations will be a valuable tool for future cosmological analyses based on large sky surveys. The cardinal mock will be released upon publication at \url{https://chunhaoto.com/cardinalsim}.

\end{abstract}

\setcounter{footnote}{1}

\section{Introduction}

Over the past two decades, large galaxy surveys have systematically mapped hundreds of millions of galaxies with unprecedented precision, allowing us to establish the standard cosmological model that describes the universe's evolution over billions of years. However, analyzing these data to their full potential requires advanced theoretical models and excellent control of systematics. Achieving these requirements is challenging because most of the information lies in the scales where the theory is highly non-perturbative. Furthermore, one galaxy survey can be analyzed with multiple cosmological probes, which could share the same sources of systematics. Consistently modeling these systematics in different cosmological probes is essential to yielding unbiased cosmological constraints. Finally, developing accurate theoretical models is more challenging when considering blind analyses, in which the data is transformed to obscure actual cosmological signals during the development of the models. 

Synthetic sky catalogs, also known as mock catalogs or mocks, provide a valuable tool for quantifying systematics and developing analysis techniques. They consist of plausible universes that can serve as a sandbox for researchers to test and develop methods for analyzing survey data. Accomplishing this task places several requirements on the synthetic catalogs. First, one wishes to use these catalogs to control systematics so that they are much smaller than the statistical uncertainties of the data. Therefore, the volume of the mocks has to be larger (ideally, much larger) than the volumes probed by the targeted surveys. Second, the galaxies in the mocks should be realistic, although the level of realism required depends on the specific surveys and analysis techniques used. Third, fast generation of new mocks is desirable. When analyzing survey data, new techniques might be developed, and new systematics might be found. These developments might require new mocks that meet newly defined requirements. Further, fast mock generation allows the creation of various plausible realizations, allowing one to marginalize over uncertain physical processes.

Many techniques have been developed over the past two decades to generate synthetic catalogs \citep[see e.g.][for a review]{risaawesomepaper}. Ideally, one would want to simulate galaxies directly from numerical solutions of coupled dark matter and baryon evolution to generate a realistic galaxy catalog. Unfortunately, while significant progress has been made over the past two decades, this method is still too computationally demanding to produce synthetic galaxy catalogs larger than the volume observed by galaxy surveys \citep[see e.g.][ for a review]{2020NatRP...2...42V}.
On the other hand, several practical alternative methods have been developed to simulate galaxy formation processes using phenomenological models. Ordered from least computationally demanding to most computationally demanding, these alternative methods include: 
\begin{enumerate}
    \item the halo occupation model \citep[HOD, ][]{2002ApJ...575..587B, zhenghod}, where one adopts phenomenological models to describe the statistical relations of galaxy properties and properties of the largest dark matter halos hosting these galaxies;
    \item the subhalo abundance matching model \citep[SHAM, ][]{2004ApJ...609...35K, 2006ApJ...647..201C}, where one relates galaxy properties to subhalo properties via simple rankings;
    \item semi-analytic models \citep[SAMs, see e.g.][ for reviews]{2006RPPh...69.3101B, 2015ARA&A..53...51S}, where one simulates galaxy formation physics using analytical prescriptions and integrates galaxy properties through halo merger histories. 
\end{enumerate} 
Combinations of these alternative methods have led to a blossoming of synthetic catalogs that meet the aforementioned requirements for galaxy survey cosmology analyses \citep[e.g.][]{Mice, euclid, JoeBuzzard, cosmodc2}. The most direct predecessor of this work is the Buzzard simulations \citep{JoeBuzzard}, which combine subhalo abundance matching \citep{lehmann}  and low-resolution large volume N-body simulations using a machine learning-based technique  \citep[\textsc{Addgals}, ][]{addgals}. The Buzzard simulations produce realistic galaxy properties, allowing one to run target selections (such as \redmagic{} galaxy selections, \citealt{Redmagic}) on the simulations in the same way as survey data. Further, the simulations are relatively computationally inexpensive, making it possible to generate large numbers of realizations of survey data. Because of these features, the Buzzard simulations have facilitated end-to-end validations of cosmological analysis pipelines from main galaxy catalogs to cosmological constraints (e.g. \citealt{Niall}; \citealt*{4x2pt1}; \citealt{y3buzzard}; \citealt{2022JCAP...02..007W}; \citealt{2022JCAP...07..041C}). 

While the Buzzard simulations and other catalogs built with the {\sc Addgals} \footnote{We note that there are two existing versions of Buzzard: the Buzzard Flock containing $18$ DESY1 realizations, and Buzzard v2.0, containing $18$ DESY3 realizations. These catalogs are based on 
$18$ realizations of the Chinchilla N-body simulations, each of which is generated from a different random seed. In the remainder of this paper, we use "the Buzzard simulation" to refer to one of the Buzzard v2.0 mocks that shares the same N-body simulation as the Cardinal.} have facilitated analyses in multiple cosmological surveys \citep[see][and references therein]{addgals}, the galaxy clustering on scales less than $1\,h^{-1}\rm{Mpc}$ in this simulation is smaller than the SDSS measurements up to $50$ percent. This suppressed galaxy clustering significantly impacts the properties of optically selected clusters, where one relies on the overdensity of red-sequence galaxies at $<1\,h^{-1}\rm{Mpc}$ scales to identify galaxy clusters. Specifically, the number of \redmapper{} galaxy clusters at a given richness ($\lambda$) in the Buzzard simulation is a factor of three to four smaller than the observed value in the Dark Energy Survey Year 1 data \citep{JoeBuzzard, DES_cluster_cosmology}. The lack of \redmapper{}  cluster problem is not unique in the Buzzard simulation. CosmoDC2 \citep{cosmodc2}, the only alternative simulation that currently has been tested with its \redmapper{} cluster properties, also underpredicts the richness of optically selected clusters unless one artificially boosts red-sequence galaxies in cluster environments. While the additional boosting solves the lack of clusters problem \citep{cosmodc2}, it creates discontinuities between galaxy properties in cluster environments and in the field.  

The unrealistic galaxy population in cluster environments is a critical limitation for studies that use these simulations to quantify the performance of optically selected clusters (\citealt{2019MNRAS.487.2900S, DES_cluster_cosmology, 2021MNRAS.505...33M}; \citealt*{4x2pt1}; \citealt{Heidiselection, 2022arXiv220208211Z}). This shortcoming can be partly mitigated by abundance matching: one compares the Nth richest clusters in the data to those in the simulations. The abundance matching technique makes the comparison of simulations and data sensitive only to the rank of richness instead of its absolute value, thereby reducing the problem of mismatched galaxy abundances in clusters. However, these effects cannot be calibrated reliably from simulations in which cluster richnesses are a factor of two lower than observed values.

The exact reason for this lack of cluster galaxies in the Buzzard simulation was previously unknown. \cite{JoeBuzzard} and \cite{addgals} hypothesize that it arises from artificial disruptions of subhalos that experience close pericentric passages \citep{2018MNRAS.475.4066V}. Following this line, \cite{campaper} developed a new SHAM model that includes orphan galaxies to address this problem. However, while the SHAM model with orphan galaxies can fit the galaxy clustering measured in SDSS in several stellar mass bins, no model was found to fit all three stellar mass bins considered in that work simultaneously. Further, \cite{campaper} also found that the color assignment model used in \cite{addgals} and \cite{y3buzzard} can lead to an underestimation of red galaxy clustering in the lowest stellar mass bins ($\rm{log} M^* \in [9.8, 10.2]\ h^{-1}M_\odot$). Because red galaxies more likely reside in cluster environments, this reduced red galaxy clustering can also lead to a lack of cluster galaxies. 

\begin{figure*}
    \centering
    \includegraphics[width=1.0\textwidth]{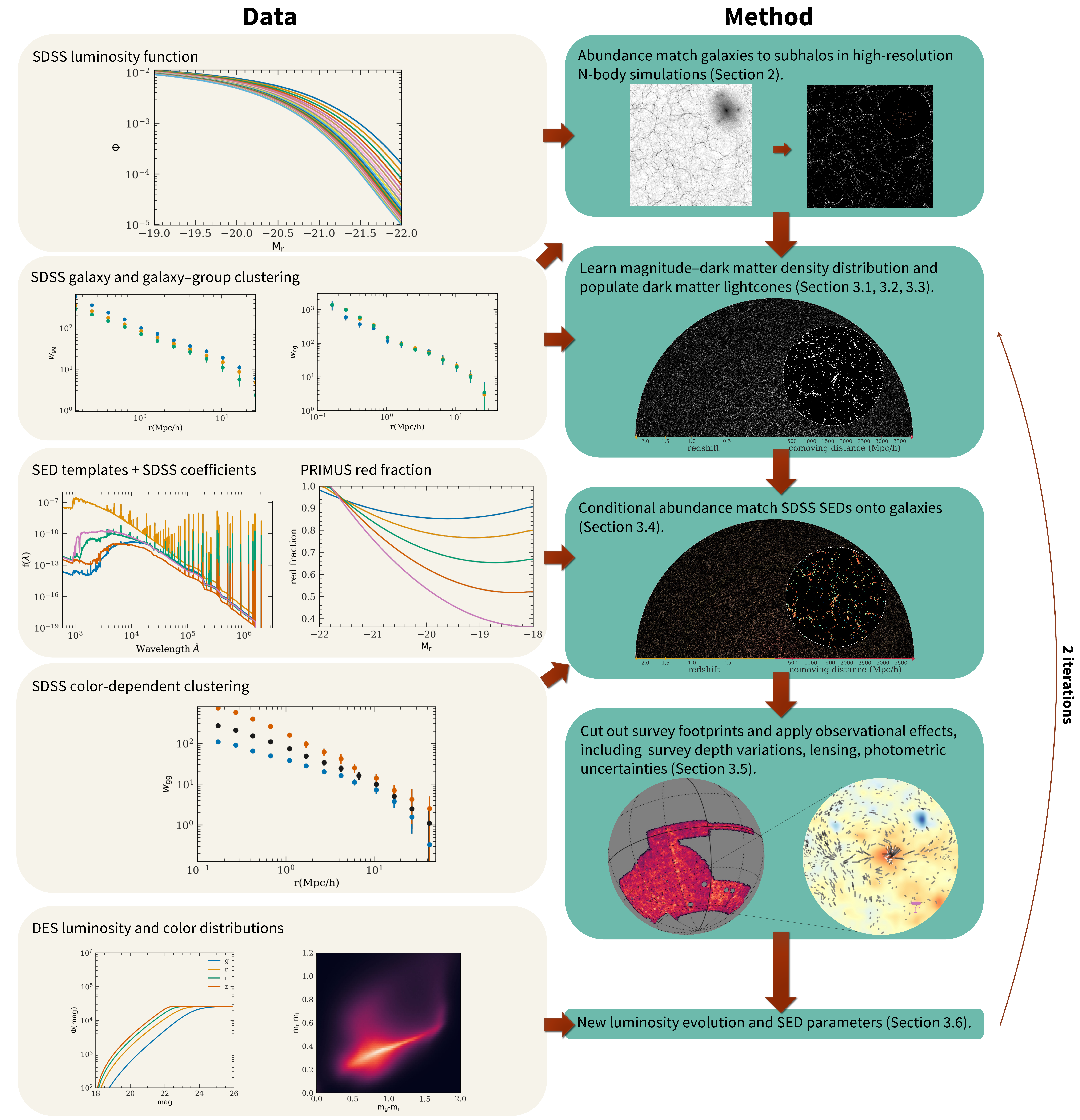}
    \caption{Flowchart of the algorithm. The left column shows the observational inputs, and the right column shows the algorithm. The algorithm can be broadly categorized into five steps. First, we develop an extended Subhalo Abundance Matching (SHAM) model to populate galaxies on N-body simulations with resolved structure (section \ref{sec:SHAM}). Second, we measure the statistical relations of galaxies' luminosities and local density tracer $R_\delta$ in the SHAM galaxy catalogs. We then paint luminosity onto particles of a lightcone simulation using the measured relation (section \ref{sec:luminosity}). Third, we use conditional abundance matching techniques to assign colors to galaxies in the mock using observed galaxy colors in SDSS (section \ref{sec:paintcolor}). Fourth, we apply observational effects in the mocks, including lensing, photometric uncertainties, and varying survey depths (section \ref{sec:observationeffect}). Finally, we use DES's photometric galaxy catalog to fine-tune the galaxy colors (section \ref{sec:camcolor}).}
    \label{fig:cardinalflow}
\end{figure*}

In this paper, we solve the lack of cluster galaxies problem in the Buzzard simulation by quantifying and addressing both contributing factors: (1) artificial subhalo disruption in the SHAM model and (2) the color assignment model. Our new model can simultaneously fit galaxy clustering and group--galaxy cross-correlations measured at three luminosity thresholds and also fit color-dependent galaxy clustering. We propagate this model through the \textsc{Addgals} algorithm \citep{addgals} and generate the Cardinal simulations. Figure \ref{fig:cardinalflow} shows the flowchart that summarizes the key steps of generating Cardinal. A list of improvements from Buzzard v2.0 to Cardinal is also presented in appendix \ref{app:improvement}. Finally, we compare properties of \redmagic{} galaxies and \redmapper{} clusters in Cardinal and DES-Y3 data \citep{Y3gold} and find excellent agreement. %

This paper is organized as follows. In section \ref{sec:SHAM}, we detail the construction of the new SHAM models that Cardinal is based on. In section \ref{sec:addgals}, we detail the steps of generating Cardinal using the new SHAM model. Specifically, the improved color assignment method is presented in section \ref{sec:paintcolor}. In section \ref{sec:camcolor}, we address the remaining problem in our color assignment models, including the lack of redshift evolution in training spectra and the inadequacy of summarizing colors using current SED templates. In section \ref{sec:compare}, we compare the properties of \redmagic{} galaxies and \redmapper{} clusters in Cardinal and DES-Y3 data. Finally, we conclude in section \ref{sec:conclu} with a discussion on future improvements.

\section{A SHAM-based galaxy--halo connection model}
\label{sec:SHAM}
We use a modified subhalo abundance matching (SHAM) algorithm to construct the training data for the galaxy--halo connection model. We describe the data and simulations used to construct the SHAM model below. 
\subsection{Calibrating Data}
\label{sec:measurementsham}
We use the NYU Value-Added Galaxy Catalog (VAGC, \citealp{VAGC})  constructed from SDSS DR7 \citep{2009ApJS..182..543A} main galaxy catalog to constrain the parameters in the SHAM model. We consider three volume-limited galaxy samples: $\Mr<-19$, $\Mr<-20$, and $\Mr<-21$ with $0.026<z<0.067$, $0.026<z<0.106$, and $0.026<z<0.106$ respectively. We limit our analysis to the north galactic cap (NGC) to avoid modeling differences in target selection between north and south galactic caps. From these three volume-limited samples, we measure the projected correlation function given by 
\begin{eqnarray}
\label{eq:corr2}
w_{\rm{gg}}(r) = 2\int_{0}^{\pi_{\rm{max}}}  \xi(r, \pi) \,d\pi, 
\end{eqnarray}
where $\pi$ is the line-of-sight distance between pairs, $r$ is the distance between pairs perpendicular to the line-of-sight, and $\pi_{\rm{max}}$ is $40\,h^{-1}\rm{Mpc}$. We measure $ \xi(r, \pi)$ using the Landy--Szalay estimator \citep{1993ApJ...412...64L}, with 12 logarithmically spaced bins between $r=0.13\,h^{-1}\rm{Mpc}$ to $r=32.6\,h^{-1}\rm{Mpc}$ and 40 linearly spaced bins in $\pi$. The smallest scale is chosen to avoid systematics caused by the size of SDSS fibers. 

In addition to galaxy clustering ($w_{\rm{gg}}$), we also use the cross-correlation between galaxy groups and galaxies ($w_{\rm{cg}}$) to constrain the SHAM model. \cite{BW2002} suggest that the group multiplicity function provides complementary information of galaxy--halo connections relative to galaxy clustering (see also \citealt{2018MNRAS.478.1042S}). Given the number densities of galaxies, group multiplicity functions are simply integrations of galaxy--galaxy group cross-correlations across spatial separations. Therefore, we include the galaxy--galaxy group cross-correlations to better constrain SHAM parameters important to galaxy occupations in cluster environments. We first measure galaxy groups in VAGC catalogs using the Self-Calibrated Galaxy Group Finder \citep{2020arXiv200712200T, 2021ApJ...923..154T}. We restrict the measurement to galaxies with $0.026<z<0.067$ and $\Mr<-19$ to ensure the completeness of galaxies. We remove all color information used in the  Self-Calibrated Galaxy Group Finder because this information does not exist in the mock galaxy catalogs constructed from the SHAM model. This might degrade the performance of the group finders, but it allows apples-to-apples comparison between the measurements in data and mocks. With the group catalogs, we select groups with mass greater than $5\times10^{13}\, h^{-1}\Msun$, the mass range of $\lambda>20$ \redmapper{} clusters. We then cross-correlate the centers of these groups with galaxies at $0.026<z<0.067$ and $\Mr>-19$, $\Mr>-20$, and $\Mr>-21$, using equation \ref{eq:corr2}. 

We measure the covariance matrix of the data using the jackknife resampling technique. We first employ a Kmeans algorithm implemented in \textsc{treecorr}\citep{treecorr} on random points to separate the sky into 128 uniform patches. We then perform the measurements by excluding one patch at a time. The covariance matrix is then estimated as, 
\begin{eqnarray}
\rm{cov (d_i, d_j)}= \frac{N-1}{N} \sum_{k=1}^{k=N}\left (d_{i,k}-\langle x_i\rangle\right)\left(d_{j,k}-\langle x_j\rangle\right),
\end{eqnarray}
where $d_i$ is the $i^{th}$ element of the data vector, N is $128$, $d_{i,k}$ is the $i^{th}$ element of the data from measurements that exclude the $k^{th}$ patch, and $\langle x_i\rangle$ is the mean of the element over $N=128$ patches. Jackknife estimations introduce noise in the covariance matrix, which leads to biases in the inversion of the covariance. Thus, one has to regularize the covariance matrix before inverting it. Here, we adopt an approach similar to \cite{universemachine}. In short, we perform an eigenvalue decomposition of the jackknife-estimated covariance matrix to obtain eigenvalues $D_n$ and the associated eigenvectors $\vect{v}_n$. Due to various possible sources of noise (such as variations of sky backgrounds, variations of fiber assignment efficiency, and systematics in galaxy photometry), we do not expect the error estimated to be better than 10 percent. We therefore rank order the eigenvalues and find the eigenvalues whose square rooted values are below 0.1 of the data projected onto the eigenspace. We then replace those eigenvalues with $0.1$  of the data projected onto the eigenspace and multiply the new eigenvalues with the eigenvectors to form a regularized covariance matrix. 
\subsection{Simulations and models}

\begin{table*}
\begin{tabular}{l l l l}
\hline
\hline
Name  & $L_{\rm box}$  & $M_{\rm part}$ & $\epsilon_{\rm{plummer}}$ \\
\hline

\multicolumn{4}{c}{\textbf{SHAM galaxy catalogs generation and tests}} \\ 
\Tbox{} &  $400\,h^{-1}\rm{Mpc}$ &$5.9\times 10^8\, h^{-1}M_\odot$& $5.5\,h^{-1}\kpc$\\
SMDPL & $400\,h^{-1}\rm{Mpc}$ &$9.6\times 10^7\, h^{-1}M_\odot$&$1.5\,h^{-1}\kpc$\\
\hline
\multicolumn{4}{c}{\textbf{Lightcone mock generation}}\\
L1 $(z<0.315)$&  $1.05\,h^{-1}\rm{Gpc}$ &$3.3\times 10^{10}\,h^{-1}M_\odot$& $20\,h^{-1}\kpc$\\
L2 $(0.315<z<0.955)$& 2.6 $h^{-1}\rm{Gpc}$ &$1.6\times 10^{11} h^{-1}M_\odot$& $35\,h^{-1}\kpc$\\
L3 $(0.955<z<2.35)$&  $4\,h^{-1}\rm{Gpc}$ &$5.9\times 10^{11}\,h^{-1}M_\odot$& $53\,h^{-1}\kpc$\\
\hline 
\end{tabular}
\centering
\caption{\label{tab:simsummary} Descriptions of the simulations used for this analysis, including the size of the box ($L_{\rm box}$), the mass resolution of the particles ($M_{\rm{part}}$), and the force softening length ($\epsilon_{\rm{plummer}}$).}
\end{table*}

\subsubsection{Simulations}
To generate the training model, we use the \Tbox{} dark matter simulation, which have volume  $(400\, h^{-1}\rm{Mpc})^3$ with particle resolution $5.9\times10^8 h^{-1}M_\odot$. The simulation is generated using \textsc{L-GADGET2} \citep{gadget} with a $\Lambda$CDM cosmology that has $\Omegam=0.286, \Omegab=0.047, \sigma_8=0.82, n_{\rm{s}}=0.96, h=0.7$, and three massless neutrino species with $\rm{N}_{eff}=3.046$. Halo finding is performed using \textsc{Rockstar} \citep{Rockstar}, and merger trees were generated using \textsc{Consistent Tree} \citep{consistentree}. We refer to \cite{addgals} for details of this simulation. Given that our galaxy samples are selected at low redshift, we use the snapshot corresponding to $z=0$ for the work described in this section.

Evidence has shown that subhalos in dark matter-only simulations are susceptible to physical and unphysical disruptions \citep{Klypin1999, 2008ApJ...678....6W, 2018MNRAS.475.4066V}. We account for this effect using a prescription similar to \cite{universemachine}, which adds disrupted subhalos back to the simulation. We first identify subhalos that are no longer detected by \textsc{Rockstar} in each snapshot. Then, we locate the host halos that contain these subhalos within their virial radius and use the semi-analytic model from \cite{universemachine} to simulate the evolution of the subhalos' position, mass, and maximal circular velocity. These procedures produce a catalog containing standard halos that can be found and tracked using \textsc{Rockstar} and disrupted subhalos (also known as orphans) that have properties calculated semi-analytically. 

\subsubsection{Models}
\label{sec:SHAMmodel}
Using the subhalo abundance matching technique, we populate galaxies on subhalos (tracked subhalos and orphans). Here, we first describe the general concept of this technique and then describe the extensions we develop in this paper. In the most basic form, subhalo abundance matching assigns each subhalo with a luminosity by enforcing the relation, 
\begin{eqnarray}
\label{eq:basicsham}
n(L>x) = n(X_h>y), 
\end{eqnarray}
where $n(L>x)$ represents the number density of galaxies with luminosity greater than $x$ and $n(X_h>y)$ represents the number density of subhalos with properties $X_h$ greater than $y$. Following \cite{lehmann}, we adopt $X_h$ as $v_\alpha$ defined as,  
\begin{eqnarray}
\label{eq:valpha}
v_\alpha = v_{\rm{vir}} \left(\frac{v_{\rm{max}}}{v_{\rm{vir}}}\right)^\alpha, 
\end{eqnarray}
where $v_{\rm{vir}}$ is the virial velocity of the halos, $v_{\rm{max}}$ is the maximum circular velocity, and $\alpha$ is a free parameter. These quantities are evaluated at the epoch when the halo's mass is at the maximum to avoid complicated physics subhalos experienced when falling into a big halo. The free parameter $\alpha$ allows additional flexibility for the subhalo abundance matching model. In particular, $\left(\frac{v_{\rm{max}}}{v_{\rm{vir}}}\right)$ can be viewed as a proxy of halo concentration. The free parameter $\alpha$ controls the dependence of galaxy luminosity on halo concentrations. 

The $n(L>x)$ in equation \ref{eq:basicsham} is estimated by fitting a modified double-Schechter function with a Gaussian tail to the luminosity function measured in SDSS DR7 galaxy catalogs. For details of constructing $n(L>x)$, we refer the readers to appendix E.1.~of \cite{JoeBuzzard}. We cannot directly apply the estimated $n(L>x)$ in equation \ref{eq:basicsham} because the relations between galaxy luminosity and halo properties are stochastic. This stochasticity comes from observational uncertainties, complicated astrophysical processes that affect galaxy evolution within dark matter halos, and additional halo properties that are correlated with galaxy luminosities. One can model these complicated processes as \begin{eqnarray}
\label{eq:conv}
n(L>x) = \int P(L|L^\prime) n(L^\prime>x) \, dL^\prime, 
\end{eqnarray}
where $n(L>x)$ is the measured luminosity function. The intrinsic galaxy luminosity $L^\prime$ in the above expression is determined solely by the selected halo properties $X_h$ by comparing the number density of galaxies with luminosity above $L^\prime$, $n(x>L^\prime)$, to the number density of halos with the selected property above $X_h$, $n(x>X_h)$.  In most of the subhalo abundance matching work \citep[e.g.][]{reddick2013, lehmann, campaper, Contreras2021}, one parametrizes $P(L|L^\prime)$ as a log-normal distribution with mean $L=L^\prime$ and scatter $\sigma$. With this assumption, equation \ref{eq:conv} can be viewed as a convolution problem and can be solved using standard deconvolution algorithms. However, numerous lines of evidence based on observations and simulations have shown that the scatter in $P(L|L^\prime)$ might depend on halo mass \citep[see][for review]{risaawesomepaper}. \cite{risaawesomepaper} shows that the value of this scatter is a constant at the high mass end and increases at the low mass end. We therefore parametrize $P(L|L^\prime)$ as a log-normal distribution with scatter, 
\begin{equation}
\label{eq:shamscatter}
    \sigma(L^\prime) = \rm{max}(\sigma_{v}+\sigma_{vs}(-2.5 \rm{log} (L^\prime)+\sigma_{vp}), \sigma_{v}),
\end{equation}
where $\sigma_{v}$, $\sigma_{vs}$, and $\sigma_{vp}$ are free parameters, and $\rm{max}(A,B)$ represents the maximum of A and B. 
Further, observational constraints based on galaxy groups \citep{2021ApJ...923..154T} and satellite kinematics  
\citep{2019MNRAS.487.3112L} constrain this scatter to be less than $0.4$ with 95 percent confidence over the range of luminosities considered here. We, therefore, set an additional prior on $\sigma(L^\prime)$ to be below $0.4$. With a functional form of $P(L|L^\prime)$, we then estimate $n(L^\prime>x)$ based on the measured $n(L>x)$ using the algorithm presented in appendix \ref{app:SHAMscatter}.
\begin{table*}
\begin{tabular}{l l l l l}
\hline
\hline
Parameter  & Prior  & Best-fit value& Description & Relevant equations \\
\hline
\multicolumn{4}{c}{\textbf{Halo properties for SHAM}} \\ 
$\alpha$ & flat(0.0, 1.0) & $0.16^{+0.12}_{-0.26}$ &Halo properties used in SHAM & Equation \ref{eq:valpha}\\
\hline
\multicolumn{4}{c}{\textbf{Scatter in luminosity halo mass relation}} \\ 
$\sigma_v$  & flat(0.1, 0.25) & $0.20\pm 0.017$& Scatter in luminosity halo mass relation at $\Mr=\sigma_{vp}$ &
\multirow{3}{*}{Equation  \ref{eq:shamscatter}} 
\\
$\sigma_{cs}$ & flat(-0.01, 0.3) & $0.17^{+0.05}_{-0.06}$ & Luminosity dependence of the scatter  \\
$\sigma_{vp}$ & flat(-20.5, -20.0)& $-20.39^{+0.07}_{-0.16}$ & Pivot point of the scatter\\
\hline
\multicolumn{4}{c}{\textbf{Subhalo disruption}} \\ 
$T_l$ & flat(0.01, 1.0) & $0.01\pm 0.082 $ &Asymptotic value of $T_{{\rm{dis}}}$ at $v_\alpha=0$ & \multirow{4}{*}{Equations  \ref{eq:SHAM_disrupt1} and \ref{eq:SHAM_disrupt2}} \\
$T_h$& flat(0.2, 1.6) & $0.78^{+0.17}_{-0.27} $ &Asymptotic value of $T_{{\rm{dis}}}$ at $v_\alpha=\infty$\\
$v_m$& flat(1.9, 3.3) & $2.00^{+0.23}_{-0.16}$ & Value of log $v_\alpha (km/s)$ when $T_{{\rm{dis}}}=0.5(T_l+T_h)$ \\
$\sigma_d$& flat(0.1, 1.0) & $0.47^{+0.17}_{-0.33} $ & Steepness of the transition\\

\hline 
\end{tabular}
\centering
\caption{\label{tab:shamparam} Parameters and priors of the extended SHAM model described in section \ref{sec:SHAMmodel}}
\end{table*}

When subhalos (tracked subhalos and orphans) fall onto big halos, they might be tidally disrupted, the gas within them could be stripped, and the galaxy living inside them might be destroyed. Thus, we must allow additional flexibility in the model to capture these physical processes. Previous work has been done on mitigating these issues by applying cuts in halo properties on tracked subhalos \citep{reddick2013}, orphans \citep{universemachine, campaper}, or both \citep{Contreras2021}. In this paper, we treat tracked subhalos and orphans equally. This approach has the benefit that the result depends less on the resolution of the simulations. We mitigate the issues of physical disruptions using a procedure similar to \cite{universemachine} and \cite{campaper}. For each subhalo, we compare the maximum circular velocity at the current time ($v_{\rm{max}, now}$) to the maximum circular velocity at the time when the halo mass is at the maximum  ($v_{\rm{max}, M_{\rm{peak}}}$). When  $v_{\rm{max}, now}$ of a subhalo is much smaller than  $v_{\rm{max}, M_{\rm{peak}}}$, the subhalo is likely tidally stripped and is less likely to host a galaxy. We therefore set a threshold of the ratio between $v_{\rm{max}, now}$ and $v_{\rm{max}, M_{\rm{peak}}}$, below which the subhalos do not host galaxies. Further, because galaxies with different luminosities will have different time scales of dynamical frictions and resilience to tidal disruptions, we allow this threshold to depend  on $v_\alpha$, the combination of halo properties used in subhalo abundance matching. With these insights, our parametrization of the probability that a subhalo is physically disrupted reads
\begin{eqnarray}
\label{eq:SHAM_disrupt1}
&P(\rm{disrupt}) = \Theta\left(T_{{\rm{dis}}}(v_\alpha)-\frac{v_{\rm{max}, now}}{v_{\rm{max}, M_{\rm{peak}}}}\right), \\
\label{eq:SHAM_disrupt2}
&T_{{\rm{dis}}}(v_\alpha)= T_l+(T_h-T_l)\left(0.5+0.5{{\rm{erf}}}\left(\frac{\log_{10} v_\alpha-v_m}{\sqrt{2}\sigma_d}\right)\right)
\end{eqnarray}
where $\Theta$ is the Heaviside step function. In the above expression, $T_l$ and $T_h$ are free parameters, with $T_{{\rm{dis}}}(v_\alpha)$ interpolating between asymptotic behaviors  at high and low $v_\alpha$ ends, $v_m$ governs where the transition from $T_l$ to $T_h$ occurs, and $\sigma_d$ controls how steep the transition is.  Although Equations \ref{eq:SHAM_disrupt1} and \ref{eq:SHAM_disrupt2} may appear complex, the underlying physical behavior is simple. For halos with large $v_\alpha$, $T_h$ sets the threshold of $v_{\rm{max}, now}/v_{\rm{max}, M_{\rm{peak}}}$ to determine whether the halos are tidally disrupted. Conversely, for halos with small $v_\alpha$, $T_l$ determines the threshold. Equation \ref{eq:SHAM_disrupt2} ensures a smooth transition between small and large $v_\alpha$ values, and two additional parameters control the location and slope of the transition. 

To summarize, our extended SHAM model has $8$ free parameters, whose values are given in table \ref{tab:shamparam}. Given these parameters, we can generate a galaxy catalog using the following procedure. We first employ equation \ref{eq:valpha} to calculate $v_\alpha$ of each subhalo in a halo catalog, including tracked subhalos and halos identified by  \textsc{Rockstar} and orphan subhalos generated using a semi-analytic model from \cite{universemachine}. We then remove subhalos according to equations \ref{eq:SHAM_disrupt1} and \ref{eq:SHAM_disrupt2}. For each of the remaining subhalos, we populate a galaxy at the position of the subhalo with luminosity determined by equation \ref{eq:conv}.

\begin{figure*}
    \centering
    \includegraphics[width=0.9\textwidth]{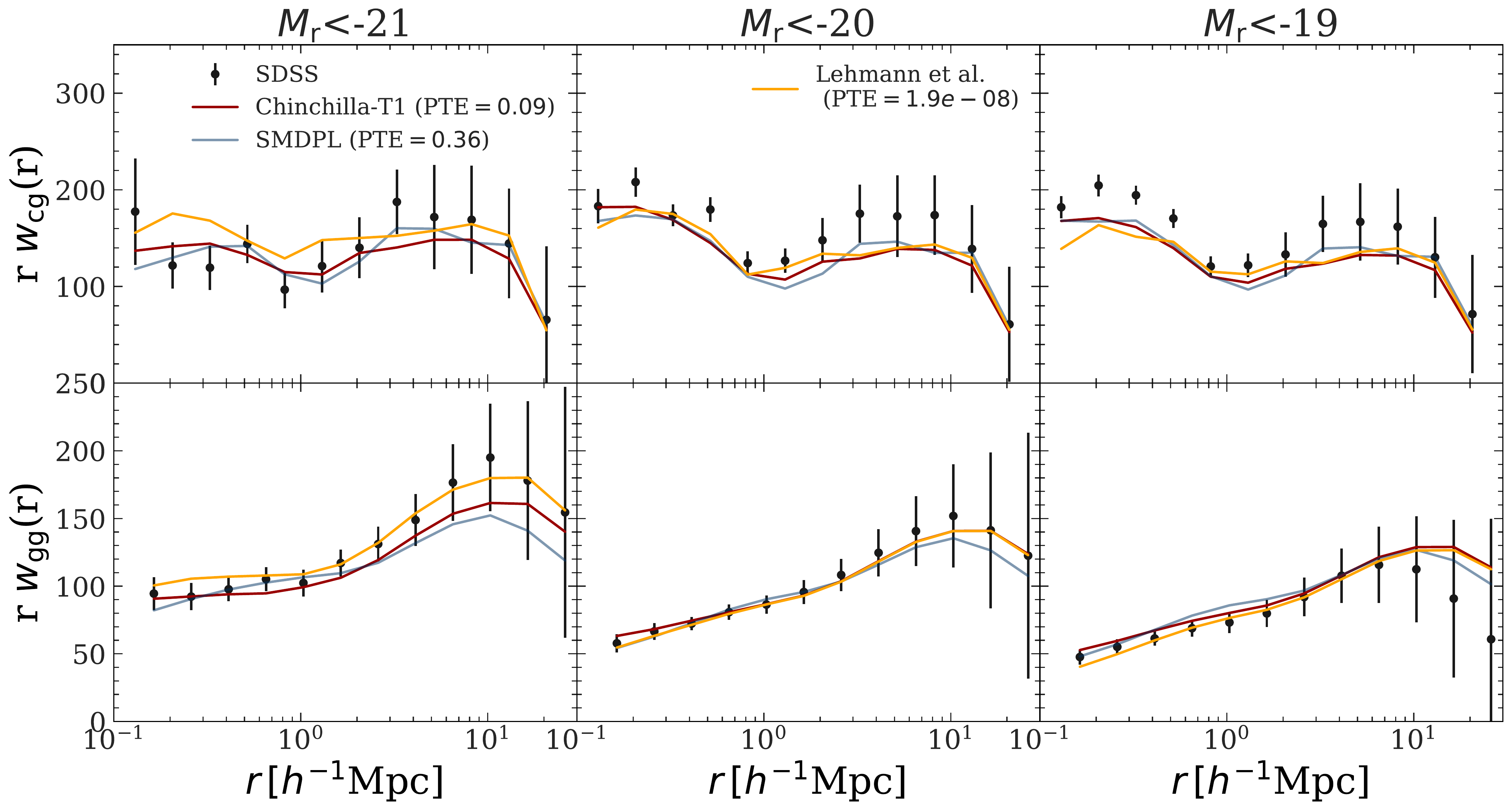}
    \caption{Comparison of the data and the best-fit SHAM model. Different columns correspond to different magnitude cuts in galaxies. The top row shows galaxy--galaxy group cross-correlations, and the bottom shows galaxy auto-correlations. Black dots are measurements using SDSS data, while error bars show the $1\sigma$ uncertainties. Red lines show the best fit SHAM models based on the \Tbox{} simulation. Shaded regions show one-sigma uncertainties. The blue line shows the best-fit SHAM models based on the SMDPL simulations. For comparison, orange lines show predictions of ~\protect\cite{lehmann}'s best-fit model, which does not consider mass-dependent scatter in luminosity--halo mass relations and orphan subhalos. PTE value in the legend corresponds to the Probability to Exceed.}
    \label{fig:SHAM_data}
\end{figure*}

\begin{figure*}
    \centering
    \includegraphics[width=0.9\textwidth]{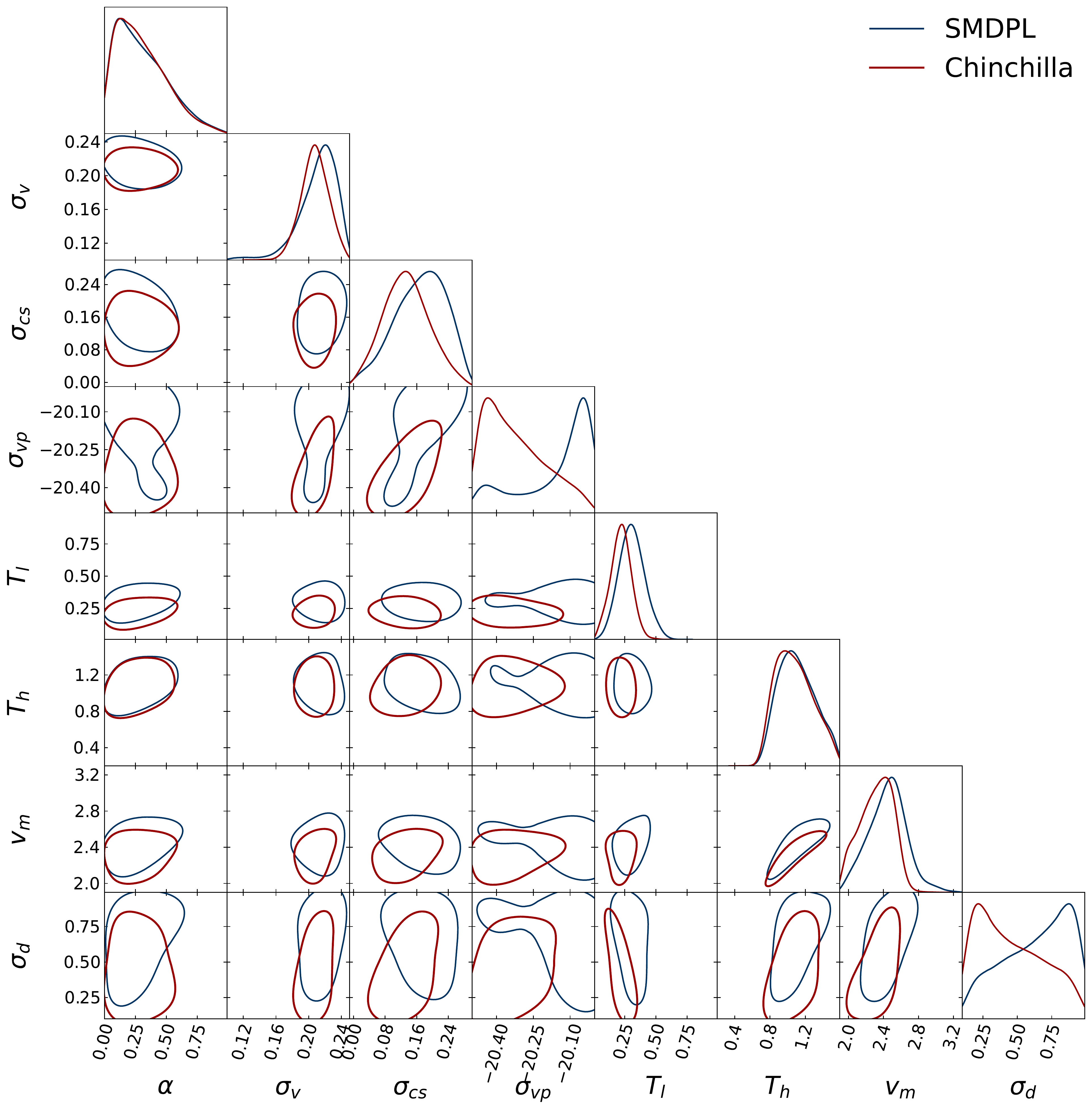}
    \label{fig:SHAM_contour}
    \caption{Posteriors of the extended SHAM model given SDSS galaxy clustering and group galaxy cross-correlations. Contours show $68$ percent confidence region of the posteriors given $w_{\rm{gg}}$ and $w_{\rm{cg}}$. Red lines show constraints based on \Tbox{}, and blue lines show constraints based on SMDPL. Details of these simulations can be found in table \ref{tab:simsummary}. Descriptions of these parameters are given in table \ref{tab:shamparam}.}
\end{figure*}

\subsubsection{Measurements in simulations}
We select galaxies in simulations using the same luminosity cut as the data. To account for the effect of redshift-space distortions, we first transform the coordinates of halos along the line of sight using the velocity of the halos. We then apply the periodic boundary condition on the transformed coordinates and calculate $w_{\rm{gg}}(r)$ using the same radial binning as the data. Finally, the $w_{\rm{gg}}(r)$ is estimated using the natural estimator, with an analytic calculation of the random--random pairs. To minimize cosmic variance, we repeat the above process by choosing line-of-sight direction along $x$, $y$, and $z$ axes of the box and taking the average of the measurements. Further, we minimize the stochasticities due to the Monte Carlo nature of the SHAM model by repopulating the simulations $10$ times with the same parameters. 

Because the simulation box size is small, cosmic variance cannot be ignored in the total error budget. We estimate this cosmic variance using the jackknife technique similar to the procedure described in section \ref{sec:measurementsham}. The main difference is that there is no survey mask in the simulations, so we generate $125$ jackknife subsamples by equally partitioning the box. Further, the jackknife subsampling breaks the periodic boundary conditions. One has to consider this when analytically estimating the random--random terms in the natural estimator. Here, we adopt the approach described in \cite{2021ApJ...921...59H} to estimate the random--random terms for each jackknife subsample. 
We estimate the covariance matrix from this cosmic variance using the best-fit parameters to the data presented in section \ref{sec:SHAMresult}. This covariance matrix is then added to the observed covariance matrix while performing the likelihood inferences. 

Regarding the galaxy group samples, we run the Self-Calibrated Galaxy Group Finder \citep{2020arXiv200712200T, 2021ApJ...923..154T} on the simulated galaxies with the same setting as run on SDSS data. As pointed out before, in both runs on simulations and data, we remove steps that use color information to ensure an apples-to-apples comparison.

\subsection{SHAM result and discussions}
\label{sec:SHAMresult}
We fit the model described in section \ref{sec:SHAMmodel} to the data described in section \ref{sec:measurementsham} assuming a Gaussian likelihood with the priors described in table \ref{tab:shamparam}. The challenge is that the fitting is hugely time-consuming because each step of the likelihood inference requires populating halos $10$ times and computing the correlation function $30$ times. We tackle this challenge by first building an emulator of the SHAM model and using it for likelihood inferences. We detail the procedure of constructing this emulator in appendix \ref{app:surrogate}. Then, the likelihood inference are performed using an ensemble slicing sampling method implemented in \textsc{zeus} \citep{karamanis2021zeus, karamanis2020ensemble}. 

Figure \ref{fig:SHAM_data} compares the best-fit model and the data. The minimum $\chi^2$ is $79.4$ with the degree of freedom $64$, yielding a Probability-to-Exceed (PTE) $0.09$. Table \ref{tab:shamparam} shows the best-fit parameters. Overall, our model describes the data well, but a small difference can be seen at $r<0.3\, h^{-1}\rm{Mpc}$ of $w_{\rm{cg}}$ for the faintest galaxy sample. We want to determine whether the difference we have observed is due to limitations of the N-body simulations, such as force softening or finite resolution. To do this, we repeat the entire SHAM analysis using the Small MultiDark Planck (SMDPL) N-body simulation \citep{SMDPL}, which has a higher resolution, a different softening scale, and a different fiducial cosmology than the previous simulation. This difference persists, as shown in the blue line in figure \ref{fig:SHAM_data}. One possibility is that this difference comes from tensions between $w_{\rm{cg}}$ and $w_{\rm{gg}}$. Small-scale $w_{\rm{gg}}$ is dominated by the one-halo term, which has significant contributions from galaxies in high-mass halos. In figure \ref{fig:SHAM_data}, one can see that the small-scale $w_{\rm{gg}}$ for the faintest galaxy samples prefers a slightly smaller one-halo term than the model, while the small-scale $w_{\rm{cg}}$ prefers the opposite. This mild  tension might be related to the findings in \cite{2013MNRAS.433..659H}, where they find tensions between group multiplicity functions, which are an integrated version of $w_{\rm{cg}}$,  and galaxy clustering, under the assumption of the basic SHAM model. This tension is much smaller in our extended SHAM model, and the constraining power of the data cannot distinguish it from statistical fluctuations. We therefore leave further investigations to future work.

Figure \ref{fig:SHAM_contour} shows the posteriors of SHAM parameter constraints based on \Tbox{} (red) and SMDPL (blue). Most of the SHAM parameters based on \Tbox{} and SMDPL are consistent, indicating the robustness of the result to details of $N$-body simulations, including cosmology, resolution, and force softening. The only parameter that is slightly inconsistent is $\sigma_{vp}$. \Tbox{} (the lower resolution simulations) prefers a brighter (more negative) value  than SMDPL (the higher resolution simulations). One possible explanation is that SMDPL has more low-mass halos than \Tbox{}. For a given magnitude bin, the missing low-mass halos in \Tbox{} can be compensated by a larger scatter. Thus, \Tbox{} has a brighter pivot point for the scatter. In both \Tbox{} and SMDPL, the inclusion of orphan galaxies is strongly preferred. For the galaxy sample considered in this work, $\sim 20$ percent are orphan galaxies for \Tbox{} and $\sim15$ percent for SMDPL. Another interesting result is that both \Tbox{} and SMDPL prefer a varying scatter in the luminosity-$v_\alpha$ relation at the $1\sigma$ level. The positive value of $\sigma_{cs}$ indicates that the scatter increases for lower mass halos. This is consistent with results based on group finders \citep{2021ApJ...923..154T}, satellite kinematics \citep{2019MNRAS.487.3112L}, and galaxy clustering \citep{2018MNRAS.481.5470X}.

\section{Populating galaxies in low-resolution simulations}
\label{sec:addgals}
The SHAM model presented in section \ref{sec:SHAM} allows us to create high-fidelity galaxy catalogs based on high-resolution simulations. However, the high-resolution simulations typically have a volume much smaller than the volume accessible with current and upcoming galaxy surveys, making them insufficient to validate models with the required accuracy. In this section, we describe the formalism to transfer the knowledge learned in the SHAM catalogs to populate galaxies in large simulations with low resolution. This way, one can generate multiple realizations of mock galaxy catalogs with a modest computational expense. 
\subsection{Simulations}
We use lightcones with an area $10,313$ square degree constructed from the L1, L2, and L3 simulations detailed in table \ref{tab:simsummary}. These simulations are generated with the same cosmological parameters as \Tbox{}. Details of the lightcone construction were presented in appendix B1 of \cite{JoeBuzzard}.

\subsection{Targets}
We aim to generate mock that support sciences in large galaxy surveys, such as the Dark Energy Survey (DES) and Vera Rubin Observatories' Legacy Survey of Space and Time (LSST). The various science cases in these surveys place stringent constraints on the galaxy properties in the simulations. In this paper, we mainly focus on the two main samples in DES: \redmapper{} clusters \citep{Redmapper1,redmappersv} and \redmagic{} galaxies \citep{Redmagic}. As shown in \cite{JoeBuzzard}, these two samples place the most stringent constraints on the galaxy models using DES data. We provide a brief description of these two samples below. 

\subsubsection{\redmapper{} clusters}
\label{sec:redmapper}
In optical surveys, galaxy clusters appear to be spatial and redshift concentrations of red-sequence galaxies. The relatively tight color--absolute magnitude relations of red-sequence galaxies enable one to identify galaxy clusters using galaxies without redshift information. The primary algorithm used by Dark Energy Survey collaboration\citep{redmappersv} and the LSST Dark Energy Science Collaboration \citep{2022OJAp....5E...1K} to select clusters is \redmapper{} \citep{Redmapper1}, which employs a matched filter algorithm to select overdensities of red-sequence galaxies. %
Here, we briefly summarize the algorithm. 
First, the \redmapper{} algorithm uses spectroscopic data to construct a red sequence template empirically.  It then computes the redshift of each galaxy by matching its color to the template. Second, \redmapper{} identifies bright and red galaxies as cluster centers and determines the probability of each galaxy being a member ($p_{\rm{mem}}$) by comparing its spatial distribution, color, and luminosity to a model. Third, the algorithm removes clusters that have $p_{\rm{mem}}>0.5$ of another cluster and repeats the above processes. Finally, \redmapper{} assigns a richness value ($\lambda$) to each cluster, calculated by summing $p_{\rm{mem}}$ of each member galaxy. 
This richness value is used as a primary cluster mass proxy (\citealt{Tomclusterlensing, DES_cluster_cosmology}; \citealt*{4x2pt2}) in cosmological analyses given their expected tight relation to halo masses \citep{redmapper2, redmapper3}. \redmapper{} also provides the most probable redshift of each galaxy cluster ($z_\lambda$) based on the colors of its member galaxies. 

\subsubsection{\redmagic{} galaxies}
\label{sec:redmagic}
The red-sequence model derived from the \redmapper{} clusters can be used to select galaxy samples with excellent photometric redshift uncertainties ($\sigma_z\approx 2$ percent). To achieve this, one first constructs a color model using \redmapper{} member galaxies with high membership probabilities ($p_{\rm{mem}}$). The redshift of each galaxy can be estimated by maximizing the consistency of galaxy colors and color models. One can then select bright galaxies with colors consistent with the color model. The galaxy samples selected in this way are called \redmagic{} \citep{Redmagic} and are one of the primary lens samples in the Dark Energy Survey cosmology analyses \citep{DESY1KP,Y3kp}. Based on the consistency with the color model, each \redmagic{} galaxy is associated with a redshift probability distribution $p(z_{\rm{redmagic}})$.

\subsection{Painting galaxy luminosities onto dark matter particles}
\label{sec:luminosity}
\begin{figure}
    \centering
    \includegraphics[width=0.5\textwidth]{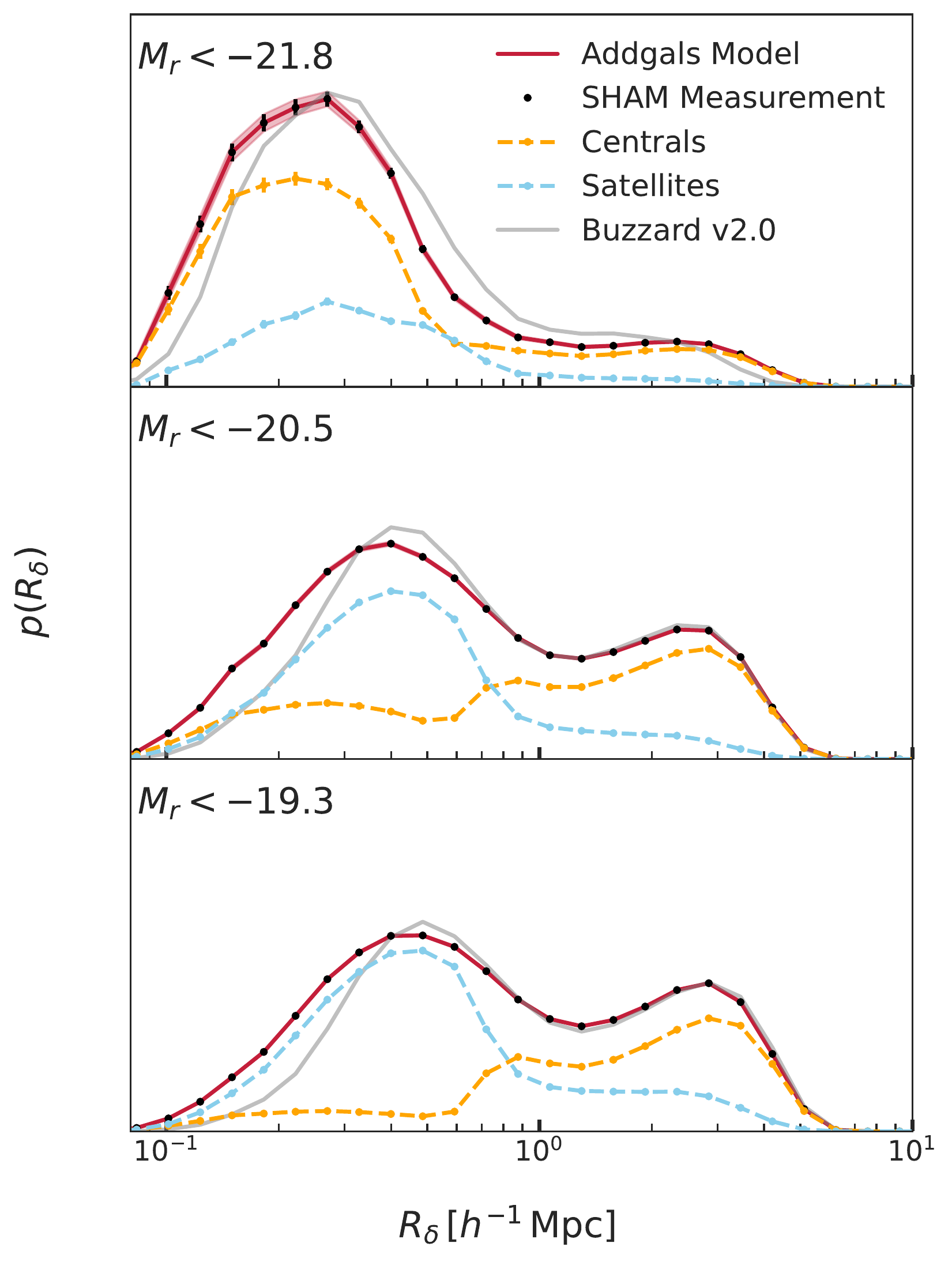}
    \caption{Distribution of 
 $R_\delta$, the radius that encloses a dark matter mass of $1.3\times 10^{13} \,h^{-1}M_\odot$,  at $z=0$. Each panel shows a magnitude cut, with the brightest samples on the top and the faintest samples on the bottom. Black dots show measurements in SHAM catalogs with error bars showing $1\sigma$ Poisson error. Red lines show measurements of $R_\delta$ using mock galaxies in lightcone simulations. We further break the $R_\delta$ distributions in mocks into centrals (blue) and satellites (orange). As a comparison, the previous version Buzzard v2.0 \citep{addgals} is shown as gray lines.}
    \label{fig:addgal_rdel}
\end{figure}

We use the \textsc{Addgals} algorithm to populate the dark matter lightcones presented in \cite{JoeBuzzard} and \cite{addgals} with galaxies. The algorithm is detailed in \cite{addgals}. Here, we briefly summarize the general formalism and highlight modifications. 

\textsc{Addgals} populates galaxies in low-resolution $N$-body simulations based on three distributions:
\begin{enumerate}
    \item The distribution of dark matter overdensities around galaxies given their absolute $r$-band magnitude $\Mr$ and redshift $z$, $P(R_\delta|\Mr<x, z)$. The local dark matter overdensity is estimated using $R_\delta$, distances to $k$ nearest dark matter particles such that the enclosed dark matter mass is $1.3\times 10^{13} \, h^{-1}M_{\odot}$, 
    \item   A galaxy luminosity function $\phi (\Mr, z)$, and
    \item  The distribution of central galaxy absolute magnitudes given their host halos' virial mass ($M_{\rm{vir}}$) and redshift, $P(\Mr|M_{\rm{vir}}, z)$. 
\end{enumerate}

In this work, the galaxy luminosity function is the same as the one used to create SHAM models at $z=0$. At $z>0$, the Schechter function's characteristic luminosity ($L^*$) is shifted based on a third-order polynomial. The parameters of this polynomial are constrained based on DES-Y1 data \citep{DESY1KP}. Details of constructing this luminosity function are given in \cite{JoeBuzzard}. 

Given the redshift-dependent luminosity function, we populate galaxies on each snapshot of \Tbox{} simulations using the best-fit SHAM model presented in section \ref{sec:SHAM}. We then use these galaxies to determine $P(\Mr|M_{\rm{vir}}, z)$ for central galaxies in resolved halos, defined as halos containing more than $200$ particles, and $P(R_\delta|\Mr<x, z)$ for satellite and central galaxies in unresolved halos.

To determine the distribution of central galaxy absolute magnitudes at fixed host halo virial mass $M_{\rm{vir}}$ and redshift, we assume that $P(\Mr|M_{\rm{vir}}, z)$ is a Gaussian distribution with a mass-dependent scatter. The mass dependency of the scatter is determined by fitting a straight line to the measured $\Mr$ scatter of $M_{\rm{vir}}>10^{13}\,h^{-1}M_\odot$ halos in the SHAM galaxy catalogs. The mean of the Gaussian distribution is given by 
\begin{eqnarray}
\langle \Mr (M_{{\rm{vir}}}) \rangle &=& A-2.5\left(a \log\left (\frac{M_{{\rm{vir}}}}{b}\right)\right)\\
&&-\frac{1}{c}\log\left (1+ \left(\frac{M_{{\rm{vir}}}}{b}\right)^{(c \times d)}\right),
\end{eqnarray}
where $A, a, b, c, d$ are free parameters determined at each snapshot of the SHAM galaxy catalogs. Using the best-fit function, we populate central galaxies on resolved halos in the low-resolution lightcone simulations. 

Similar to $P(\Mr|M_{\rm{vir}}, z)$, we determine $P(R_\delta|\Mr<x, z)$ in the SHAM galaxy catalogs at each snapshot. We model $P(R_\delta|\Mr<x, z)$ as a log-normal and normal distribution sums, given by
\begin{equation}
\label{eq:rdelta}
    P(R_\delta|\Mr<x, z) = (1-p) \frac{e^{-(\rm{ln}(R_\delta)-\mu_c)^2/2\sigma_c^2}}{R_\delta\sqrt{2\pi}\sigma_c}+p\frac{e^{-(R_\delta-\mu_f)^2/2\sigma_f^2}}{\sqrt{2\pi}\sigma_f}, 
\end{equation}
where $p, \mu_c, \mu_f, \sigma_c, \sigma_f$ are free parameters and are measured in each snapshot of the SHAM galaxy catalogs at grids of magnitude thresholds from $\Mr=-18$ to $\Mr=-24$. Note that $\sigma_f$ has length units, and $\sigma_c$ is dimensionless. The additional $R_\delta$ in the denominator of the first term ensures the consistency of the units. We then build a Gaussian process emulator of these parameters to enable accurate interpolations. We detail the emulator construction in appendix \ref{app:gaussianprocess}. 

With $P(\Mr|M_{\rm{vir}}, z)$ and $P(R_\delta|\Mr<x, z)$, we paint $r$-band luminosities onto dark matter particles in the lightcone simulations. For the resolved halos, we paint luminosities at the center of each halo using the learned $P(\Mr|M_{\rm{vir}}, z)$. The resolved halos are defined as halos with $M_{\rm{vir}}>6\times10^{12} \,h^{-1}M_\odot$ for the L1 and L2 boxes and $M_{\rm{vir}}>1\times10^{13} \,h^{-1}M_\odot$ for the L3 box. These choices of halo masses were justified in \cite{addgals}. For unresolved galaxies, we paint luminosities onto dark matter particles. We first generate random realizations of galaxies' redshifts by inverse transform sampling the measured redshift distributions of dark matter particles in the lightcone simulations. Next, we generate random realizations of galaxies' $r$-band luminosities ($\Mr$) from the luminosity function $\phi(x, z)$ after subtracting the number densities of resolved halos. By doing so, we avoid double-painting central galaxies. We then convert the cumulative conditional probability $P(R_\delta|\Mr<x, z)$ to $P(R_\delta|\Mr, z)$ using finite difference estimation:
\begin{eqnarray}
    P(R_\delta|\Mr=x, z)=& \frac{N(x+\delta x)P(R_\delta|\Mr<x+\delta x, z)}{M}- \frac{ N(x)P(R_\delta|\Mr<x, z)}{M},
\end{eqnarray}
where M is a normalization constant and $N(x)=\int_{-\infty}^x\phi(x, z)$, the cumulative luminosity function.  Each galaxy is randomly assigned an $R_\delta$ based on its $\Mr$ and $z$ by inverse transform sampling of $P(R_\delta|\Mr, z)$. We arrive at a list of galaxy $(R_\delta, \Mr, z)$ values. The remaining task is to put galaxies in the correct positions. In the N-body lightcones, we measure $R_\delta$ for each dark matter particle. We then divide the dark matter lightcone and galaxy samples into $\Delta z=0.01$ bins. For each redshift bin, we assign galaxies to dark matter particles using the closest match of $R_\delta$ in orders of galaxies' brightness. 

Figure \ref{fig:addgal_rdel} shows the comparison of $R_\delta$ distributions of the painted galaxies (red line) onto those measured in the SHAM model (black dots). The agreement between the red line and the black dots indicates that equation \ref{eq:rdelta} provides a reasonable description of the measurement and the galaxy assignment algorithm works reasonably well. Similar to \cite{addgals}, the $R_\delta$ distributions show double bump features for the two faintest galaxy thresholds. Central galaxies dominate one bump, and satellite galaxies dominate the other. This double bump feature of $R_\delta$ distributions indicates the effectiveness of $R_\delta$ on separating centrals and satellites at a given luminosity. We further compare the $R_\delta$ distribution in this work with the previous version \citep[Buzzard v2.0;][]{addgals}. We find that $R_\delta$ is shifted to smaller values at a given luminosity than Buzzard v2.0, indicating a more significant satellite fraction at a given luminosity. This is likely because we include orphan satellites in the SHAM model while Buzzard v2.0 did not. 
\begin{figure*}
    \centering
    \includegraphics[width=0.9\textwidth]{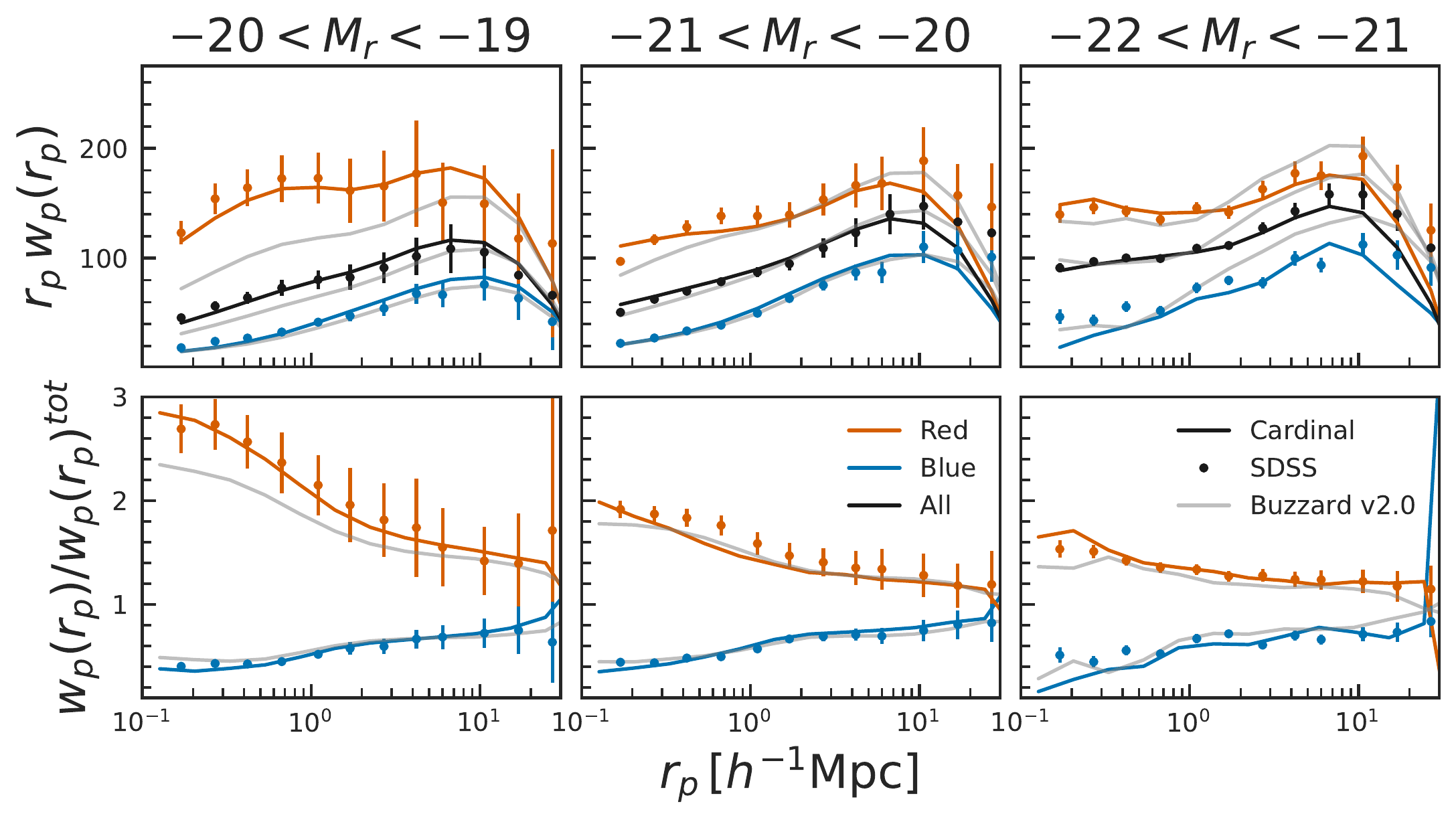}
    \caption{Comparison of color-dependent clustering measured in Cardinal (line) and data (dots with error bars, \citealt{zehavi2011}). Errorbar shows $1\sigma$ uncertainties. Red lines and dots correspond to red galaxies, while blue lines and dots correspond to blue galaxies. Different columns correspond to different magnitude bins in galaxies. For comparison, results from Buzzard v2.0 are shown in gray.}
    \label{fig:addgal_color}
\end{figure*}
\subsection{Painting colors onto galaxies}
\label{sec:paintcolor}
So far, we have generated a galaxy mock with a realistic spatial distribution and rest-frame $r$-band luminosity distribution. This section describes an algorithm for assigning colors to these galaxies. 
\subsubsection{Overall color distribution}
\label{sec:oveallcolor}
Following \cite{addgals}, we assume that galaxy SEDs can be described by the product of five coefficients and five KCORRECT spectral templates \citep{kcorrect03, kcorrect07}. One can then directly assign the measured KCORRECT coefficients in real data to galaxies in the simulations. This approach has a couple of benefits. First, the small number of coefficients for each galaxy reduces the requirements of memory and computational power to generate and store mock catalogs. Second, the product of coefficients and KCORRECT templates provides full SED information for each galaxy, from which one can compute the observed magnitude by applying band shifts and the observed bandpasses without regenerating galaxy SEDs. Third, the direct assignment from data guarantees reasonable matches between mocks and data. However, this approach requires a pool of measured KCORRECT coefficients representative of the targeted survey data. Generating this pool of KCORRECT coefficients requires representative spectroscopic redshifts down to the photometric survey depths. This is usually unachievable. 

Following \cite{JoeBuzzard}, we adopt a hierarchical approach to associate  KCORRECT coefficients of a SDSS galaxy to a simulated galaxy. We first generate a representative sample at $z<0.2$ down to DES survey depth using galaxies with $z=(0.005, 0.2)$ in the SDSS DR7 VAGC catalog \citep{SDSS}. Each galaxy has five KCORRECT coefficients according to \cite{kcorrect03} and \cite{kcorrect07}.  We then use PRIMUS \citep{2011ApJ...741....8C} galaxies to quantify the redshift evolution of this coefficient pool. Specifically, we employ PRIMUS galaxies to calculate the ratio $w_{r} (\Mr, z)$ of the probability of being red at $z$ to the probability at $z<0.2$. We define galaxies as red when $^{0.1}g-r>0.15-0.03 \Mr$. With $w_{r} (\Mr, z)$ at hand, we can generate KCORRECT coefficients for each simulated galaxy using a rejection sampling algorithm. We first bin SDSS galaxy samples and simulated galaxies using their rest-frame $r$-band luminosities ($\Mr$) with widths $\Delta \Mr= 0.1 (22.5+\Mr)$. For each simulated galaxy in the $\Delta \Mr$  bin, we first remove the SDSS galaxy samples in the same bin if
\begin{eqnarray}
        rand &>& w_{r} (\Mr, z) p\left(^{0.1}g-r>0.15-0.03 \Mr|\Mr \in \Delta \Mr \right), \nonumber\\
        p  &=& \frac{N\left(^{0.1}g-r>0.15-0.03 \Mr \land \Mr \in \Delta \Mr \right)}{N\left(\Mr \in \Delta \Mr \right)},
\end{eqnarray}
where rand is a random number drawn from a uniform distribution from 0 to 1, $p$ is an empirically measured red fraction in SDSS galaxy samples, and $z$ is the redshift of the simulated galaxy. We associate the KCORRECT coefficients of a galaxy in the remaining SDSS galaxy sample to the simulated galaxy by selecting the one with the closest value of $\Mr$.   If there is no SDSS galaxy in the $\Delta \Mr$ bin, we expand the width of the bin to $\Delta \Mr= 0.1 (22.5+\Mr)^2$ and repeat the rejection sampling step. By doing so, we paint KCORRECT coefficients to each simulated galaxy. With KCORRECT coefficients for each simulated galaxy, we can compute the galaxy's rest-frame $g-r$ color by convolving the SED generated from these KCORRECT coefficients and the SDSS's bandpass. The simulated galaxies will have the same color--$\Mr$ relation in the SDSS galaxy samples and the same red fraction--redshift relation as constrained by the  PRIMUS galaxy samples. These steps have been validated in \cite{addgals} and \cite{JoeBuzzard}.

\subsubsection{Environmental dependent galaxy color distribution }
\label{sec:encolor}
Once we create mock galaxies with realistic overall color distribution, we shuffle galaxy colors to create environmental dependencies of the color. We use the conditional abundance matching technique \citep{2013MNRAS.436.2286M,2013MNRAS.435.1313H} and the rest-frame $g-r$ color to perform the shuffling.  Specifically, we assign the SED that corresponds to the rest-frame $g-r$ color to a galaxy such that the following equation is satisfied, 
\begin{equation}
\label{eq:colorcam}
    p(<\rm{rank}(g-r)| \Mr) = p(<\rm{rank}(e_p) |\Mr),  
\end{equation}
where $e_p$ is a proxy to quantify galaxy environments. \cite{campaper} have found that $R_h$, the distance to the nearest massive halos, provides a good proxy of galaxy environments. Using $R_h$ and the conditional abundance matching technique, they find comparable color-dependent galaxy clustering to SDSS measurements \citep{addgals, campaper}. However, this proxy is inadequate for galaxies in galaxy clusters. First, more massive halos are bigger. On average, galaxies that live in a more massive halo would have larger $R_h$ than those living in small mass halos. Therefore using $R_h$ as a color proxy would make galaxies \textit{bluer} in \textit{more massive} halos. Second, for galaxies living in massive clusters, $R_h$ is the distance to the central galaxy. Using $R_h$ as a color indicator would create a strong radial color profile in a massive halo that is independent of cluster mass. This strong correlation breaks the self-similarity of galaxy clusters of different masses and lacks observational support. One hint of these problems shows in \cite{addgals}'s comparison of red galaxy clustering at $r<1\,h^{-1}\rm{Mpc}$ in the faintest magnitude bin (see also the grey line in figure \ref{fig:addgal_color}). The ratio of red galaxy clustering amplitude to all galaxies is low compared to the data. Since most of the pairs that contribute to small-scale clustering are from galaxies residing in large-mass halos, the small clustering ratio indicates that the galaxies in clusters are too blue. 

Given the aforementioned shortcomings of $R_h$, we define a new galaxy environment proxy $e_{hc}$. $e_{hc}$ is constructed with two insights. First, assuming galaxy clusters with different masses are self-similar, the distance to a cluster should be measured in units of the cluster's radius. In practice, we use the ratio of the distance to a cluster and the cluster's virial radius ($R_{\rm{vir}}$) to some power ($c_0$) as the color proxy. The additional power of the virial radius allows the possibility that more massive halos are stronger at quenching galaxies. The second insight is that the galaxy color gradient is shallower at the inner part of the clusters compared to the outskirts \citep{2021ApJ...923...37A}.
Further, given the success of producing reasonable large-scale red galaxy clustering using $R_h$ \citep{addgals}, we would like the new environment proxy $e_{hc}$ to be similar to $R_h$ on large scales. Therefore, we design a mapping from $R_h$ to $e_{hc}$ such that  $e_{hc}$  approaches $R_h+c$ when $R_h$ is infinity and becomes independent of $R_h$ when $R_h$ is zero. $c$ is an arbitrary constant irrelevant to color assignments because only the ranks of $R_h$ are related to galaxy colors (equation \ref{eq:colorcam}). Given these two insights, the environment proxy has the following functional form,
\begin{eqnarray}
\label{eq:camchto}
    x &=& d_h/R_{\rm{vir}}^{c_0} \nonumber\\
    e_{hc} &=&  0.5x+c_1\left( \frac{x-1}{2c_1}{\rm{erf}}\left(\frac{x-1}{2c_1}+\frac{\exp\left(-\left(\frac{x-1}{2c_1}\right)^2\right)}{\sqrt{\pi}}\right) \right), 
\end{eqnarray}
where $d_h$ is the distance to the nearest massive halo with mass greater than $M_{ch}$, and $M_{ch}, c_0, c_1$ are three free parameters. $c_1$ controls the sharpness of transition from $de_{hc}/dx=0$ to $de_{hc}/dx=1$. 

One remaining problem of equation \ref{eq:camchto} is that all central galaxies with $M>M_{ch}$ have $e_{hc}=0$, making them the reddest galaxies at a given magnitude. Given that $M_{ch}$ is usually $10^{12.8} \,h^{-1}M_\odot$, assigning all central galaxies with $M>M_{ch}$ the reddest SEDs will contradict observations \citep[e.g.][]{w12}. We must make some central galaxies living in low-mass halos blue. The challenge is that our lightcone has different resolutions at different redshifts, and simple mass cut might transfer this reshift-dependent resolution onto galaxy colors.
Fortunately, $R_\delta$ gives a nice separation of centrals and satellites that is less sensitive to the resolution of the simulations (see figure \ref{fig:addgal_rdel}). Specifically, using \Tbox{}, we find that none of the satellite galaxies in $M>10^{12.8} \,h^{-1}M_\odot$ halos and none of the central galaxies with $M>2\times 10^{13} \,h^{-1}M_\odot$ have $R_\delta$ greater than two. That is, galaxies with $R_\delta>2$ are likely to be low-mass isolated centrals  whose color should be preferentially blue. Built on these insights, we can increase the value of the environmental proxy $e_{hc}$ of a galaxy living in a low-density environment ($R_\delta>2$) and  likely to be red ($e_{hc}<1$) in the original algorithm. In this way, these galaxies will be bluer because of a larger environmental proxy value. We, therefore, increase galaxies' $e_{hc}$ by $c_2$ if $R_\delta>2$ and $e_{hc}<1$. 

Finally, we allow the possibility that color ranks ($\rm{rank}(g-r)$)  and galaxy environment proxy ranks ($\rm{rank}(e_p)$) are not perfectly correlated. Instead of implementing equation \ref{eq:colorcam}, we implement 
\begin{equation}
\label{eq:salmon}
    p(<\rm{rank}(g-r)| \Mr) = p(<\tilde{\rm{rank}}(e_p) |\Mr).
\end{equation}
In the above equation, $\tilde{\rm{rank}}(e_p)$ is constructed such that it is an unbiased estimator of $\rm{rank}(g-r)$ and is correlated with $\rm{rank}(g-r)$ with correlation coefficient $r_c$. We construct $\tilde{\rm{rank}}(e_p)$ using \textsc{halotools} \citep{halotool}. \cite{campaper} find that galaxies with different stellar mass prefer a different $r_c$ using SDSS color dependent clustering measurements \citep{zehavi2011}. Motivated by their findings, we parametrize $r_c$ to depend on $\Mr$ via 
\begin{equation}
\label{eq:corr}
    r_c = c_l+(c_h-c_l)\left(0.5+0.5{\rm{erf}}\left(\frac{\Mr-c_m}{\sqrt{2} c_\sigma}\right)\right), 
\end{equation}
where $c_l$, $c_h$ are free parameters that govern the asymptotic behaviors at low and high $\Mr$, and $c_m$, $c_\sigma$ govern the transition $\Mr$ and the steepness of this transition. Equation \ref{eq:corr} has the same functional form as equation \ref{eq:SHAM_disrupt2}. Again the underlying physical behavior is pretty simple. For bright galaxies (small $M_r$), $r_c$ is determined by $c_l$. Conversely, for faint galaxies,  $r_c$ is determined by $c_h$. Equation \ref{eq:corr} simply ensures a smooth transition from bright to faint objects, and two additional parameters control the location and slope of the transition.

In summary,  our color assignment model has eight parameters: 
\begin{enumerate}
    \item $M_{ch}$: a parameter defines the mass threshold of halos to which we calculate distances ($d_h$) of each galaxy.
    \item $c_0$ and $c_1$: two free parameters control the mapping of $d_h$ to the environment proxy $e_{hc}$. 
    \item $c_2$: a parameter is added to $e_{hc}$ of low-mass centrals to make it bluer.
    \item $c_l, c_h, c_m, c_\sigma$: parameters control luminosity dependence of the correlation between colors and environment proxies.
\end{enumerate} 
We use color-dependent clustering measured in \cite{zehavi2011} to constrain these parameters. Specifically, to separate the uncertainties of the color model and the galaxy clustering model, we use the ratio of galaxy clustering of red and blue galaxies to all galaxies to constrain the color model parameters. A simple downhill simplex algorithm is used to find the best-fit parameters. At each step of the downhill simplex algorithm, we shuffle the galaxy SEDs in the mocks using equation \ref{eq:salmon}, %
regenerate galaxy colors using SEDs and SDSS filters, select galaxy samples according to  \cite{zehavi2011}, calculate galaxy clustering using \textsc{Corrfunc} \citep{corrfunc}, and compare the measurements with the data assuming a Gaussian likelihood. Finally, our best-fit model has $\chi^2=84.9$ with degree-of-freedom $58$, corresponding to a PTE value of $0.01$. %
In Figure \ref{fig:addgal_color}, we compare the best-fit model to the measurement. We find that our best-fit model can accurately describe the data where the predicted clustering differs from that of Buzzard v2.0. The differences between our model and Buzzard v2.0 are especially pronounced for the lowest luminosity sample. 
\subsection{Observational effect}

\label{sec:observationeffect}
\subsubsection{Photometric noise}
So far, we have generated a quarter-sky galaxy lightcone with realistic color, luminosity, and spatial distributions. We add galaxy shapes and sizes following methods described in \cite{JoeBuzzard}. Finally, we apply lensing effects using the ray tracing code \textsc{Calclens} \citep{calclens}. Specifically, we apply deflection, rotation, shear, and magnification for all galaxies to alter their positions, shapes, and photometry. 

We cut out two DES-Y3 regions from this lightcone and rotate them into the DES-Y3 footprint \citep{Y3gold}. We then apply several observational effects on the mock catalogs. First, we apply the survey masks on the mock using the DES-Y3 survey mask that is in the form of a healpix map with $N_{\rm side}$=4096, corresponding to a resolution of $0.73~\rm{arcmin}^2$. For each healpix pixel, we randomly select galaxy samples according to the FRACGOOD value, which describes the amount of masking within the healpix pixel.   Second, we add photometry noise to each galaxy, a process that will be detailed later. This step is essential because magnitude noise can drastically affect the number of galaxies near survey depth limits due to Eddington biases. For analyses that use galaxies near survey depth limits, which is the case for most weak lensing analyses, properly modeling noise is essential. 
Finally, we cut out galaxies with observed magnitude fainter than the survey depth. 

The remaining task is to add a realistic magnitude error on each galaxy. We follow the prescription described in \citep{JoeBuzzard, addgals} that uses the effective exposure time ($t_{\rm{eff}}$) and $10\sigma$ limiting magnitudes ($m_{\rm{lim}}$) from survey data. Here, we briefly summarize the prescriptions. First, we assume that the observed total number of photons of a galaxy comes from two contributions: photons from galaxies $I_{\rm{gal}}$ and photons from the noise, such as sky, readout noise, etc. $I_{\rm{sky}}$. The $I_{\rm{gal}}$ can simply be related to galaxy magnitude ($m_{\rm{gal}}$) via, 
\begin{equation}
    I_{\rm{gal}} = 10^{-0.4 (m_{\rm{gal}}-\rm{ZP})} \times t_{\rm{eff}}, 
\end{equation}
where $ZP=22.5$. Following \cite{2015arXiv150900870R}, $I_{\rm{sky}}$ can be empirically determined from data. \cite{Y3gold} provides the 
$10\sigma$ limiting magnitudes ($m_{\rm{lim}}$) and the effective exposure time ($t_{\rm{eff}}$) in the form of healpix maps with a resolution of $0.73~\rm{arcmin}^2$. Using this information,  we calculate $ I_{\rm{sky}}$ for each healpix pixel by
\begin{equation}
    I_{\rm{sky}} =   \frac{10^{-0.8(m_{\rm{lim}}-\rm{ZP})}\times t_{\rm{eff}}}{100}-10^{-0.4(m_{\rm{lim}}-\rm{ZP})}.
\end{equation}

We assume the observed number of photons follows a Poisson distribution. The noisy observed flux $F_{\rm{obs}}$ is then given by 
\begin{equation}
\label{eq:noisy}
    F_{\rm{obs}} = \frac{\rm{Poisson(I_{\rm{gal}}+I_{\rm{sky}})}-I_{\rm{sky}}}{t_{\rm{eff}}}, 
\end{equation}
where Poisson denotes a random draw from Poisson distribution. The second term in the above equation is to mimic the process of sky background subtractions in observations. We note that this implementation is different from \cite{JoeBuzzard} and \cite{addgals}, where the authors performed a random draw from a Gaussian distribution with a scatter as the square root of the mean. The Gaussian approximation is valid for high signal-to-noise galaxy samples but could lead to a bias for low signal-to-noise galaxies. The noise of the observed flux $\sigma_{\rm{flux}}$  is given by $\sqrt{I_{\rm{gal}}+I_{\rm{sky}}}/t_{\rm{eff}}$. The associated magnitude and error are
\begin{eqnarray}
    m_{\rm{obs}} &=& \rm{ZP} - 2.5 \rm{log}_{10} F_{\rm{obs}} \nonumber\\
    m_{\rm{err}, \rm{obs}} &=& \frac{2.5}{\rm{log}(10)}\frac{\sigma_{\rm{flux}}}{F_{\rm{obs}}}. 
\end{eqnarray}

\begin{figure}
    \centering
    \includegraphics[width=0.5\textwidth]{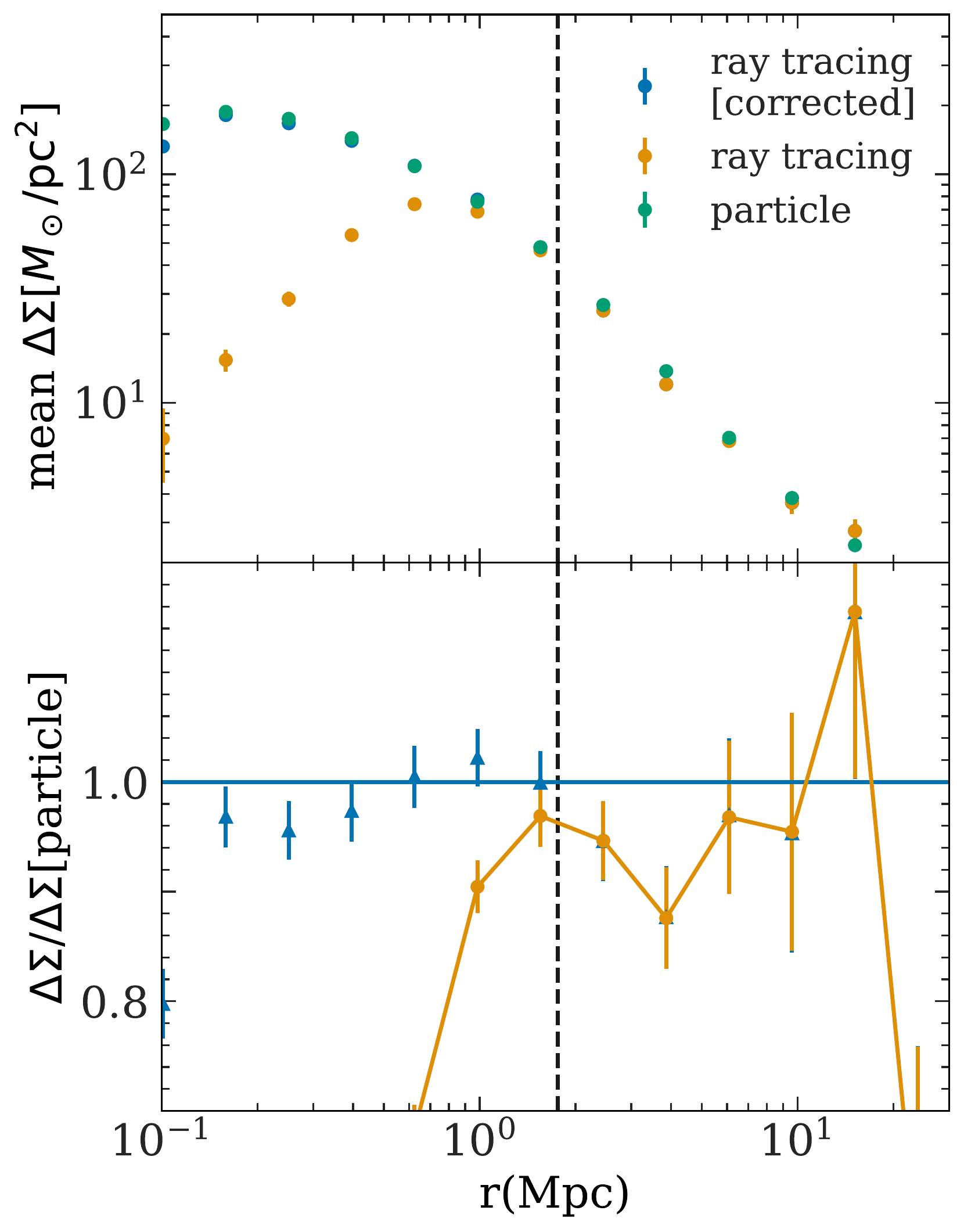}
    \caption{ Mean $\Delta \Sigma$ profiles measured around halos with  $M_{{\rm{vir}}}>2\times 10^{14} \,h^{-1}M_\odot$ and $z=[0.6,0.62]$. The $x-$axis is measured in physical coordinates. The green dots are derived from dark matter particle--halo cross-correlations. The orange dots are calculated from ray-tracing shapes (and are similar to previous catalog versions). The blue dots show our improved model, calculated from corrected ray-tracing shears. Error bars are estimated with $10000$ bootstrap resampling.
    The black vertical dashed lines show ten times the pixel size for ray-tracings.} 
    \label{fig:lensing}
\end{figure}
\subsubsection{Gravitational lensing}
\label{sec:lensing}
The matter along the line of sight will distort the light from galaxies. This distortion will modify the position of galaxies, distort their shapes, and magnify their brightness. We include these effects by performing full-sky multi-plane ray-tracings using \textsc{Calclens} \citep{calclens}. This step has been detailed in appendix C of \cite{JoeBuzzard}. Here, we provide a brief summary. First, we decompose the full-sky particle lightcones into healpix maps with $n_{\rm{side}}=8192$, corresponding to a resolution of $0.46$ arcmins. Next, for each pixel, we divide the particle lightcones into equally-spaced lens planes from $z=0$ to $z=2.4$ with separations of $25\, h^{-1}\rm{Mpc}$. Then, for each of the simulated galaxies, we calculate the deflection angle and distortion matrix of the light at each lens plane. To achieve this, we first calculate the lensing potential from integrated dark matter density fields by solving a two-dimensional Poisson equation \citep{jain2000}. Next, we calculate the deflection angle using the lens equation \citep{2009A&A...497..335T}, which relates the deflection angles to the first derivative of the lensing potential. Then, the distortion matrix is calculated using the second derivative of the lensing potential \citep{jain2000, Hilbert2009, calclens}. Finally, these deflection angles and distortion matrices at each plane are combined to produce the final total deflections, shears, and magnifications \citep[equations 10 and 12 of ][]{calclens}. 

We validate the ray-tracing shears by computing the tangential shear $\gamma_t$ around halos with  $M_{\rm{vir}}>2\times 10^{14} \,h^{-1}M_\odot$ and $z=[0.6,0.62]$.  This tangential shear can be related to the excess surface density ($\Delta \Sigma$) of the matter around halos, which can be measured directly using particles in the simulation. We relate the tangential shear to $\Delta \Sigma$ using the estimator presented in \cite{sheldon2004}, which reads
\begin{equation}
\label{eq:lensing}    \Delta \Sigma (r) = \frac{\sum_i^{N_{\rm{halos}}} \sum_j^{N_{\rm{Sources}}}\gamma_t^{i,j}\Sigma^{i,j}_{\rm{crit}}}{\sum_i^{N_{\rm{halos}}} \sum_j^{N_{\rm{Sources}}}\left(\Sigma^{i,j}_{\rm{crit}}\right)^2}, 
\end{equation}
where $N_{\rm{halos}}$ is the number of halos in the bin, $N_{\rm{Sources}}$ is the number of galaxies around halos at distance $r$, $\gamma_t^{i,j}$ is the tangential shear of of galaxy $j$ around halos $i$. To avoid contamination, we only use galaxies with a redshift of $0.1$ greater than the redshift of the halos. In the above equation $\Sigma^{i,j}_{\rm{crit}}$ is given by 
\begin{equation}
    \Sigma^{i,j}_{\rm{crit}} = \frac{c^2}{4\pi G}\frac{D_s}{D_h D_{hs}},
\end{equation}
where $D_h, D_S$, and $D_{hs}$ are angular diameter distances to halos, galaxies, and between halos and galaxies. 
The $\Delta \Sigma$ can also be calculated from the matter-halo cross-correlation functions $\xi_{hm}$ using 
\begin{eqnarray}
\label{eq:lensingp}  
    \Delta \Sigma (r) &=& \bar{\rho}\frac{4}{r^2} \int_0^r x \,dx \int_0^{\infty} d\chi\, \xi_{hm} \left(\sqrt{x^2+\chi^2}\right)\nonumber \\ 
    &&- 2\bar{\rho} \int_0^{\infty} d\chi\, \xi_{hm} \left(\sqrt{r^2+\chi^2}\right).
\end{eqnarray}

Figure \ref{fig:lensing} shows the comparison of $\Delta\Sigma$ measured via ray tracing (equation \ref{eq:lensing}, orange dots) and via direct measurements of particles (equation \ref{eq:lensingp}, green dots). We find that the ray tracing and particle calculations agree well on large scales but deviate on small scales. This deviation has been shown in literature \citep{2022OJAp....5E...1K, 2017ApJ...850...24T} and likely comes from the finite angular resolution of $0.46$ arcmins when we calculate the lensing potentials for ray tracing. This resolution problem in lensing can be problematic for cluster lensing analyses when most of the signal lies below $1\,h^{-1}\rm{Mpc}$. Because of the resolution problem, cluster lensing studies have relied on the dark matter particle--halo cross-correlation \citep{Heidiselection, DES_cluster_cosmology}. However, \cite{HeidiCov} points out that while the particle--halo cross-correlation produces the right mean $\Delta \Sigma$, it significantly underestimates the halo-to-halo variance of $\Delta \Sigma$ at large scales. This is because particle--halo cross-correlations ignore the line-of-sight structure's contributions in the lensing profile \footnote{We note that one could include line-of-sight contribution into particle--halo cross-correlations by calculating two-dimensional projected correlations as implemented in Halotools \citep{halotool}. However, as shown in \cite{HeidiCov}, one must consider particles with distances to clusters much greater than $100$ $h^{-1}\rm{Mpc}$ to avoid underestimating variances, which might be computationally challenging. }. 

We empirically correct the shears of each galaxy using the measured particle--halo cross-correlations. We first bin the halos in simulations with redshift from $z=0.18$ to $z=0.67$ and the virial mass $M_{\rm{halo}}$ above $10^{13} \,h^{-1}M_\odot$ into $\Delta z=0.02$  and $\Delta \rm{log}_{10}(M)=0.15 
 \,h^{-1}M_\odot$ bins. Next,  we calculate $\Delta \Sigma$ for halos in each bin using particle--halo cross-correlations ($\Delta \Sigma_p$) using equation \ref{eq:lensingp} and ray-tracing derived shears using equation \ref{eq:lensing} ($\Delta \Sigma_\gamma$). We calculate the differences between the two and apply a correction on the two shear components of each galaxy $\gamma_1$ and $\gamma_2$. Specifically, our algorithm reads

\begin{algorithm}[H]
\begin{algorithmic}
\For{Galaxies with redshift $z$}
\For{Halos $h$ with mass $M_h>10^{13} \,h^{-1}M_\odot$, redshift $z_h<z$, and distance to galaxies $r_h<4.6$ arcmins (10 times the resolution of the ray-tracings)}
\State $\Sigma^{h,g}_{\rm{crit}} \gets \frac{c^2}{4\pi G}\frac{D_g}{D_h D_{hg}}$
\State $\Delta \gamma_t \gets$ $(\Delta \Sigma_p-\Delta \Sigma_\gamma)/\Sigma^{h,g}_{\rm{crit}}$
\State $\Delta \gamma_{1,h}= \Delta \gamma_t/\left(\cos(2\phi)+\sin(2\phi)\tan(2\phi) \right)$
\State $\Delta \gamma_{2,h}= \Delta \gamma_{1,h} \tan(2\phi)$
\EndFor
\State $\gamma_{1} \gets \gamma_1 + \sum_h \Delta \gamma_{1,h}$
\State $\gamma_{2} \gets \gamma_2 + \sum_h \Delta \gamma_{2,h}$
\EndFor
\end{algorithmic}
\end{algorithm}
In the above algorithm, $D_g$, $D_h$, $D_{h,g}$ are angular diameter distances to galaxies, halos, and between halos and galaxies. Index $h$ goes through haloes with mass greater than $10^{13} \,h^{-1}M_\odot$ along the line of sight with angular separations less than $4.6$ arcmins of the galaxies. The two equations relating $\Delta \gamma_t$ to $\Delta \gamma_1$  and $\Delta \gamma_2$  are derived assuming $\Delta \gamma_\times=0$. This is motivated by the fact that gravitational lensing due to the localized mass distribution does not generate $\gamma_\times$ modes. The blue dots in figure \ref{fig:lensing} show the $\Delta \Sigma$ measurements using the corrected galaxy shears. We find that it is consistent with the derivation from particle--halo cross-correlations, indicating the effectiveness of our algorithm. In appendix \ref{app:cosmicshear}, we further show that this correction of galaxy shears has well below $0.1$ percent impact on cosmic shear ($\xi_{+/-}$) at large scales. At small scales, $\xi_{+/-}$ based on corrected galaxy shears is more consistent with theoretical predictions.

\begin{figure*}
    \centering
    \includegraphics[width=0.95\textwidth]{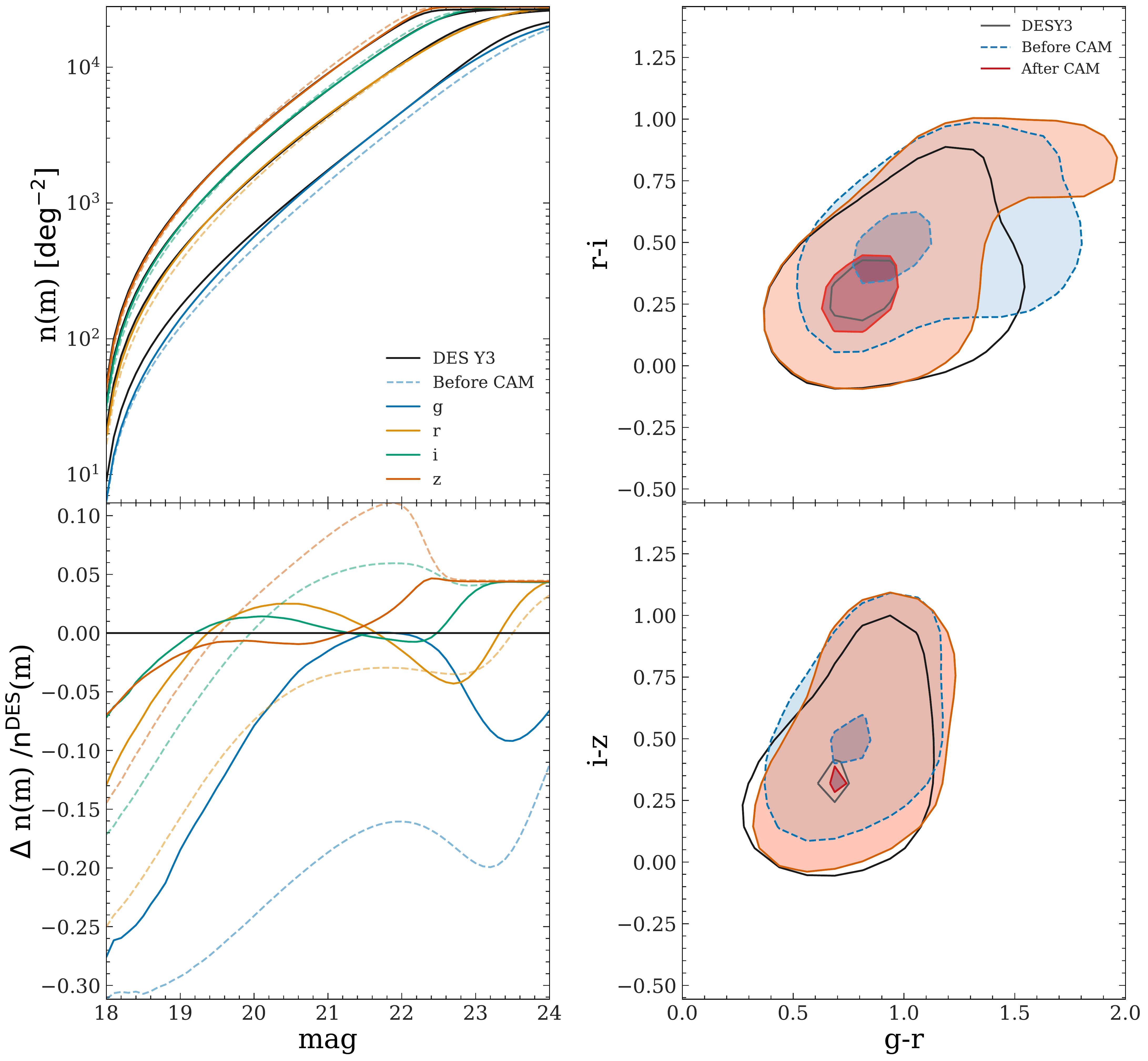}
    \caption{Comparison of apparent magnitude between Cardinal and DES Y3 data. The top left panel compares overall magnitude distributions, and the bottom left panel shows the fractional differences. The top right panel compares $g-r$ and $r-i$ distributions, and the bottom right compares $g-r$ and $i-z$ distributions. The black lines show measurements of DES Y3 data, the dashed lines show Cardinal before applying the conditional abundance matching scheme (detailed in section \ref{sec:camcolor}), and the solid lines show Cardinal after applying the conditional abundance matching scheme. Contours show $1\sigma$ and $2\sigma$ boundaries of the density distributions. The new conditional abundance matching scheme greatly improves the consistency between mocks and data. 
    } 
    \label{fig:1dcomp}
\end{figure*}

\begin{figure}
    \centering
    \includegraphics[width=0.5\textwidth]{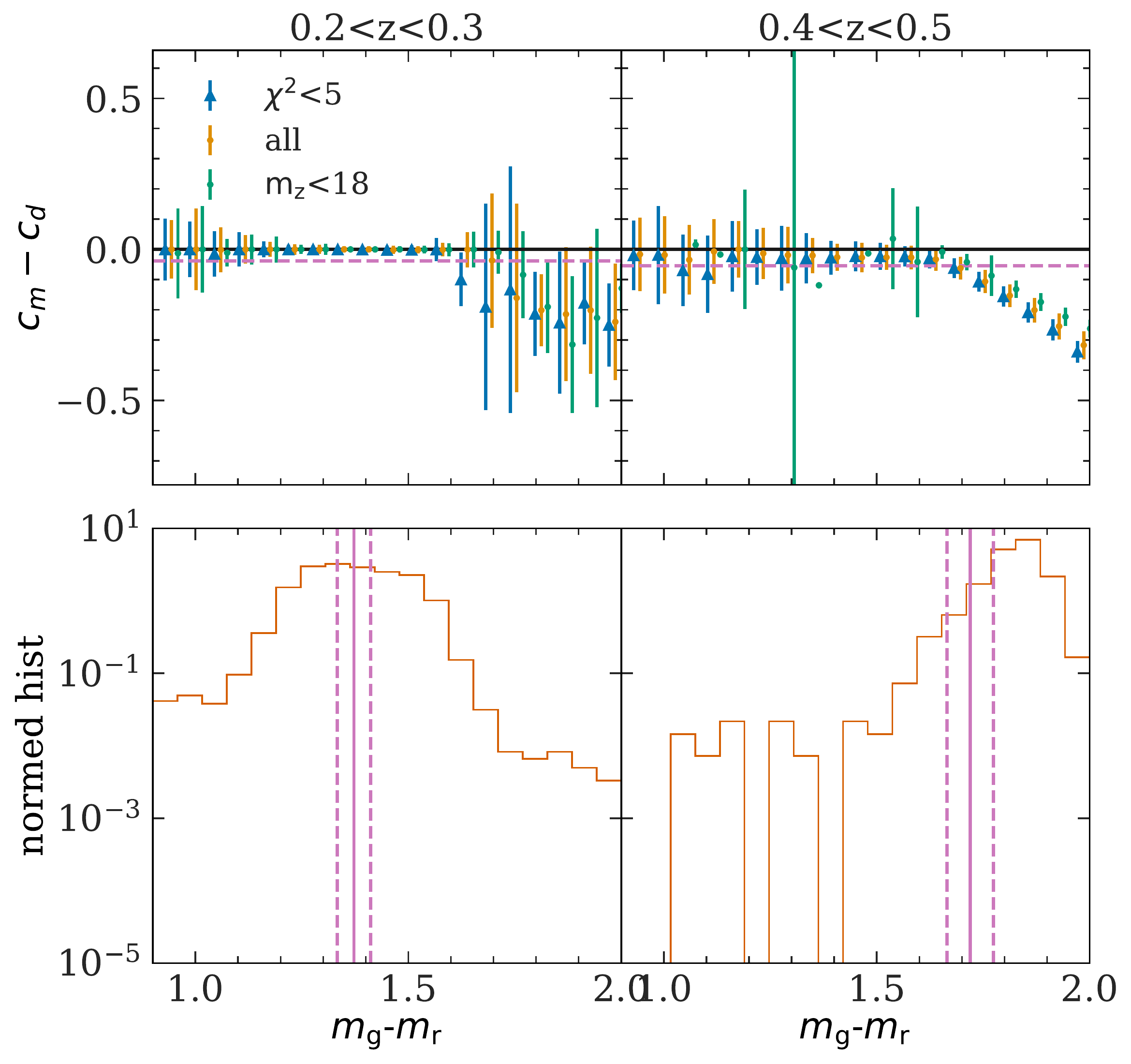}
    \caption{Color ($g-r$) residuals as a function of $g-r$ color ($c_d$). $c_m$ is the $g-r$ color constucted using $k$-correction templates and best-fit coefficients. This calculation uses \redmapper{} member galaxies with \redmapper{} run on DES-Y3 data. Each column corresponds to a redshift slice using redshifts of host galaxy clusters. Top panels: dots show the median $c_m-c_d$ in each color bin, while error bars show $68$ percent scatter. Magenta lines show $1\sigma$ widths of red sequence reported by redMaPPer{}. We split the samples into two subsets: galaxies that are bright with $m_z<18$ (green), and galaxies that best match \redmapper{}'s red-sequence template (blue). Both subsets are artificially shifted horizontally to improve clarity. Bottom panels: histograms show the measured color distributions of all galaxies in the calculation. Magenta lines show the mean (solid) and width (dashed) of the red sequence reported by \redmapper{}. For all samples, the reconstructed color is mostly unbiased at $z=0.2$-$0.3$. At $z=0.4$-$0.5$, the reconstructed color is biased for red galaxies and mostly unbiased for blue galaxies. } 
    \label{fig:kcorr}
\end{figure}
\subsection{Conditional abundance matching color}
\label{sec:camcolor}
\begin{figure}
    \centering
    \includegraphics[width=0.5\textwidth]{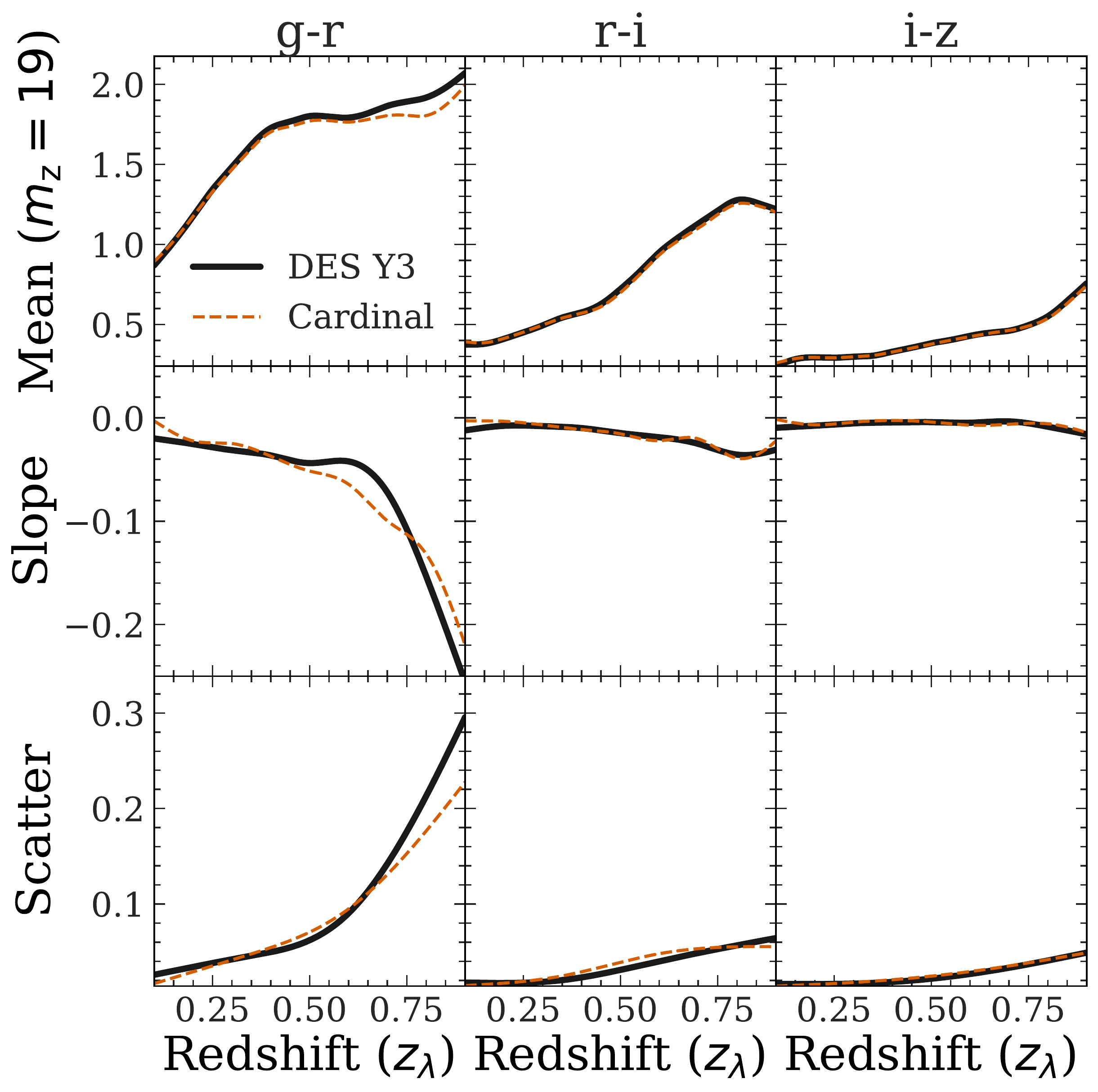}
    \caption{Comparison of red-sequence galaxies in Cardinal (red) and DES Y3 data (black) estimated by \redmapper{}. Three columns correspond to $g-r$, $r-i$, $i-z$ colors, respectively. The top row shows the mean colors at $m_z=19$ as a function of redshift. The middle row shows the slope of colors-$m_{\rm{z}}$ relation as a function of redshifts estimated by \redmapper{}. The bottom row shows the scatter of red-sequence colors. The largest discrepancy occurs at high redshift $g-r$ colors, where the data is particularly noisy.} 
    \label{fig:app:redsequence}
\end{figure}

\begin{figure*}
    \centering
         \centering

         \centering
         \includegraphics[width=0.99\textwidth]{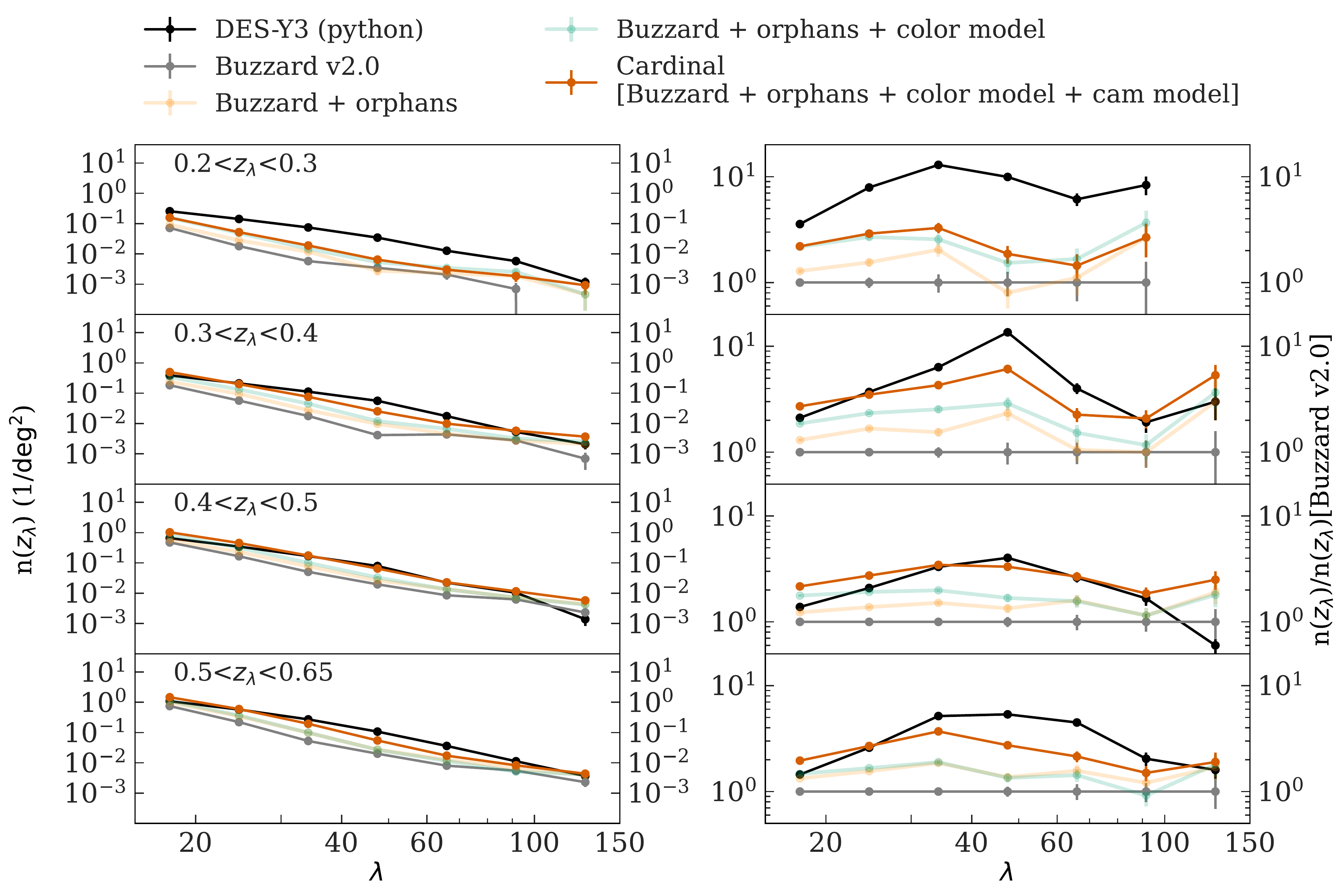}

     \hfill

         \centering
          \includegraphics[width=0.99\textwidth]{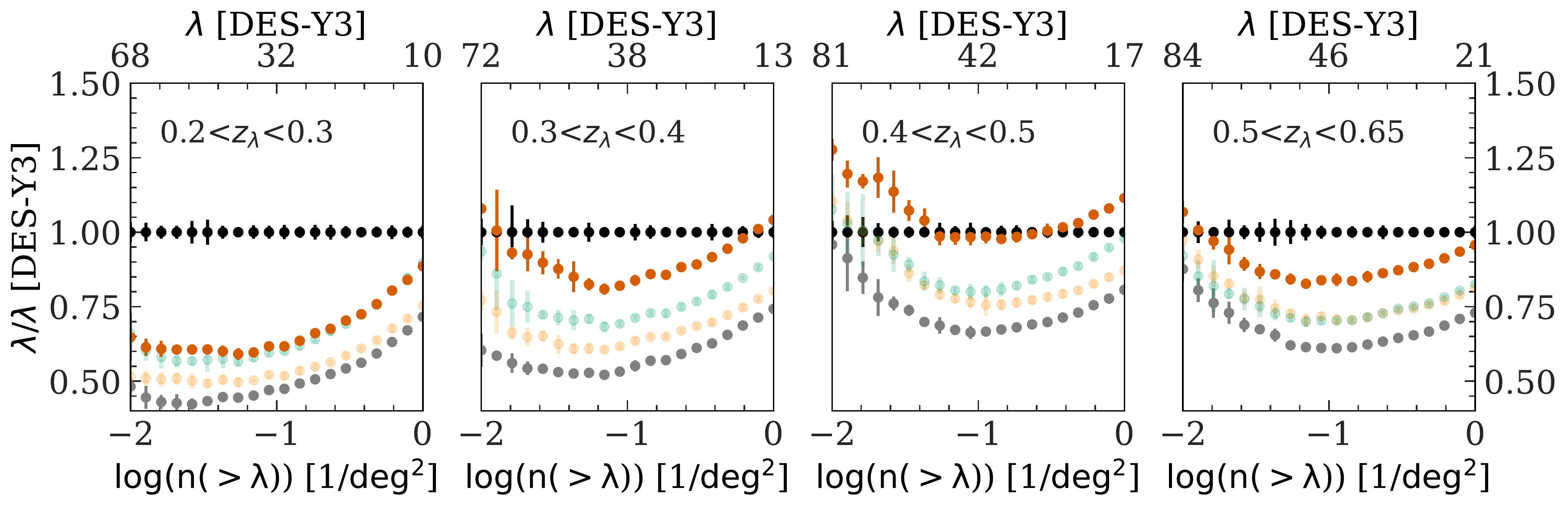}

      \caption{Comparison of cluster number density in Cardinal and DES-Y3 data. The top left plot shows \redmapper{} cluster abundances per $\rm{deg}^2$ as a function of richness ($\lambda$) and redshift ($z_\lambda$). The top right plot shows the relative cluster number density of each realization relative to Buzzard v2.0.   Four different rows in the top two panels show four redshift bins based on $z_\lambda$, the most probable redshifts of clusters estimated by \redmapper{}. The black lines correspond to the DES-Y3 data, and the blue lines correspond to Buzzard v2.0 \citep{y3buzzard}. We note that in both cases, we re-run \redmapper{} using the same version of \redmapper{} as used in Cardinal to ensure apples-to-apples comparisons. The red lines show the number of \redmapper{} identified clusters in Cardinal. Evidently, one can see that the cluster abundances in Cardinal are much more consistent with the data. The green and orange lines show two different realizations that are intermediate steps from Buzzard v2.0 to Cardinal. Specifically, the orange line corresponds to a realization where the only difference to Buzzard v2.0 is including orphans in the SHAM model (detailed in section \ref{sec:SHAM}). The green line represents one step forward by changing the galaxy color assignment models detailed in section \ref{sec:paintcolor}. Error bars show the Poisson noise. For further comparison, the bottom panel shows the richness ratio as a function of cumulative cluster abundance between simulations and DES-Y3 data. The top $x$-axis shows the corresponding richness of DES-Y3 data at a given cumulative number density.  Error bars are estimated with $50$ jackknife resampling.
    }

    \label{fig:clusterabundance}
\end{figure*}

In this section, we first diagnose mismatches between modeled and observed color distributions. We then describe our schemes for correcting these mismatches. This correction has several moving parts, but our tests (see figures \ref{fig:app:redsequence} and \ref{fig:clusterabundance}) show that it has the desired effect.

Figure \ref{fig:1dcomp} compares the total magnitude distributions and color--color distributions of galaxies in the mock (dashed lines) and the data (solid black lines). In general, the galaxies in the mock have redder colors. This trend has also been shown in figure 5 of \cite{JoeBuzzard}, where the authors compared Buzzard to COSMOS data. As described in section \ref{sec:oveallcolor}, the overall color distribution is determined by two different factors: (a) KCORRECT spectral templates in \citep{kcorrect07}, and (b) SDSS KCORRECT coefficients with redshift evolution controlled by PRIMUS' red fraction measurements. To identify the exact cause of this mismatch of colors, we first test the effectiveness of KCORRECT spectral templates in describing colors of high redshift galaxies. We calculate KCORRECT coefficients of redMaPPer member galaxies in clusters with a richness greater than $20$ measured in the DES-Y3 data. In this calculation, we use the redshift of the host clusters to minimize the error of photometric redshifts. We then reconstruct the magnitude of galaxies using these KCORRECT coefficients. Figure \ref{fig:kcorr} compares the $g-r$ color measured using the reconstructed magnitudes to those measured with the input magnitudes. We choose to show $g-r$ for simplicity but find that the trend is consistent between different colors. Figure \ref{fig:kcorr} shows that the KCORRECT reconstructed color is unbiased for the low redshift bin. However, for the high redshift bin, while the reconstructed color of blue galaxies seems to be mostly unbiased, the reconstructed color is increasingly biased for redder galaxies. This indicates that summarizing colors with KCORRECT spectral templates is valid for low redshift galaxies but biases the colors of high redshift red galaxies. We further test this finding using $31$ COSMOS SED templates \citep{2009ApJ...690.1236I}, and find a similar result.
Interestingly, this bias makes red galaxies bluer, which contradicts the trend in figure \ref{fig:1dcomp}. Therefore, we conclude that the trend in figure \ref{fig:1dcomp} is likely due to the insufficiency of the combination of SDSS KCORRECT coefficients and PRIMUS' red fraction measurements for describing the colors of high-redshift galaxies. 

In summary, for red galaxies, two competing effects affect their colors: the insufficiency of SDSS KCORRECT coefficients and PRIMUS' red fraction make them too red, and summarizing the color with KCORRECT spectral templates makes them too blue. The cancellation of these two competing effects on the colors of red galaxies can potentially explain why the mean color of red-sequence galaxies in Buzzard \citep{JoeBuzzard} is consistent with data. In contrast, the overall colors of galaxies are biased red. Further, while the cancellation of these two competing effects can make the colors of red galaxies unbiased, it could boost the total number of red galaxies because blue galaxies are more abundant than red-sequence galaxies. Figure 5 of \cite{JoeBuzzard} shows that the number of red galaxies in Buzzard is larger than in the COSMOS data. To further understand this hypothesis, we present a toy model in appendix \ref{app:colorbiassim}, showing that our hypothesis can qualitatively reproduce Buzzard's $g-r$ color distribution shown in figure 5 of \cite{JoeBuzzard}.                 

A more comprehensive and physical solution to these two causes of color mismatches between mocks and data is important but requires further research. Here we provide an empirical solution using DES-Y3 photometric data. The DES-Y3 data constrain galaxies' overall color and apparent magnitude distributions. Because DES-Y3 data do not have redshift information for each galaxy, we must make some assumptions to calibrate our simulations with photometric data. We make the following assumptions: 
\begin{enumerate}
    \item The ranks of galaxy colors given observed magnitude in mocks are valid.
    \item The relative colors of galaxies in different redshifts are correct.
\end{enumerate} With these two assumptions, we can then employ the DES-Y3 data to tune the multi-dimensional color and magnitude distributions of galaxies in Cardinal using the conditional abundance matching technique. We first match the observed $z$ band magnitude distributions in Cardinal and data because the $z$ band is the reference magnitude used to select \redmagic{} galaxies and \redmapper{} clusters. To achieve this, we retune the third-order polynomial parameters that govern the redshift evolution of $L^*$ by matching the luminosity functions of the mock galaxy catalogs to the data. We use a downhill simplex algorithm to minimize a Gaussian likelihood with Poisson errors. Second, we match the color distributions by enforcing the equality of the following probabilities between mocks and data, 
\begin{equation}
\label{eq:abcolor}
    P(<c_i | m_z, c_{j<i}), 
\end{equation}
where $c_i=[g-r, r-i, i-z]$ denotes the three colors used for lens galaxies and cluster selections. We use the method implemented in \textsc{halotools} \citep{halotool} to perform this matching. With this second step, we can ensure the matches of overall color distributions between mocks and data. The above algorithm has one important caveat. The red galaxies live in a tight color--magnitude space known as the red sequence. Because we match overall galaxy colors and magnitude, this process will widen the color--magnitude relations of red galaxies. In appendix \ref{app:matchredm}, we present an algorithm that uses \redmapper{} to reduce this broadening.

So far, we have obtained galaxy catalogs with realistic magnitude and color distributions. Unfortunately, this catalog likely has incorrect environmentally-dependent galaxy colors because of additional noise introduced in the various conditional abundance matching processes. Despite this problem, we can still use this catalog to generate a new color template that captures the redshift dependence of color distributions more accurately than the original SDSS catalogs. Naively, we could rebuild the SED template coefficients using the catalog. But, as shown in figure \ref{fig:kcorr}, we find that summarizing galaxy colors using SED templates could lead to a bias of colors for red galaxies up to $0.3$ dex. %
We, therefore, adopt a different approach. Instead of building new SED template coefficients using the abundance-matched catalog, we build an observed color table as a function of the galaxy's observed  $z$-band magnitude $m_z$ and redshifts. We then use this table and repeat steps starting from section \ref{sec:encolor} to build a new mock catalog.

While we use a special treatment of red-sequence galaxies in the conditional abundance matching method (detailed in appendix \ref{app:matchredm}), the additional noise caused by this process 
makes the width of the red sequence too wide compared to the data. This too-wide red sequence could lead to overly-pessimistic estimations of photometric errors of red galaxies and could cause massive background contamination of optical cluster identifications. To solve this problem, we further improve the  consistency of the red sequence in mocks and the data using the algorithm presented in appendix \ref{app:matchred}.  The resulting red sequence in Cardinal is compared with DES-Y3 data in figure \ref{fig:app:redsequence}. We find that the red sequences in simulations and data are very consistent. 

After the color adjustment, we add observational noise to generate the new mock catalog using the method described in section \ref{sec:observationeffect}. The new mock catalog's overall color and magnitude distributions are shown as solid lines in figure \ref{fig:1dcomp}. The agreement between mocks and data is greatly improved. In most bands, we achieve a fractional error in galaxy luminosity functions $\sim5$ percent.

\begin{figure}
    \centering
    \includegraphics[width=0.5\textwidth]{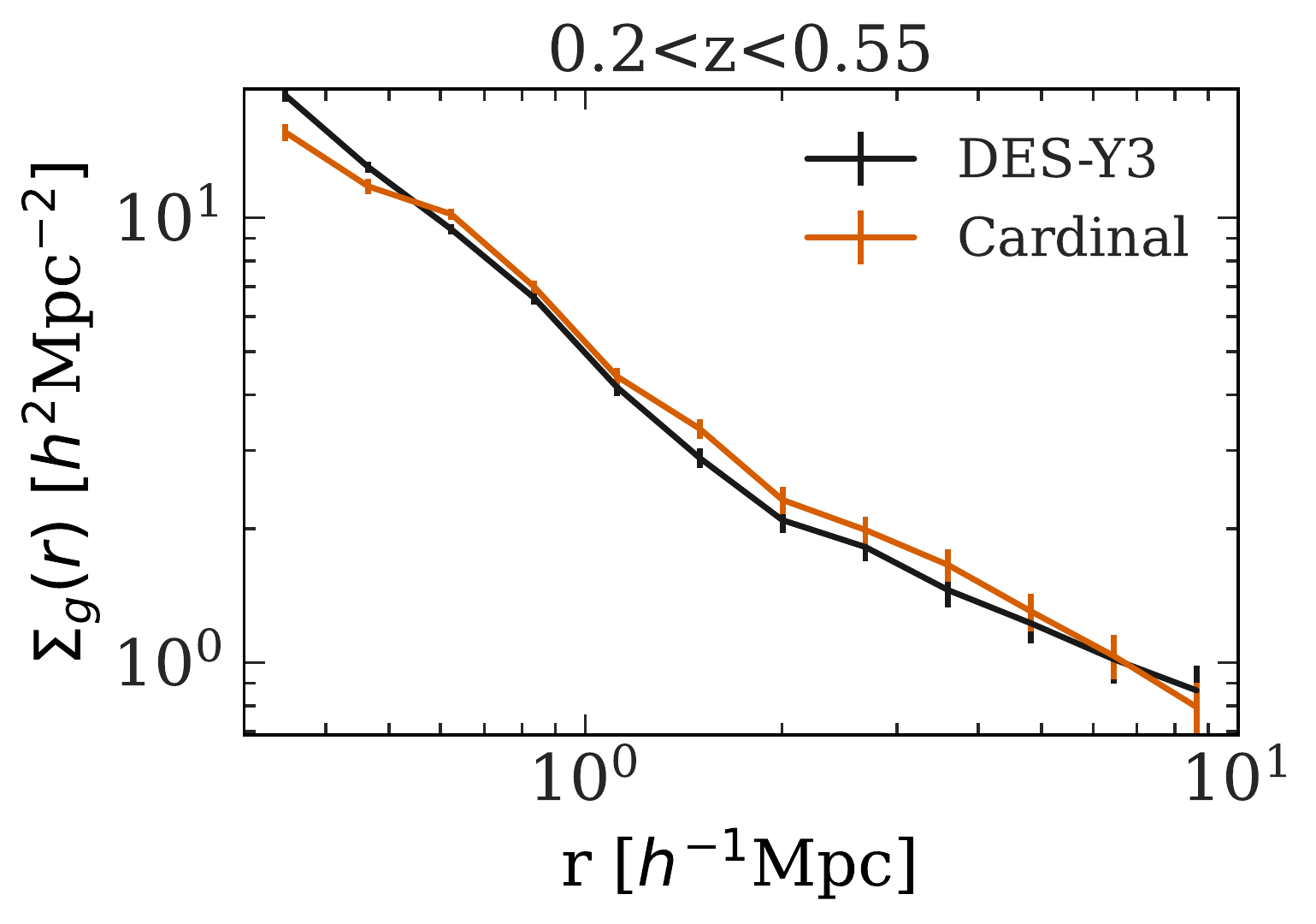}
    \caption{Stacked surface densities of galaxies around \redmapper{} clusters with $\lambda>20$ and redshift $0.2$ to $0.55$. Errorbars show $1\sigma$ uncertainties estimated using $64$ jackknife resampling. Black dots show measurements for DES-Y3 galaxies and red dots show measurements for Cardinal.}
    \label{fig:radialprofile}
\end{figure}

\section{Comparison to DES Y3 observations}
\label{sec:compare}
Now that we have a realistic galaxy catalog with observed magnitudes down to survey depth limits, we now proceed to select various cosmic structure tracers and compare their properties with the DES-Y3 data.

\subsection{\redmapper{} clusters}
In this section, we compare clusters in mocks and DES-Y3 data. As described in section \ref{sec:redmapper}, \redmapper{} is the main cluster sample in DES cluster cosmology analyses. 
Observationally, \redmapper{} clusters are selected based on their richness ($\lambda$). For example, DES-Y1 cosmology analyses (\citealt{DES_cluster_cosmology}, \citealt*{4x2pt2}) consider $\lambda>20$ \redmapper{} clusters as cosmological samples. 

The simplest way to generate \redmapper{} clusters in mocks is to select dark matter halos containing more than $20$  bright red-sequence galaxies or some other value to match the observed abundance. %
However, the richness value of \redmapper{} clusters in the data contains significant components from correlated large-scale structure along the line of sight due to redshift uncertainties \citep{Matteoprojection, 2022arXiv220503233S}. %
Further \cite{DES_cluster_cosmology}, \cite{Tomomi}, \cite*{4x2pt1}, and \cite{Heidiselection} show that these projected components in richness can cause biases in the two-point correlation functions of \redmapper{} clusters, including cluster lensing,  cluster--galaxy cross-correlations, and cluster clustering. Therefore, counting the numbers of red-sequence galaxies within three-dimensional distance to halos in simulations is not the same as \redmapper{} richness, making comparisons with observed \redmapper{} clusters hard to interpret. 

\cite{Matteoprojection, Tomomi, Heidiselection} present an improved way to simulate \redmapper{} richness by counting the numbers of red-sequence galaxies within cylinders along the line of sight. This approach provides a numerically efficient way to simulate \redmapper{} richness that does not require lightcone simulations. The problem with this approach is that \redmapper{} does not weigh galaxies uniformly along the line of sight; instead, galaxies that are further away from clusters along the line of sight usually have smaller $p_{\rm{mem}}$. Thus, counting galaxies within a cylinder is likely to overestimate the richness of clusters. One way to remedy this problem is to measure the $p_{\rm{mem}}$ distributions of galaxies as a function of spectroscopic redshifts. However, obtaining spectroscopic redshifts for galaxies down to $0.2L^*$ is challenging. 

Finally, one could run \redmapper{} algorithm on realistic mock galaxy catalogs to ensure apples-to-apples comparison of simulated and observed clusters. However, the richness value of \redmapper{} clusters depends on member galaxies' colors, magnitudes, and positions, making it hard to reproduce in simulations. Therefore, \redmapper{} has only been applied to a limited number of mock catalogs \citep{cosmodc2, JoeBuzzard}. 

Given the realistic red-sequence galaxies in Cardinal, we adopt this last approach to generate \redmapper{} clusters. We select galaxy clusters using \redmapper{} v0.8.4 \footnote{\url{https://github.com/erykoff/redmapper}}, which includes several improvements relative to its predecessors \citep{Redmapper1, redmappersv, Tomclusterlensing}. This includes a complete adaptation of the code from IDL to python and improved models of red-sequence galaxies. This additional improvement will affect the richness of \redmapper{} clusters; therefore, we run the same \redmapper{} on both Buzzard v2.0 \citep{y3buzzard} and publicly available DES-Y3 data \citep{Y3gold} to ensure apples-to-apples comparisons. Figure \ref{fig:clusterabundance} compares the cluster abundances as a function of richness in four different redshift bins measured in Cardinal and DES-Y3 data. We find that the data and Cardinal generally agree except for the lowest redshift bins. This might indicate differences between the redshift evolution of the richness--mass relations in Cardinal and the data. %
In comparison, the blue lines in figure \ref{fig:clusterabundance} show the cluster abundances measured in Buzzard v2.0. Cardinal demonstrates a much better agreement with the data in all redshift bins. There are three main differences between the Cardinal model and the Buzzard model, which are 
\begin{enumerate}
\item The training SHAM model in Cardinal includes orphan subhalos as detailed in section \ref{sec:SHAM}. 
    \item The color assignment model that produces environment-dependent galaxy color distributions is improved. We detailed this step in section \ref{sec:paintcolor}. 
    \item We include an additional abundance matching step to address the remaining color inconsistencies due to the insufficient training spectra and SED templates. We detailed this step in section \ref{sec:camcolor}.
\end{enumerate}
To better understand how these steps affect the cluster abundances, we run \redmapper{} on two additional realizations. First, we only update the training SHAM model in Cardinal but fix other color assignment models the same as Buzzard v2.0. The \redmapper{} clusters identified in this realization are shown as the orange line in figure \ref{fig:clusterabundance}. We can then compare this with clusters in Buzzard v2.0 to see the impact of SHAM models on the \redmapper{} cluster abundances. We note that this comparison depends on redshift and richness. For the simplicity of the argument, let us focus on the $\lambda = [30, 40]$ and $z_\lambda=[0.4, 0.5]$ bin. We find that including orphan models can boost cluster abundance with richness $\lambda>20$ by roughly $50$ percent, which is insufficient to account for the factor of three differences in cluster abundances between Buzzard v2.0 and data. Second, we update the training SHAM and color assignment models for environment-dependent galaxy color distributions. The redmapper{} clusters in this realization are shown as the green lines in figure \ref{fig:clusterabundance}. Combining the color assignment models and the SHAM model can boost the richness by $\sim 100$ percent. Finally, comparisons of the full Cardinal model and the green lines show that the additional color abundance matching step can boost the richness by another $\sim 100$ percent, making the cluster abundances in Cardinal agree with the DES-Y3 data. We, therefore, conclude that all three modifications of the Buzzard v2.0 model are needed to reproduce the observed cluster abundances. In appendix \ref{app:conditional}, we further compare the conditional luminosity functions of satellites and centrals of \redmapper{} clusters in Cardinal, Buzzard v2.0, and DES-Y3 data. 

Figure \ref{fig:radialprofile} further compares the stacked galaxy surface density around \redmapper{} clusters in DES-Y3 data and Cardinal, calculated using the method detailed in appendix \ref{app:radial profile}. We find decent agreements between Cardinal and DES-Y3 data. Compared to Buzzard v2.0 (e.g. figure 8 in \citealt{addgals}), the one-halo regime agrees with the data much better. This is likely due to the better match of richness values for a given cluster number density in Cardinal. On the other hand, the consistency on large scales is very interesting. As shown in \cite*{4x2pt2, 4x2pt1}, \redmapper{}--\redmagic{} cross-correlations in Buzzard has a much larger large-scale bias compared to DES-Y1 data. If this excess of large-scale bias is related to galaxy cluster samples, figure \ref{fig:radialprofile} suggests that Cardinal will not have this problem. We leave further investigations on this aspect in future work.

\begin{figure}
    \centering
    \includegraphics[width=0.5\textwidth]{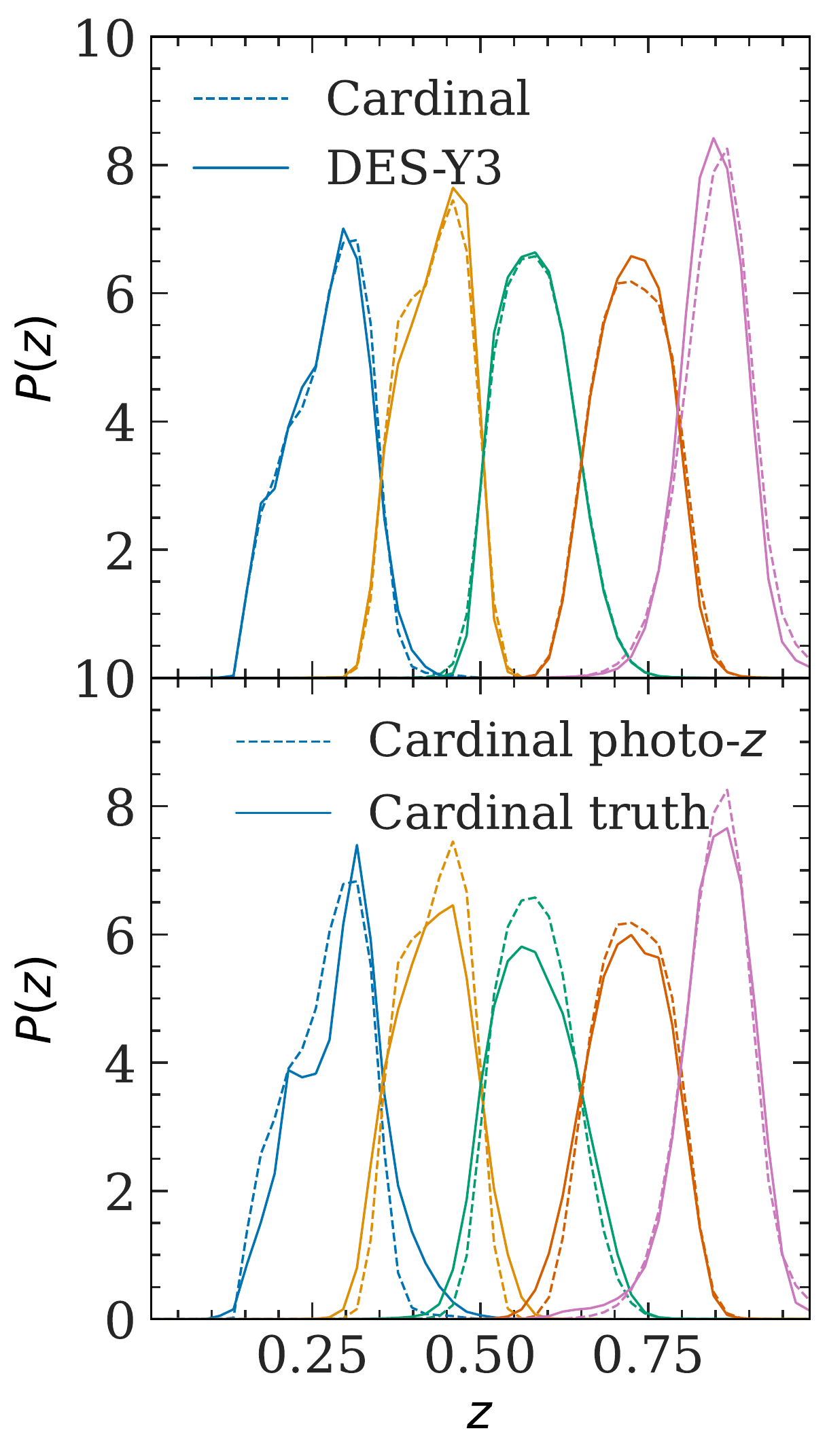}
    \caption{Comparison of redshift distributions. The top panel shows redshift distributions estimated by \redmagic{} in Cardinal (dashed) and data (solid). The bottom panel compares true (solid) and photometric redshift distributions estimated by \redmagic{} (dashed) in Cardinal. Different colors show different tomographic bins. } 
    \label{fig:redmagicredshift}
\end{figure}

\begin{figure*}
    \centering
    \includegraphics[width=0.99\textwidth]{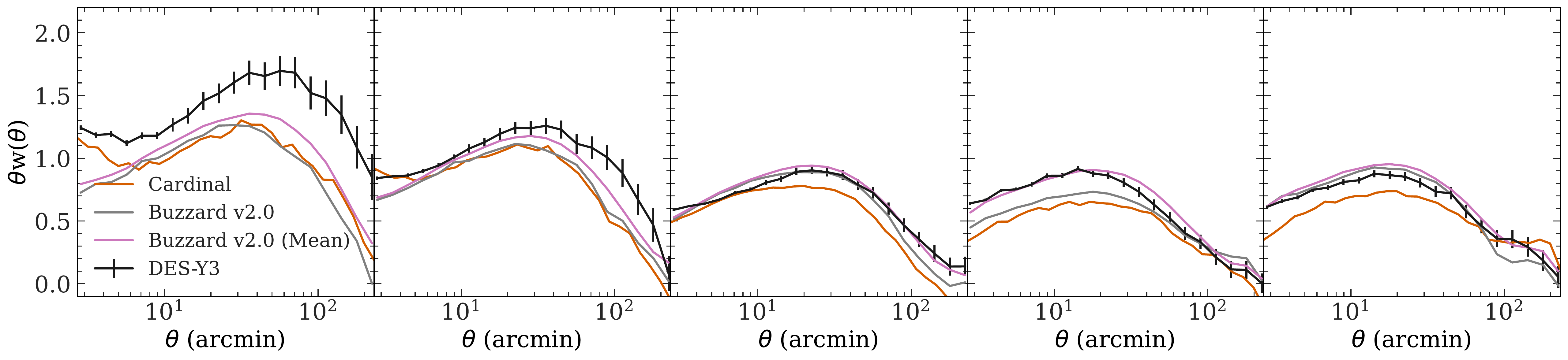}
    \caption{Comparison of \redmagic{} clustering in Cardinal (red lines) and DES Y3 data (black lines). Five different panels show the five tomographic bins based on $z_{\rm{redmagic}}$, with edges at $[0.15, 0.35, 0.5, 0.65, 0.8, 0.9]$. Error bars show $1 \sigma$ confidence region. For comparison, blue lines show the galaxy clustering of Buzzard v2.0 using the same dark matter simulations as Cardinal. The magenta lines show galaxy clustering of Buzzard v2.0 averaged over 18 DES-Y3 realizations \citep{y3buzzard}.}
    \label{fig:redmagicclustering}
\end{figure*}
\subsection{\redmagic{} galaxies}
As described in section \ref{sec:redmagic}, \redmagic{} galaxies are one of the main galaxy samples in the DES cosmological analyses \citep{DESY1KP, Y3kp}. Given the realistic galaxy cluster properties in Cardinal, we apply the \redmagic{} algorithm on Cardinal galaxies in the same way as \redmagic{} run on the DES-Y3 data. Specifically, \redmagic{} galaxies are selected using redshifts of \redmapper{} clusters instead of the true redshifts in the simulations. This allows us to have realistic galaxy selection-related systematics in Cardinal. Following the DES-Y3 cosmology analysis, we further bin \redmagic{} galaxies into five tomographic bins with edges, $[0.15, 0.35, 0.5, 0.65, 0.8, 0.9]$, using $z_{\rm{redmagic}}$, the redshift that maximizes  $p(z_{\rm{redmagic}})$. We then Monte-Carlo sample $p(z_{\rm{redmagic}})$ for each redshift bin to estimate their redshift distributions. 

Figure \ref{fig:redmagicredshift} compares the redshift distributions of \redmagic{} galaxies in Cardinal and the data. We find that the \redmagic{} estimated redshift distributions are very similar to those in DES-Y3 data. This agreement is non-trivial. Although the same algorithm produces both samples, this algorithm is applied to two different galaxy catalogs. This similarity in redshift distributions demonstrates the realism of red galaxy properties in Cardinal. In the bottom panel of figure \ref{fig:redmagicredshift}, we compare the redshift distributions estimated by \redmagic{} and the true redshift distributions in the simulation. While the redshift distributions estimated by \redmagic{} are slightly narrower compared to the true redshift distributions, the overall shapes of redshift distributions are very consistent between \redmagic{}'s estimations and the true distributions. 

We proceed to access the clustering properties of \redmagic{} in Cardinal. We first measure the galaxy correlation functions $w(\theta)$ using the weighted version of the Landy-Szalay estimator \citep{1993ApJ...412...64L}, 
\begin{eqnarray}
\label{eq:ls}
        w(\theta_{\rm{min}}<\theta<\theta_{\rm{max}}) &=& \sum_{i,j} \frac{w_i w_j}{\sum_i w_i(\sum_j w_j-1)}- 2\Sigma_{i,R}\frac{w_i}{\sum_i w_i N_R}\nonumber\\
        && +
    \sum_{RR} \frac{1}{N_R (N_R-1)}, 
\end{eqnarray}
where $i, j$ goes through each galaxy pair that have angular speparation $\theta_{\rm{min}}<\theta<\theta_{\rm{max}}$. The $w_i$ and $w_j$ are systematic weights associated with galaxies to remove spurious correlations of galaxies and survey depths. We detail the construction of these weights in appendix \ref{app:syswight}. The index R corresponds to randoms that describe the survey footprint. In this analysis, we make the number of randoms ($N_R$) $10$ times greater than the number of galaxies in each tomographic bin to remove additional shot noise caused by the randoms. The $i, R$ and $RR$ in equation \ref{eq:ls} go through each galaxy--random and random--random pairs that are separated by $\theta$. We compare the measured $w(\theta)$ of Cardinal, Buzzard v2.0, and DES-Y3 data in figure \ref{fig:redmagicclustering}. To avoid cosmic variance, we compare Cardinal and Buzzard v2.0 generated with the same dark matter simulation. In the lower two redshift bins, the \redmagic{} clustering in Cardinal has a stronger one-halo term than Buzzard v2.0. This is likely due to the combination of changes in the color assignment and the inclusion of orphan galaxies. This is consistent with the finding compared to SDSS galaxies, which are galaxies at $z\simeq 0.1$ (figure \ref{fig:addgal_color}). For all redshift bins, the clustering in Cardinal is somewhat smaller than that of the DES-Y3 data. To better understand the origin of this inconsistency, we compare the clustering of \redmagic{} in Cardinal, and Buzzard v2.0 generated with the same dark matter simulation. We find that for the first and second redshift bins, the clustering of \redmagic{} in Cardinal is consistent with Buzzard v2.0. For the third redshift bin, the deficit of clustering is likely due to the slightly worse photometric redshift performances in Cardinal compared to Buzzard v2.0. For the fourth and fifth redshift bins, we find that the \redmagic{} samples in these redshift ranges are slightly fainter than Buzzard v2.0 and data, which could potentially explain the differences in clustering.

\section{Conclusions and outlook}
\label{sec:conclu}
Large simulations with realistic galaxies are increasingly essential in precision cosmological analyses based on large surveys. One of the widely-used multi-purpose simulations is the Buzzard simulation, which produces DES, DESI, and LSST-like mock catalogs out to $z=2.35$ to a depth of $m_{\rm{r}}=27$ \citep{JoeBuzzard, y3buzzard, addgals}. This simulation played an essential role in the core cosmology analyses for DES and in planning and methodology development for several other surveys.
In this paper, we introduce several model improvements to the Buzzard catalogs and generate a new set of mock catalogs, the Cardinal mocks. The main improvements are listed below, with a more detailed list in appendix \ref{app:improvement}. 
\begin{enumerate}
    \item We update the subhalo abundance matching model (SHAM) used to generate the Buzzard simulation. The new SHAM model considers orphan galaxies and a flexible disruption model and incorporates mass-dependent scatter between galaxy luminosity and halo properties. For the first time, the SHAM model can simultaneously fit galaxy clustering and group--galaxy cross-correlations in three different luminosity thresholds measured in SDSS (Section \ref{sec:SHAM}). 
    \item A new color assignment model is developed to produce the environmentally dependent galaxy colors accurately. For the first time, the color assignment model can simultaneously fit color-dependent galaxy clustering in three different luminosity bins measured in SDSS (Section \ref{sec:paintcolor}). 
    \item We identify two causes of the discrepancy between the color distribution in mocks and data. These include the need for redshift evolution in the spectroscopic training data and the insufficiency of summarizing galaxy colors using current SED templates. We provide a solution that uses photometric data and conditional abundance matching techniques. Applying this conditional abundance matching scheme, the apparent magnitude and color--color distributions are much more consistent with the DES-Y3 data (Section \ref{sec:camcolor}). 
    \item We address the lack of lensing shear due to limited ray-tracing resolution. We develop a novel method that uses dark matter particle--halo cross-correlations to fix this problem. We find that the $\Delta \Sigma$ around massive halos is more consistent with expectations after this correction is applied (Section \ref{sec:lensing}). 
\end{enumerate}

We incorporate these improvements into the \textsc{Addgals} algorithm to generate the Cardinal mock, a one-quarter sky simulation out to $z=2.35$ to a depth of $m_{\rm{r}}=27$. We further cut out one DES-Y3 footprint and apply realistic DES-Y3 photometric errors and sky backgrounds. The latest \redmapper{} cluster finding algorithm and \redmagic{} lens galaxy algorithm are also run on the catalogs to produce realistic DESY3-like cluster and lens samples (summarized in figure \ref{fig:cardinalflow}). 

We compare the Cardinal mocks with DES-Y3 and SDSS data. These comparisons include the abundance of \redmapper{} clusters, the projected galaxy density profiles around \redmapper{} clusters, the redshift distribution of \redmagic{} galaxies, and galaxy clustering for various samples. In the cluster abundance comparison, we find that the Cardinal clusters have a much more consistent number density to the data as a function of richness and redshift than those identified in the Buzzard simulation (Figure \ref{fig:clusterabundance}). We further make the comparison by changing one model component at a time. We find that two main factors contribute to the long-standing issue of lower cluster number densities in the Buzzard simulation, where the cluster abundance in simulations is $10$-$25$ percent of the cluster abundance in the data at the same richness. One of these is the previously postulated lack of orphans in the SHAM model \citep{addgals, JoeBuzzard}, and another is the galaxy color assignment model. We further find galaxy profiles around \redmapper{] clusters in Cardinal are in excellent agreement with the data (figure \ref{fig:radialprofile}). As for the \redmagic{} galaxy comparison, we find that the Cardinal algorithm produces reasonable \redmagic{} galaxy properties compared to the DES-Y3 data. These include the redshift distributions of the lens samples and galaxy clustering in each tomographic bin. Finally, we compare color-dependent galaxy clustering between Cardinal and SDSS data and find excellent consistency. 

With the updated model, which includes more realistic small-scale galaxy clustering and lensing properties, we expect many applications using the Cardinal mock. We list a few potential applications below. 
\begin{enumerate}
    \item Optically-selected cluster cosmology: Cardinal is a valuable tool to quantify the selection bias in optically selected clusters, which affects observables such as weak lensing profiles because cluster selection is based on richness. This bias has been identified as the dominant systematic in optical cluster cosmology (\citealt{Tomomi}; \citealt*{4x2pt2, 4x2pt1}; \citealt{Heidiselection}). Cardinal is one of the very few mocks that the \redmapper{} cluster finder can be run on, and it is so far the \textit{only} mock that yields realistic \redmapper{} clusters as a function of richness and redshift without artificially boosting cluster member galaxies. Future work will use Cardinal simulations to understand the selection functions of \redmapper{} clusters and the level of selection biases and to motivate a flexible model to mitigate these biases.
    \item Photometric reshifts: given the more consistent galaxy color distributions to the data (figure \ref{fig:1dcomp}), Cardinal is a valuable tool to assess the uncertainties of photometric redshifts \citep{2021MNRAS.505.4249M}. For example, one could use Cardinal mock to perform controlled experiments to quantify and develop programs to mitigate various photometric redshift calibration systematics \citep[see e.g.][for a review]{2022ARA&A..60..363N}. This includes (a) how heterogeneous selections of spectroscopic redshift samples and redshift errors in deep fields of cosmological surveys can affect photometric redshift calibrations of wide-field galaxies, (b) how astrophysical systematics can affect the effectiveness of cross-correlation redshifts, such as lensing magnifications, non-linear clustering, and bias evolutions, and (c) how the sample variances due to limited area of deep fields propagate to redshift uncertainties of wide-field galaxies. 
    \item Quantify small-scale lensing systematics: the new empirical correction of the ray tracing algorithm (section \ref{sec:lensing}) makes Cardinal accurately produce one-halo term lensing profiles. Cardinal can then serve as a tool to study systematic effects of the small-scale galaxy--galaxy and cluster lensings. This includes (a) developing methods to quantify boost factor measurements of cluster lensing \citep{2019MNRAS.489.2511V} using different photometric redshift estimations and (b) validating small-scale galaxy--galaxy lensing and clustering models based on halo occupation distribution (HOD) or conditional luminosity function (CLF). 
    \item End-to-end tests of multi-probe analyses: with realistic clusters, galaxies, and lensing properties in the same simulation, Cardinal is the ideal mock to develop a combine-probe analysis pipeline that aims to perform joint analyses of auto- and cross-correlations of these observables (\citealt{Niall}; \citealt*{4x2pt2, 4x2pt1}; \citealt{y3buzzard}).
    \item Defining optimal galaxy samples for clusters, galaxies, and lensing joint analyses: one of the limitations of forward modeling multi-probe analyses is that different cosmological probes rely on different galaxy samples. Under the framework of HOD models, different galaxy samples require a different set of galaxy--halo connection parameters, making parameter inferences harder and  diluting cosmological signals. One interesting question is whether it is possible to define a common galaxy sample that serves as lens and source galaxies as well as can be used to identify galaxy clusters. Cardinal can serve as a sandbox to develop this project. 
\end{enumerate}

Cardinal can be further improved in the following directions. First, the current model for generating the colors and magnitudes of galaxies that are dependent on their environment only considers information measured in the same snapshot, even though the formation history of a galaxy plays an important role in determining its colors and brightness. The reason for this choice is purely computational: accurately determining the formation history of halos requires simulations with a high level of resolution, which in turn requires significant computing resources. However, recent developments in the hybrid Lagrangian perturbation theory model can help bypass this problem. For example, \cite{2020MNRAS.492.5754M, Nick} show that using the local density of dark matter with proper weights based on initial conditions can model galaxy clustering of mock galaxies generated using the halo occupation model down to $k~=0.7\,h^{-1}\rm{Mpc}$. The success of these models in producing small-scale galaxy clustering indicates a significant amount of information about galaxies encoded in the initial conditions beyond the dark matter density measured in the same snapshot.
Furthermore, \cite{2022MNRAS.515.2164L} find that the initial conditions play an essential role in determining the inner part of the dark matter profiles. Thus, we expect that combining initial conditions and local density will significantly improve the accuracy of the color and magnitude assignment models. Moreover, incorporating initial conditions does not place additional requirements on simulation resolution, and thereby does not require significant additional computational power. Second, we find that the SED templates cannot summarize galaxy colors with uniform accuracy across colors. In Cardinal, we use DES photometric catalogs to solve this problem empirically. However, by doing this, we lose some ability to generate mock catalogs with arbitrary photometric bands. Future work will expand the SED templates using advanced stellar population synthesis models and the Dark Energy Spectroscopic Instrument (DESI) data. Third, we empirically address the lack of redshift evolution in the training spectra using the photometric data. Although this method works well in the end, it requires significant tuning to remove additional noise that broadens the red sequences. The main problem is that cluster finders require accurate red galaxy colors down to $0.2 L^*$, which is $m_{{\rm{z}}}=21.8$ at $z=0.6$. The spectroscopic training samples must go down to the same magnitude with uniform selections. This data is unavailable even with Stage-4 spectroscopic surveys (e.g., DESI). However, the Cardinal algorithm only requires a small area coverage for these spectroscopic samples. These requirements align well with the need for training spectra for photometric redshift calibrations. We anticipate future campaigns to obtain photometric redshift training spectra will be helpful for generating more accurate simulations. 

Finally, the current approach of generating simulations is bottom-up. We first obtain a decent luminosity-dependent galaxy clustering and group--galaxy cross-correlations. We employ conditional abundance matching techniques to paint colors to ensure reasonable color-dependent galaxy clusterings. We then perform several conditional abundance matching steps to fix problems in the color model when matching to photometric datasets. This approach has been successful but may not be scalable with the increasing amount of data. A top-down approach might scale better with a plethora of data coming in. With all the datasets and targeted summary statistics in hand, one can optimize a galaxy--dark matter connection model. Here, we specifically chose galaxy--dark matter connections instead of galaxy--halo connections because halo finding and halo definitions could suffer from limited resolution. For this top-down approach to perform well, work has to be done to avoid overfitting the data with an overly flexible model and to increase the interpretability of the constrained model. For the latter, one can derive conditional probabilities from the constrained model and combine that with the bottom-up approach to gain physical insights. We leave this interesting direction for future work.

Although Cardinal was primarily developed for validating cosmological analyses, the method has the potential for broader applications. The ultimate goal of the program is to learn the physics of the universe from the vast amounts of data collected by cosmological surveys and use this knowledge to create a mock universe with higher fidelity. However, to achieve this goal, it is necessary to have high resolution to resolve important physical processes and sufficient volumes to leverage the full constraining power of the data, making the program computationally demanding. One of the critical features of Cardinal is connecting galaxy properties between high-resolution, small-volume simulations and low-resolution, large-volume simulations. While the current connection in Cardinal is limited to galaxy luminosity, one can extend this to other properties, such as rest-frame galaxy colors, sizes, ellipticities, gas densities, and temperatures. Using improved connections, replacing the SHAM training model with first principle-driven models (e.g. SAMs or hydrodynamical simulations), and employing emulator techniques, one can constrain physics of the baryonic component of our universe, the expansion histories of the universe, the growth of cosmic structures, and the interplays between these models using the full power of the survey data. At the same time, one can use this learned physics to create mocks with rich information to facilitate multi-wavelength and multi-probe cosmology analyses. However, significant challenges remain in making the first principle-driven models of baryons computationally efficient, identifying relevant physics,  and judiciously selecting informative observational data or summary statistics derived from them. 

Finally, although we have been focusing on optical surveys, the Cardinal simulations can also inform many multiwavelength studies. The dark matter halos in Cardinal are resolved down to $M_{\rm{200m}}=10^{13} \,h^{-1}M_\odot$, making the simulation sufficient to paint SZ and X-ray signals. Future work will generate realistic SZ and X-ray groups and clusters, making Cardinal useful for combined probe analyses of SZ, X-ray, and optical observations. %

\section*{Acknowledgements}
\label{sec:acknowledgements}
We are grateful to Matthew Becker, Andrew Hearin, and Daisuke Nagai for valuable comments that significantly improved this work. CHT thanks Yun-Ting Cheng for helpful comments on the draft, Daniel Gruen and Justin Myles for enlightening discussions on applications of photometric redshifts, John Moustakas for helpful discussions on SED templates, and Peter Taylor for answering various math and statistics questions. Part of the analysis was facilitated by illustris-TNG simulation \citep{2019ComAC...6....2N}. CHT acknowledges the team for making simulations publicly available.
This work received support from the 
U.S. Department of Energy under contract number DE-AC02-76SF00515 to SLAC National Accelerator Laboratory. CHT is supported by the United States Department of Energy, Office of High Energy Physics under Award Number DE-SC-0011726. JD is supported by the Lawrence Berkeley National Laboratory Chamberlain Fellowship. DHW acknowledges support of NSF grant AST-1516997. HYW is supported by the DOE award DE-SC0021916 and the NASA award 15-WFIRST15-0008. This research was partially supported by the Munich Institute for Astro-, Particle and BioPhysics (MIAPbP) which is funded by the Deutsche Forschungsgemeinschaft (DFG, German Research Foundation) under Germany´s Excellence Strategy – EXC-2094 – 390783311.

Some of the computing for this project was performed on the Sherlock cluster. We thank Stanford University and the Stanford Research Computing Center for providing computational resources and support that contributed to these research results. Additional computations in this paper were run on the CCAPP condo of the Ruby Cluster at the Ohio Supercomputer Center. 

Some of the results in this paper have been derived using the healpy and HEALPix packages. This research made use of Astropy, a community-developed core Python package for Astronomy \citep{2018AJ....156..123A, 2013A&A...558A..33A}, SciPy \citep{Virtanen_2020}, Scikit-learn \citep{scikit-learn}, NumPy \citep{harris2020array}, matplotlib, a Python library for publication-quality graphics \citep{Hunter:2007}, IPython package \citep{PER-GRA:2007}, and Gotetra \url{https://github.com/phil-mansfield/gotetra}. 

This study made use of the SDSS DR7 Archive 
for which funding has been provided by the Alfred P. Sloan Foundation, the Participating Institutions, the National Science Foundation, the U.S. Department of Energy, the National Aeronautics and Space Administration, the Japanese Monbukagakusho, the Max Planck Society, and the Higher Education Funding Council for England. The SDSS Web Site is http://www.sdss.org/. The SDSS is managed by the Astrophysical Research Consortium for the Participating Institutions. The Participating Institutions are the American Museum of Natural History, Astrophysical Institute Potsdam, University of Basel, University of Cambridge, Case Western Reserve University, University of Chicago, Drexel University, Fermilab, the Institute for Advanced Study, the Japan Participation Group, Johns Hopkins University, the Joint Institute for Nuclear Astrophysics, the Kavli Institute for Particle Astrophysics and Cosmology, the Korean Scientist Group, the Chinese Academy of Sciences (LAMOST), Los Alamos National Laboratory, the Max-Planck-Institute for Astronomy (MPIA), the Max-Planck-Institute for Astrophysics (MPA), New Mexico State University, Ohio State University, University of Pittsburgh, University of Portsmouth, Princeton University, the United States Naval Observatory, and the University of Washington.  

This project used public archival data from the Dark Energy Survey (DES). Funding for the DES Projects has been provided by the U.S. Department of Energy, the U.S. National Science Foundation, the Ministry of Science and Education of Spain, the Science and Technology FacilitiesCouncil of the United Kingdom, the Higher Education Funding Council for England, the National Center for Supercomputing Applications at the University of Illinois at Urbana-Champaign, the Kavli Institute of Cosmological Physics at the University of Chicago, the Center for Cosmology and Astro-Particle Physics at the Ohio State University, the Mitchell Institute for Fundamental Physics and Astronomy at Texas A\&M University, Financiadora de Estudos e Projetos, Funda{\c c}{\~a}o Carlos Chagas Filho de Amparo {\`a} Pesquisa do Estado do Rio de Janeiro, Conselho Nacional de Desenvolvimento Cient{\'i}fico e Tecnol{\'o}gico and the Minist{\'e}rio da Ci{\^e}ncia, Tecnologia e Inova{\c c}{\~a}o, the Deutsche Forschungsgemeinschaft, and the Collaborating Institutions in the Dark Energy Survey.
The Collaborating Institutions are Argonne National Laboratory, the University of California at Santa Cruz, the University of Cambridge, Centro de Investigaciones Energ{\'e}ticas, Medioambientales y Tecnol{\'o}gicas-Madrid, the University of Chicago, University College London, the DES-Brazil Consortium, the University of Edinburgh, the Eidgen{\"o}ssische Technische Hochschule (ETH) Z{\"u}rich,  Fermi National Accelerator Laboratory, the University of Illinois at Urbana-Champaign, the Institut de Ci{\`e}ncies de l'Espai (IEEC/CSIC), the Institut de F{\'i}sica d'Altes Energies, Lawrence Berkeley National Laboratory, the Ludwig-Maximilians Universit{\"a}t M{\"u}nchen and the associated Excellence Cluster Universe, the University of Michigan, the National Optical Astronomy Observatory, the University of Nottingham, The Ohio State University, the OzDES Membership Consortium, the University of Pennsylvania, the University of Portsmouth, SLAC National Accelerator Laboratory, Stanford University, the University of Sussex, and Texas A\&M University. Based in part on observations at Cerro Tololo Inter-American Observatory, National Optical Astronomy Observatory, which is operated by the Association of Universities for Research in Astronomy (AURA) under a cooperative agreement with the National Science Foundation.

\appendix

\section{Varying scatter in the Subhalo Abundance Matching model}
\label{app:SHAMscatter}
This section describes an algorithm to incorporate varying scatter in the Subhalo Abundance Matching Model. Given an observed luminosity function $\hat{\phi} (L)$ and a halo mass function $N(M) $, one can relate the two functions via, 
\begin{eqnarray}
\label{eq:A1}
    \hat{\phi} (L) &=& \rm{Poisson} \left (\int P(L|L^\prime) N(L^\prime)\, d L^\prime \right)\nonumber, \\
    N(L^\prime)&: =& \int D(L^\prime, M) N(M) \, dM,
\end{eqnarray}
where $\rm{Poisson}(x)$ represents a realization from the Poisson distribution with mean $x$, and $D(L, M)$ represents the mapping such that $N(>L) = N(>M)$. This mapping ($D(L, M)$) can be understood as assigning the brightest galaxy to the most massive halos. In equation \ref{eq:A1}, $L^\prime$ is the luminosity without scatter, and $P(L|L^\prime)$ represents the relations between $L^\prime$ and the observed luminosity of galaxies $L$. In most of the Subhalo Abundance Matching Models, $P(L|L^\prime)$ is assumed as a Gaussian distribution with a constant scatter. Under this assumption, equation \ref{eq:A1} can be considered a convolution problem. One can then obtain $N(L^\prime)$ from the observed $\hat{\phi} (L)$ using the Richardson-Lucy algorithm \citep{1974AJ.....79..745L}.

However, much evidence has shown that the scatter in $P(L|L^\prime)$ varies according to halo masses (see \citealt{risaawesomepaper} for a review). %
Thus, we want to incorporate this modeling flexibility into our model. We adopt $P(L|L^\prime)$ as a Gaussian distribution with a scatter $\sigma(L^\prime)$ defined as, 
\begin{eqnarray}
\sigma(L^\prime) = A+B(-2.5 \log (L^\prime)+C),
\end{eqnarray}
where A, B, and C are free parameters. With this parameterization, equation \ref{eq:A1} remains linear. The linearity allows one to solve it using the same Richardson-Lucy algorithm, even though it cannot be phrased as a convolution problem. 

Figure \ref{fig:app:abundance} compares the performance of our implementation of the Richardson-Lucy algorithm to a widely used abundance matching code \href{https://github.com/yymao/abundancematching}{AbundanceMatching}. Evidently, our implementation reaches a better accuracy than the existing code. Further, we evaluate the performance of our algorithm in the case of non-constant scatter. We find that our algorithm can achieve an accuracy of $10$ percent for the best-fit model of this paper.

\begin{figure*}
    \centering
    \includegraphics[width=0.99\textwidth]{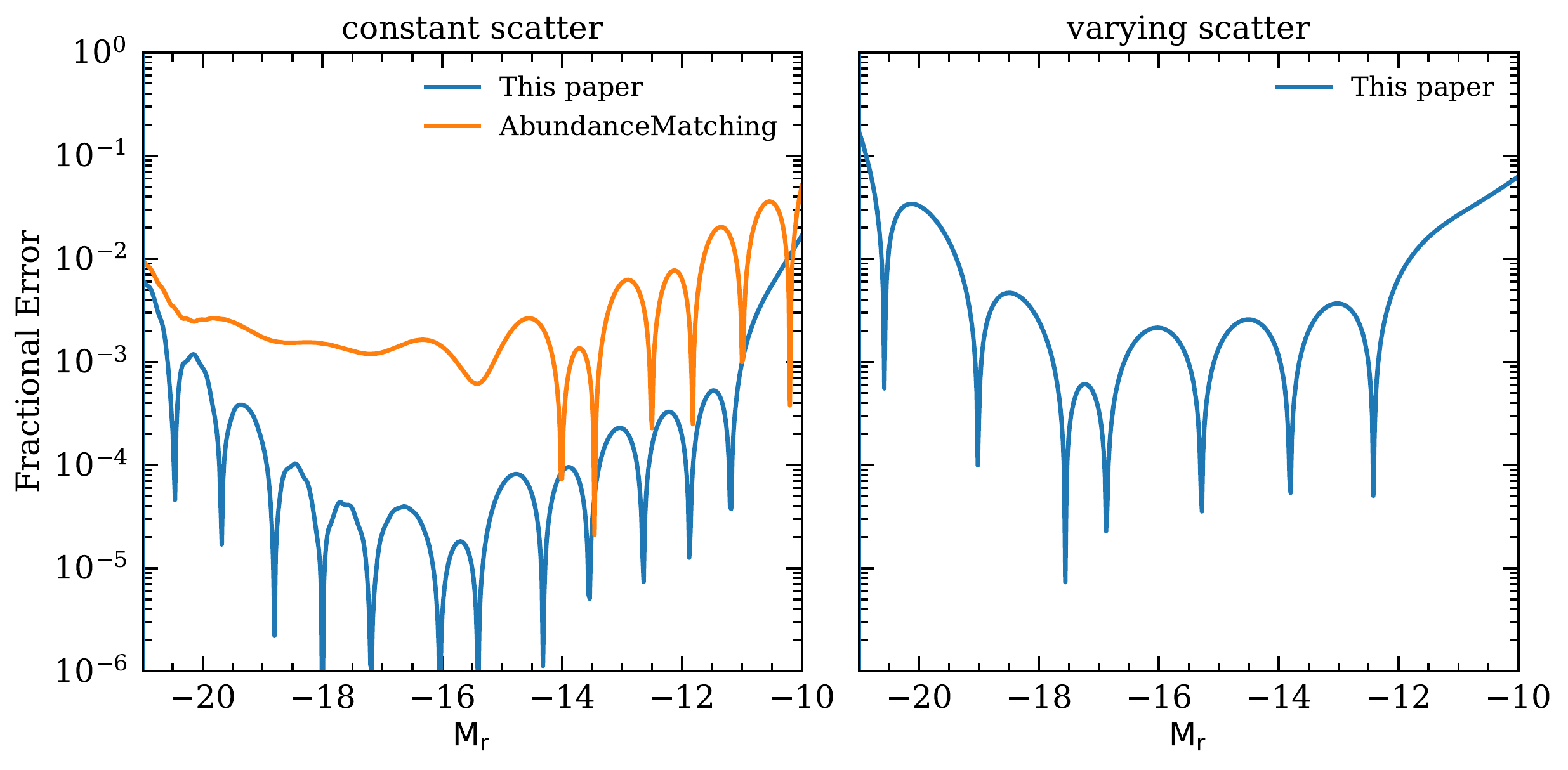}
    \caption{Performance of the algorithm. The y-axis shows the relative error caused by the Richardson-Lucy process, evaluated as $\rm{conv}\left(\rm{deconv}\left(\phi(L)\right)\right)/\phi(L)-1$, where conv shows the process in equation \ref{eq:A1}, deconv shows the Richardson-Lucy process, and $\phi(L)$ shows the original luminosity function. If $P(L|L^\prime)$ is a Gaussian distribution with a constant scatter, conv is a convolution operation and deconv is a deconvolution operation.  The $x$-axis shows $-2.5 \rm{log} L$. In the first panel, we perform the process on a constant scatter (0.2 dex), where we can compare the performance of our implementation (blue line) to other existing codes (orange). In the second panel, we perform the process on varying scatter using the best-fit parameters in this paper. }
    \label{fig:app:abundance}
\end{figure*}

\section{SHAM emulator}
\label{app:surrogate}
We build an emulator for the SHAM model to facilitate the likelihood inference. We first generate the training data by Latin hypercube sampling of the priors described in table \ref{tab:shamparam}. We then generate the data vector at each sample following the procedure described in section \ref{sec:SHAMmodel}. We build an emulator for each radial bin of the data vector. Specifically, given the training data and the associated data vector at a given radial bin, we construct the model using the Polynomial Chaos Expansion implemented in \textsc{Chaospy} \citep{FEINBERG201546}. In short, we first model the objective function that maps training data to the associated data vector as 
\begin{equation}
    f = \sum_{\alpha} c_\alpha m_\alpha, 
\end{equation}
where $c_\alpha$ are free parameters and $m_\alpha$ is a set of orthogonal polynomials up to a certain degree constructed using the discretized Stieltjes procedure \citep{stieltjes1884quelques}. We determine $c_\alpha$ by minimizing the mean squared error of the input data vector.  Because $m_\alpha$ are polynomials, this minimization can be done analytically. In practice, we find that using $1000$ training data and including polynomials up to the third degree are sufficient to achieve accuracy so that the uncertainties associated with the emulator are subdominant of the error budget. 

To assess the accuracy of the emulator, we perform the leave-one-out test: iteratively removing one training point off the training sets, training an emulator, and calculating the differences between the removed training point and the emulator's prediction. Figure \ref{fig:app:emu} shows the result of the leave-one-out test. The mean of the error is shown as white dots.  Evidently, the error is much smaller than the total error budget. We, therefore, ignore the emulator errors in the likelihood inferences.

\begin{figure*}
    \centering
    \includegraphics[width=0.99\textwidth]{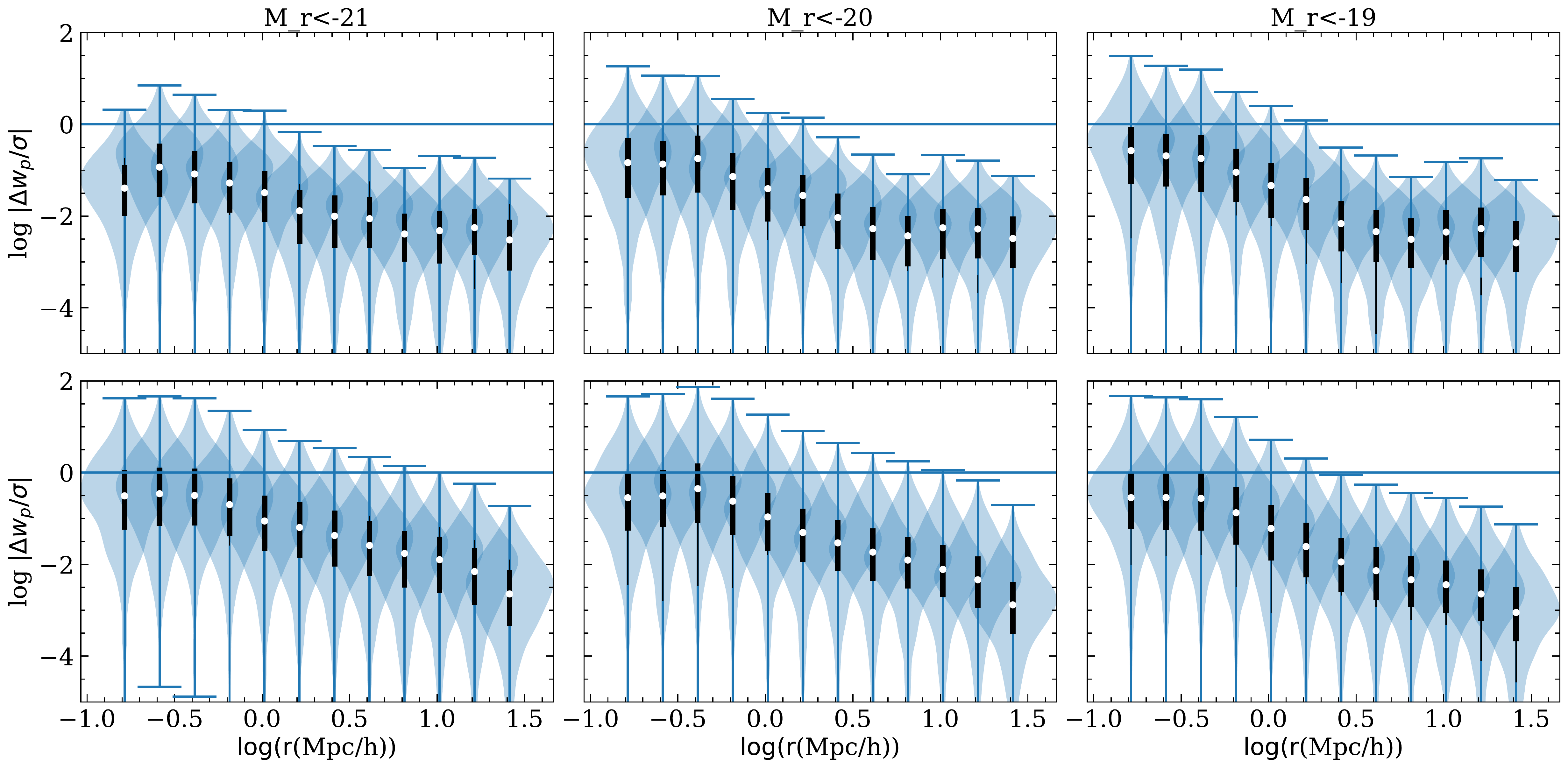}
    \caption{Accuracy tests of the SHAM emulators. The top row corresponds to galaxy--galaxy group correlations in each magnitude bin and the bottom row shows  galaxy auto correlations. Each panel shows emulator errors relative to the total error budget. Shaded blue regions are constructed by performing the leave-one-out test: removing one training data, training emulators, and calculating the error of the emulator at the removed point. The mean of the distribution is indicated by white dots and the black line shows $25$ to $75$ percent quantiles. Evidently, the emulator errors are much smaller than the total error budget in all radial bins and can therefore be ignored in the likelihood inferences.}
    \label{fig:app:emu}
\end{figure*}

\section{Gaussian Process of $P(R_\delta|\Mr<x, z)$ model}
\label{app:gaussianprocess}
Following \cite{addgals}, we employ a Gaussian process model to interpolate parameters $\thetat(\Mr, z)$ of $P(R_\delta|\Mr<x, z)$ model determined from SHAM simulation. We first generate the training sample by measuring $\thetat(\Mr, z)$ on $18$ evenly spaced magnitude bins from $\Mr=-22.5$ to $\Mr=-18$ at each snapshot of \Tbox{}. We then interpolate the training data using the Gaussian process algorithm implemented using \textsc{george} \citep{2015ITPAM..38..252A}. That is,  we model the joint probability of the targeted $\thetat_p$ at parameter space $\vect{x=(\Mr, z)}$, and the training set $\thetat_t$ as, 
\begin{equation}
\label{eq:Gauss}
    \begin{bmatrix}
\thetat_p\\
\thetat_t
\end{bmatrix} \sim \rm{Gaussian} \left(0, 
    \begin{bmatrix}
K(\vect{x}, \vect{x})+\sigma^2_n \vect{I} & K(\vect{x}, \vect{x}_t)\\
K(\vect{x}_t, \vect{x}) & K(\vect{x_t}, \vect{x}_t), 
\end{bmatrix}
\right)
\end{equation}
where $\thetat_t= (\theta_1, \theta_2, \dots, \theta_n)$ is the measured $\thetat$ at $\vect{x}_t=((\Mr, z)_1, (\Mr, z)_2, \dots, (\Mr, z)_n)$, $\vect{I}$ is an identity matrix, $\sigma_n$ are free parameters to prevent overfitting. One can then find the $\thetat_p$ that maximizes the probability and use it as the interpolated value of $\thetat$ at $\vect{x=(\Mr, z)}$. The remaining task is then to determine $K$ and $\sigma_n$. We model $K(x_1, x_2)$ combination of exponential kernel and Matérn kernel that has the following forms, 
\begin{equation}
    K = \exp \left(\frac{-\|x_1-x_2\|^2}{l_1^2}\right) + \left(1+\frac{\sqrt{3}\|x_1-x_2\|}{l_2}\right)\exp\left(-\frac{\sqrt{3}\|x_1-x_2\|}{l_2}\right),
\end{equation}
where $l_1, l_2$ are free parameters. We can then determine $l_1, l_2, \sigma_n$ by maximizing equation \ref{eq:Gauss} evaluated on the training set $\thetat_t$. Once $l_1, l_2, \sigma_n$ is determined, we can make a prediction of $\theta_t$ at other parameters by maximizing equation \ref{eq:Gauss} analytically. 

Figure \ref{fig:app:gaussian} shows the Gaussian process interpolations of the training data on grids of test points as well as the training data. We find that the  Gaussian process model agrees with the training data and does not show signs of over-fitting.

\begin{figure}
    \centering
    \includegraphics[width=0.5\textwidth]{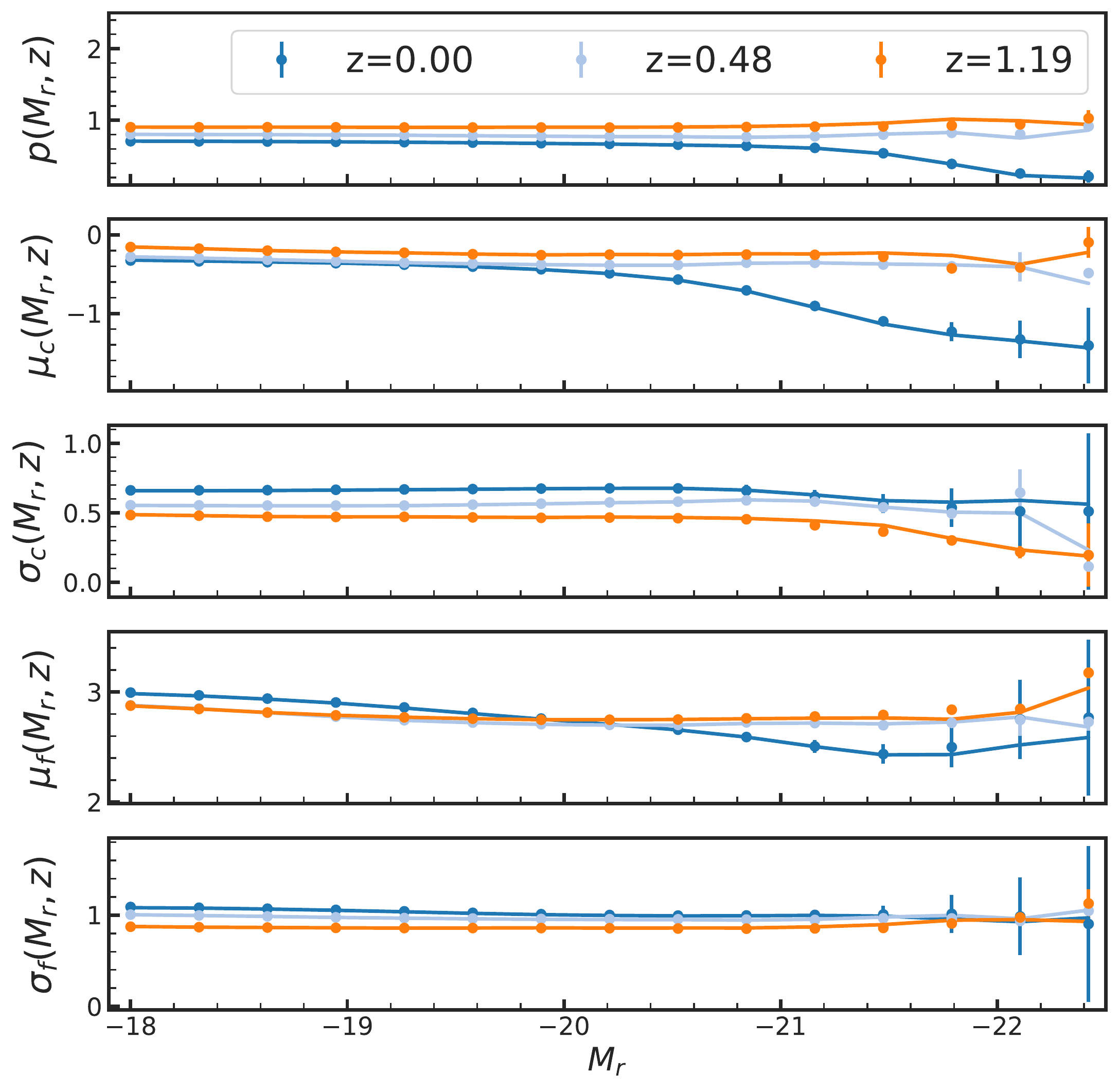}
    \caption{Parameters of $P(R_\delta | \Mr, z)$ model. The line shows the Gaussian process models for the redshift and magnitude dependence.}
    \label{fig:app:gaussian}
\end{figure}

\section{Systematic weight}
\label{app:syswight}
The observed galaxy densities can have spurious correlations due to varying observing conditions across the sky. While we do not have most of the varying observing conditions in the simulations, such as dust reddening, seeing, and stars, we have spatially varying survey depths in the simulation. These spatially varying survey depths will produce spurious correlations of galaxies. We remove this signal following the methods described in \cite{y3sys}. First, we bin the galaxy density into healpix maps with $\rm{nside}=512$ corresponding to a resolution of $0.11$ degrees. Next, we measure the relation of galaxy number density and the input survey depth $s$ used to generate Cardinal. Finally, we fit a linear function to this relation. The fitted function reads the form $F=ms+c$, where $m$ and $c$ are free parameters. The weight of each galaxy to remove the spurious correlations is then given by $1/F$. Figure \ref{fig:app:sysweight} shows the performance of this algorithm. One can see that the weights significantly suppress the correlations between galaxy density and survey depths. 

\begin{figure}
    \centering
    \includegraphics[width=0.5\textwidth]{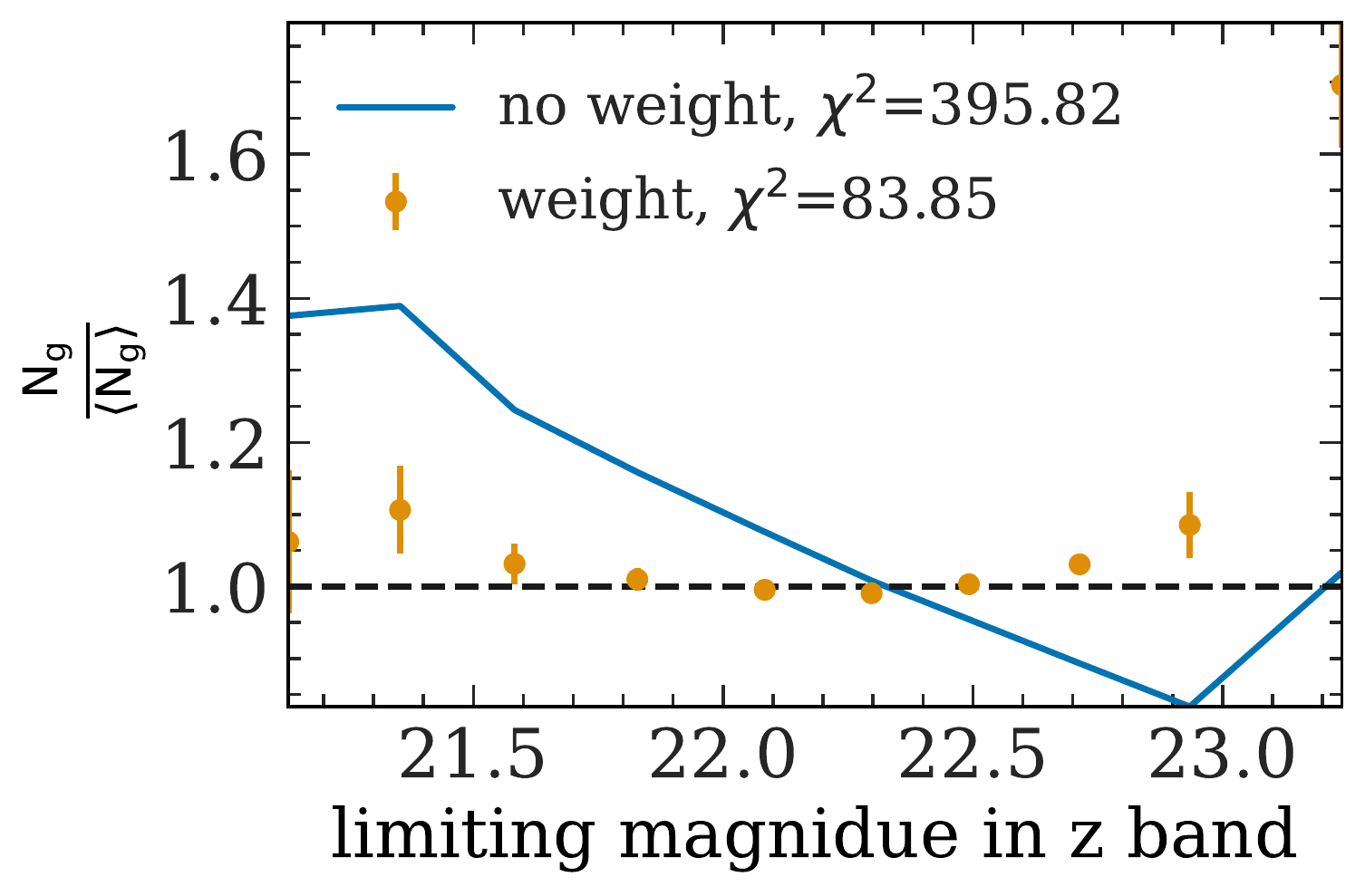}
    \caption{Performance of systematic weights on removing spurious correlations. The plot shows the observed pixel number densities evaluated on $\rm{nside}=512$ maps as a function of the limiting magnitudes of each pixel. The original \redmagic{} number density is shown as the blue line, and the weighted number density is shown as the orange dots. Error bars show the $1\sigma$ error estimated with 150 jackknife resamples.} 
    \label{fig:app:sysweight}
\end{figure}

\section{Validation of cosmic shear}
\label{app:cosmicshear}
In section \ref{sec:lensing}, we present an algorithm to fix the small-scale lensing around halos (cluster lensing and galaxy--galaxy lensing) using particle--halo cross-correlations. In this appendix, we verify that this correction has a negligible impact on large-scale cosmic shears and makes small-scale cosmic shears less affected by ray-tracing resolutions. Figure \ref{fig:app:cosmicshear} compares $\xi_{+/-}$ measured in Cardinal using corrected and original shears from ray-tracing. The corrected and uncorrected $\xi_{+/-}$ are consistent on large scales. However, on small scales, the corrected $\xi_{+/-}$ are much more consistent with the theories based on CAMB, and Cosmic Emu calculated using CCL \citep{CCL}. This indicates that the $\xi_{+/-}$ calculated using corrected shapes is less affected by the limited resolution of ray tracing. This finding is consistent with \cite{2007NJPh....9..446T},  where the authors show that the lensing effect from massive clusters ($M>10^{13} M_\odot$) significantly contributes to the one-halo term of cosmic shear power spectra.

\begin{figure*}
    \centering
    \includegraphics[width=0.99\textwidth]{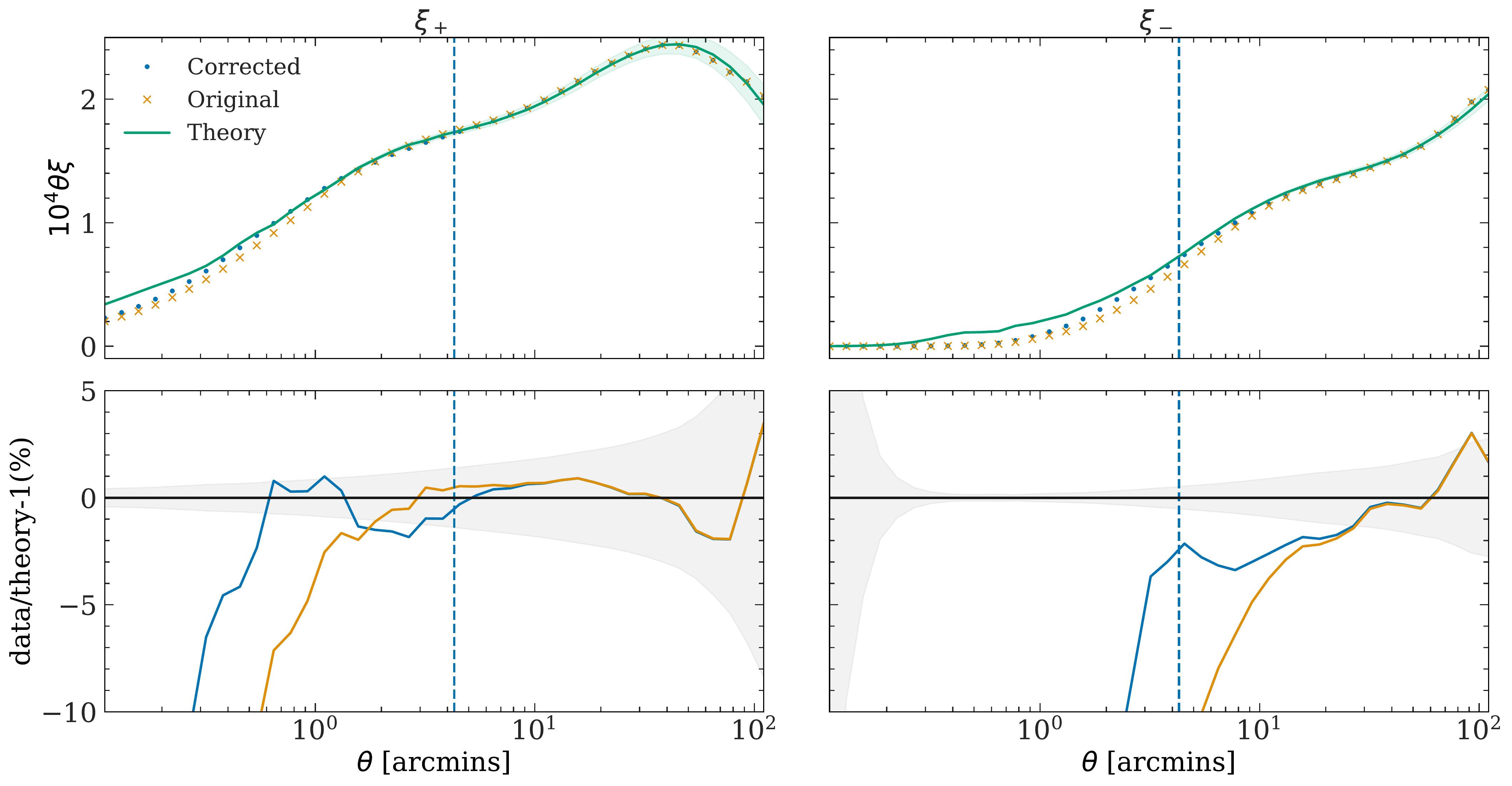}
    \caption{Comparison of $\xi_{+/-}$ calculated using original shear from ray-tracing (orange) and corrected shears (blue) of galaxies in Cardinal with $z=[0.6,0.9]$.  Green lines show theory prediction using CAMB and Cosmic Emulator. Shaded regions show $1\sigma$ error estimated with $204$ jackknife resampling. Bottom panels show fractional differences of $\xi_{+/-}$ measurements and theory. The shaded region shows $1\sigma$ error. The vertical dashed line shows ten times the ray-tracing resolution. } 
    \label{fig:app:cosmicshear}
\end{figure*}

\section{Conditional luminosity function}
\label{app:conditional}
We compare the conditional luminosity function in Cardinal, Buzzard v2.0, and DES-Y3 data. The measurements follow the prescriptions detailed in \cite{2020ApJ...897...15T}. Figure \ref{fig:app:clf} shows the result. We find that for the lower two redshift bins, the central galaxy luminosity distributions are consistent between Buzzard v2.0 and Cardinal. For the highest redshift bin, the central galaxy luminosity in Cardinal is somewhat fainter than Buzzard v2.0. However, the relative brightness of centrals and satellites in Cardinal is more consistent with the DES-Y3 data than Buzzard v2.0. This is an important feature for the performance of cluster finders, as indicated in \cite{2022OJAp....5E...1K} and \cite*{4x2pt1}. In terms of the satellites, we find that the satellite luminosity distributions in Cardinal are more consistent with the DES-Y3 data than Buzzard v2.0 on the faint end. On the bright end, Cardinal and Buzzard's satellite luminosity distributions are consistent but are fainter than the DES-Y3 data. 

\begin{figure*}
    \centering
    \includegraphics[width=0.99\textwidth]{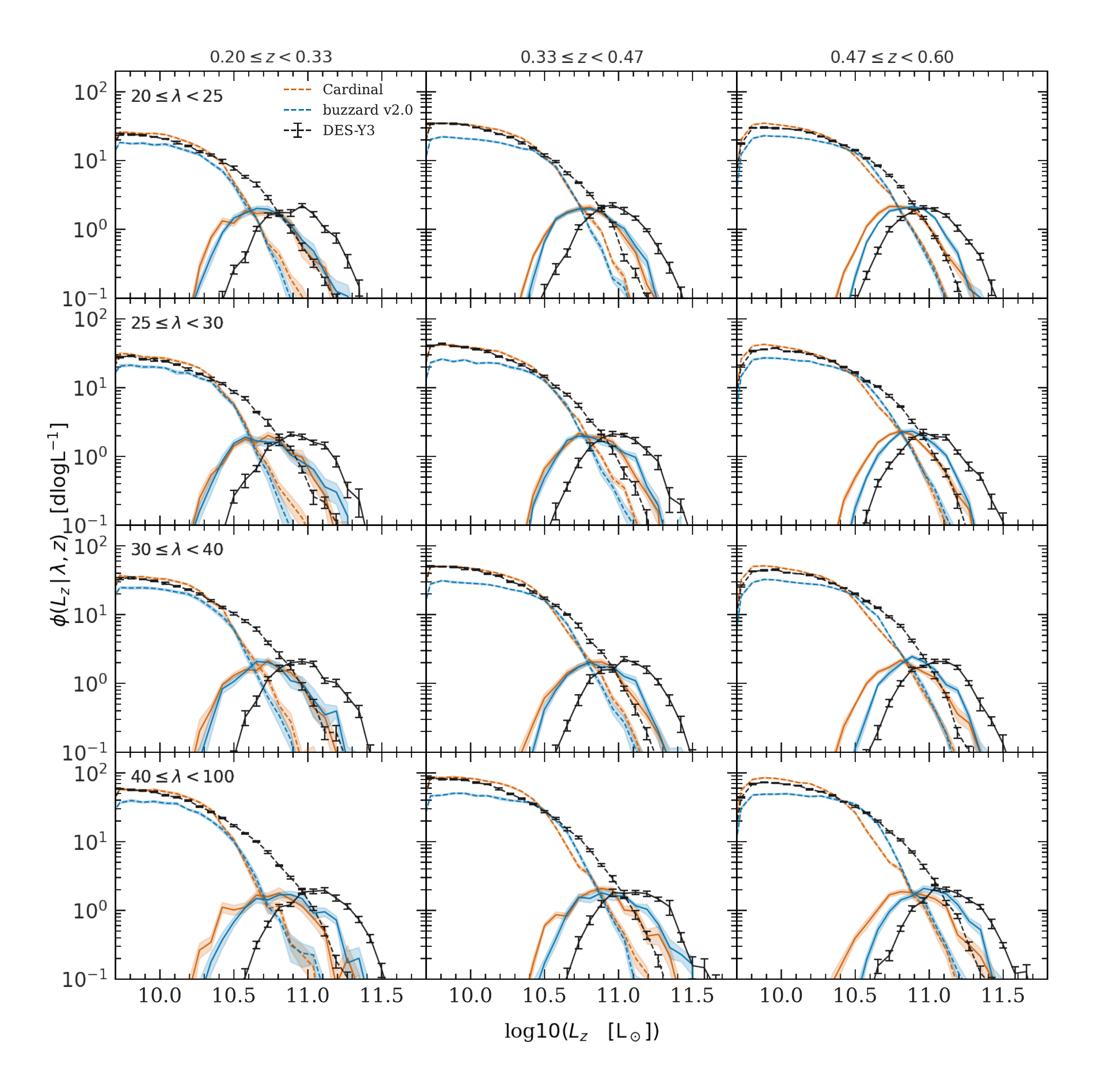}
    \caption{The conditional luminosity function for central galaxies (solid lines) and satellite galaxies (dashed lines) for \redmapper{} clusters in Buzzard v2.0 and Cardinal (blue and orange lines, respectively) and the DES-Y3 data (black lines). Different panels show the different richness and redshift bins, as indicated in the legend in each panel. Richness increases from top to bottom, and redshift increases from left to right. Error bars and shaded regions show the $1\sigma$ errors estimated from $50$ jackknife resampling.} 
    \label{fig:app:clf}
\end{figure*}
\section{Algorithm of avoiding widening red sequence with conditional abundance matching}
\label{app:matchredm}
Here, we present an algorithm that uses \redmapper{} to avoid widening the width of the red sequence in the conditional abundance matching step. First, we run the \redmapper{} cluster finder in the mocks generated before applying the conditional abundance matching. For each galaxy in the mock, we obtain a $\chi^2$, quantifying the consistency of the multidimensional colors with the empirically constructed red-sequence model, and $z_{\rm{red}}$, the redshift that minimizes the $\chi^2$ \citep[See details in ][]{Redmapper1}. For the galaxies consistent with the red sequence (i.e., small $\chi^2$), $z_{\rm{red}}$ provides a good estimator of galaxies' redshift, with a typical uncertainty $\sim 0.02$. Second, we abundance-match this $\chi^2$ measured in mocks to $\chi^2$ measured in the data by enforcing the equality of $p(<\chi^2|m_z)$. In this way, we can avoid the possibility that $\chi^2$ distributions in the mocks and data might differ. Third, we select red galaxies in the data and mocks using the abundance-matched $\chi^2$. According to \cite{redmapper4}, we select red galaxies as $\chi^2<20$. Third, we perform abundance matching separately for blue and red galaxies. For blue galaxies with $\chi^2>20$,  we enforce the equality of equation \ref{eq:abcolor} between mocks and data. For red galaxies with $\chi^2<20$, we use the  $z_{\rm{red}}$ information. To avoid the possibility that  $z_{\rm{red}}$ distributions in mocks and data differ, we first compute $z_{\rm{a, red}}$,  the abundance matched $z_{\rm{red}}$ in Cardinal by matching $p(<z_{\rm{red}}|m_z)$ between Cardinal and data. We then match the overall color distributions of mocks and data by enforcing the equality of the following equation, 
\begin{equation}
        P(<c_i | m_z, c_{j<i}, z_{\rm{a, red}}).
\end{equation}

\section{Algorithm of matching red sequence in mocks and data}
\label{app:matchred}
In this appendix, we describe an algorithm that matches red sequences in mocks and the data. We first run the \redmapper{} algorithm on mocks and data to obtain a red-sequence model. This model has the following functional form, 
\begin{eqnarray}
    &P(\boldsymbol{c}|z,m_z) \propto& \exp (-0.5\chi^2), \nonumber \\ 
    &\chi^2 =& \left( \boldsymbol{c} - \langle \boldsymbol{c} | z,m_z \rangle \right) \left(\boldsymbol{C}_{\rm{int}}(z)+\boldsymbol{C}_{\rm{err}}(z)\right)^{-1}\nonumber\\
    &&\left( \boldsymbol{c} - \langle \boldsymbol{c} | z,m_z \rangle \right)^T,
\end{eqnarray}
where $\boldsymbol{c}=[g-r, r-i, i-z]$, $\boldsymbol{C}_{\rm{int}}(z)$ is the intrinsic scatter of red sequence, 
and $\boldsymbol{C}_{\rm{err}}(z)$ is the observational noise. The mean color $\langle \boldsymbol{c} | z,m_z \rangle$ is modeled as a simple power law $c(z)+s(z) (m_z-m_p(z))$, where $c(z)$, $s(z)$, and $m_p(z)$ are free parameters. With $P(\boldsymbol{c}|z,m_z)$ models for mocks and data, we can  enforce the consistency of red sequence using the following algorithm,

\begin{algorithm}[H]

\begin{algorithmic}
\For{galaxies with redshift $z$ and observed magnitudes $m_{g,r,i,z}$}
\State $m^o_{g,r,i,z} \gets$ noisy realizations of $m_{g,r,i,z}$ using equation \ref{eq:noisy}. 
\State Calculate $\chi^2$ using $m^o_{g,r,i,z}$ and $P_m(\boldsymbol{c}|z,m_z)$ 
\If{$\chi^2<100$ and $R_n<R(z)$ and $M_n> M(z)$}
    \For{$c \in \{g-r, r-i, i-z\}$}
        \State  $c += \frac{F(1-\frac{\chi^2}{100})}{G(z)}$ $\left( \left(c-\langle \boldsymbol{c} | z,m_z \rangle_m \right)\left(\frac{s_d}{s_m}\right)^2+\langle \boldsymbol{c} | z,m_z \rangle_d -c\right)$
    \EndFor
    \State $m_g \gets c_{g-r}+c_{r-i}+c_{i-z}+m_z$ 
    \State $m_r \gets c_{r-i}+c_{i-z}+m_z$ 
    \State $m_i \gets c_{i-z}+m_z$ 
\EndIf

\EndFor

\end{algorithmic}
   
\end{algorithm}

In the above algorithm, $x_d$ represents the model evaluated using DES-Y3 data, $x_m$ represents the model evaluated using mocks, $R_n$ is the distance to the nearest resolved halo whose mass is $M_n$, and $s$ is the square root of the diagonal terms of $\boldsymbol{C}_{\rm{int}}(z)$. The $(1-\frac{\chi^2}{100})$ term is to ensure a smooth transition between red galaxies and blue galaxies. F controls the non-linear dependence of the color shift and $(1-\frac{\chi^2}{100})$, which is given by $F(x) = 0.5+0.5 {\rm{erf}}((x-0.1)/0.05)$, where ${\rm{erf}}$ is the error function. $(s_d/s_m)^2$ term ensures consistency of the scatter in the observed colors, which are proportional to the square root of the scatter of $c$ due to the Poisson draws. The $G(z)$ removes additional noise in the red sequence introduced by the CAM method. This additional noise becomes large when the galaxies' magnitude approaches the survey depth limits. Therefore, G(z) must be larger at higher redshifts when galaxies are fainter than their low redshift counterparts. We empirically find that $G(z) = 2(0.5+0.5{\rm{erf}}((z-0.8)/0.3)+1.2$ produces reasonable galaxy properties. Further, shifting all galaxies with $\chi^2<100$ will produce too many red galaxies. We, therefore, control the number of red galaxies and their clustering with two additional cuts $M_n> M(z)$ and $R_n<R(z)$. Again, because the additional noise introduced by the CAM model becomes larger at higher redshifts, $M(z)$ should decrease, and $R(z)$ should increase with redshifts. If $M(z)$ becomes too small, the clustering of red-sequence galaxies will become too small. We increase R(z) to ensure enough red galaxies and strong enough clustering. Empirically, we find 
\begin{eqnarray}
&R(z)&= 6\left(0.5+0.5 {\rm{erf}}\left(\left(z-0.54\right)/0.1\right)\right)+1.2 \, h^{-1}{\rm{Mpc}}\nonumber\\
&M(z)&= 10^{-0.25\left(0.5+0.5{\rm{erf}}\left((z-0.65)/0.05\right)\right)+12.90}\, h^{-1}M_\odot
\end{eqnarray}
give reasonable red galaxy properties.

\section{Galaxy radial profiles around clusters}
\label{app:radial profile}
This appendix details our analyses on constructing galaxy radial profiles around \redmapper{} clusters. We construct radial profiles for \redmapper{} clusters in simulation and data using exactly the same algorithm to ensure an apples-to-apples comparison. Our calculation pipeline is similar to the one presented in \cite{2018ApJ...864...83C} but has important differences in details. We emphasize that it is necessary to recalculate the galaxy profile around \redmapper{} clusters for data because \redmapper{} has been sufficiently changed.  

We wish to measure galaxy density profiles around \redmapper{} clusters using photometric data. This density profile ($\Sigma_g$) can be related to correlation functions via the following equation, 
\begin{equation}
\label{eq:sigmag}
    \Sigma_g(R) = \langle\Sigma_g\rangle w(R), 
\end{equation}
where $R$ is the comoving distance to galaxy clusters, $\langle\Sigma_g\rangle$ is the mean density of the galaxy samples, and $w(R)$ is the correlation functions between galaxies and clusters. Without knowing galaxy redshifts, we cannot evaluate equation \ref{eq:sigmag} directly. For a photometric dataset like DES, we must construct an estimator that gives equation \ref{eq:sigmag} but does not require a knowledge of galaxy redshifts. For the simplicity of the argument, let's first assume galaxy cluster samples all have the same redshifts $z_c$. 
We can then define our estimator of $\Sigma_g(R)$ as 
\begin{equation}
    \hat{\Sigma}_g(R) = \left(\frac{DD_{CT}(\theta)}{DR_{CR}(\theta)}\frac{N_R}{N_T}-1\right)\frac{N_T}{A},
\end{equation}
where $DD_{CT}$ denotes the number of pairs between clusters and galaxy samples in photometric data, $DD_{CR}$ denotes the number of pairs between clusters and random points describing survey footprints, $N_T$ is the total number of considered galaxies, $N_R$ is the total number of randoms, and $A$ is the survey area in the comoving unit at redshift $z_c$. In the above equation, $\theta$ is the angular separation of the pairs corresponding to a comoving distance $R$ at redshift $z_c$. 

We now show that $\hat{\Sigma}_g(R)$ is an unbiased estimator of $\Sigma_g(R)$. The full galaxy sample ($T$) can be divided into two groups: (1) galaxies physically associated with galaxy clusters ($g$) and (2) other galaxies ($f$). Because pair counting is an additive process, $DD_{CT}$ is then a sum of $DD_{Cg}$ (pairs between clusters and associated galaxies) and $DD_{Cf}$ (pairs between clusters and other galaxies). Further, clusters and unassociated galaxies are not clustered, so $DD_{Cf}$ is then $N_c N_f \zeta(\theta)$, where $N_f$ is the number of unassociated galaxies and $\zeta$ is some geometric factor that depends on angular separation. By the same argument $DD_{CR}$ is $N_c N_R\zeta(\theta)$. With these, we have 
\begin{eqnarray}
\label{eq:sigmaest}
    \hat{\Sigma}_g(R) &=& \left(\frac{DD_{CT}(\theta)}{DR_{CR}(\theta)}\frac{N_R}{N_T}-1\right)\frac{N_T}{A} \nonumber \\
    &=& \left(\frac{DD_{Cg}(\theta)+DD_{Cf}(\theta)}{DR_{CR}(\theta)}\frac{N_R}{N_g+N_f}-1\right)\frac{N_g+N_f}{A} \nonumber\\
    &=& \left(\frac{DD_{Cg}(\theta)+N_cN_f\zeta(\theta)}{DR_{CR}(\theta)}\frac{N_R}{N_g+N_f}-1\right)\frac{N_g+N_f}{A}\nonumber\\
     &=& \left(\frac{DD_{Cg}(\theta)}{DR_{CR}(\theta)}\frac{N_R}{N_g+N_f}+\frac{N_cN_f\zeta(\theta)}{N_c N_R\zeta(\theta)}\frac{N_R}{N_g+N_f}-1\right)\frac{N_g+N_f}{A} \nonumber\\
     &=& \left(\frac{DD_{Cg}(\theta)}{DR_{CR}(\theta)}\frac{N_R}{N_g}-1\right)\frac{N_g}{A} \nonumber\\
      &=& \left(\frac{DD_{Cg}(\theta)}{DR_{CR}(\theta)}\frac{N_R}{N_g}-1\right)\langle\Sigma_g\rangle,
\end{eqnarray}
In the above equation, we see the terms in the parenthesis are the Davis--Peebles estimator \citep{1983ApJ...267..465D} of $w(R)$. We note that the above derivation is only exact when considering Davis--Peebles estimator. Using Landy--Szalay estimator \citep{1993ApJ...412...64L} to estimate $\Sigma_g(R)$ like the one described in \cite{2018ApJ...864...83C} would include additional terms and potentially lead to biases. 

In survey data, not all clusters live at the same redshifts. We, therefore, follow the prescription described in \cite{2018ApJ...864...83C} for the full survey data. We first divide the \redmapper{} clusters with $\lambda>20$ from $z=0.2-0.55$ into $\Delta z=0.025$ bins. We estimate galaxy density profiles around clusters for each redshift bin ( $\hat{\Sigma}_g(R)_i$) using equation \ref{eq:sigmaest}. We then estimate the $\Sigma_g(R)$ using the weighted sum of $\hat{\Sigma}_g(R)_i$, with the number of clusters in each redshift bin as the weight. To avoid mixing galaxies with different luminosities across the entire redshift range, we further make a cut on galaxy samples at each redshift bin $\Mr-5\log(h)<-20$. The $\Mr$ is calculated assuming galaxies are at the same redshift as galaxy clusters. 

In Cardinal, we repeat all the above calculations. 
\begin{figure*}
    \centering
    \includegraphics[width=0.5\textwidth]{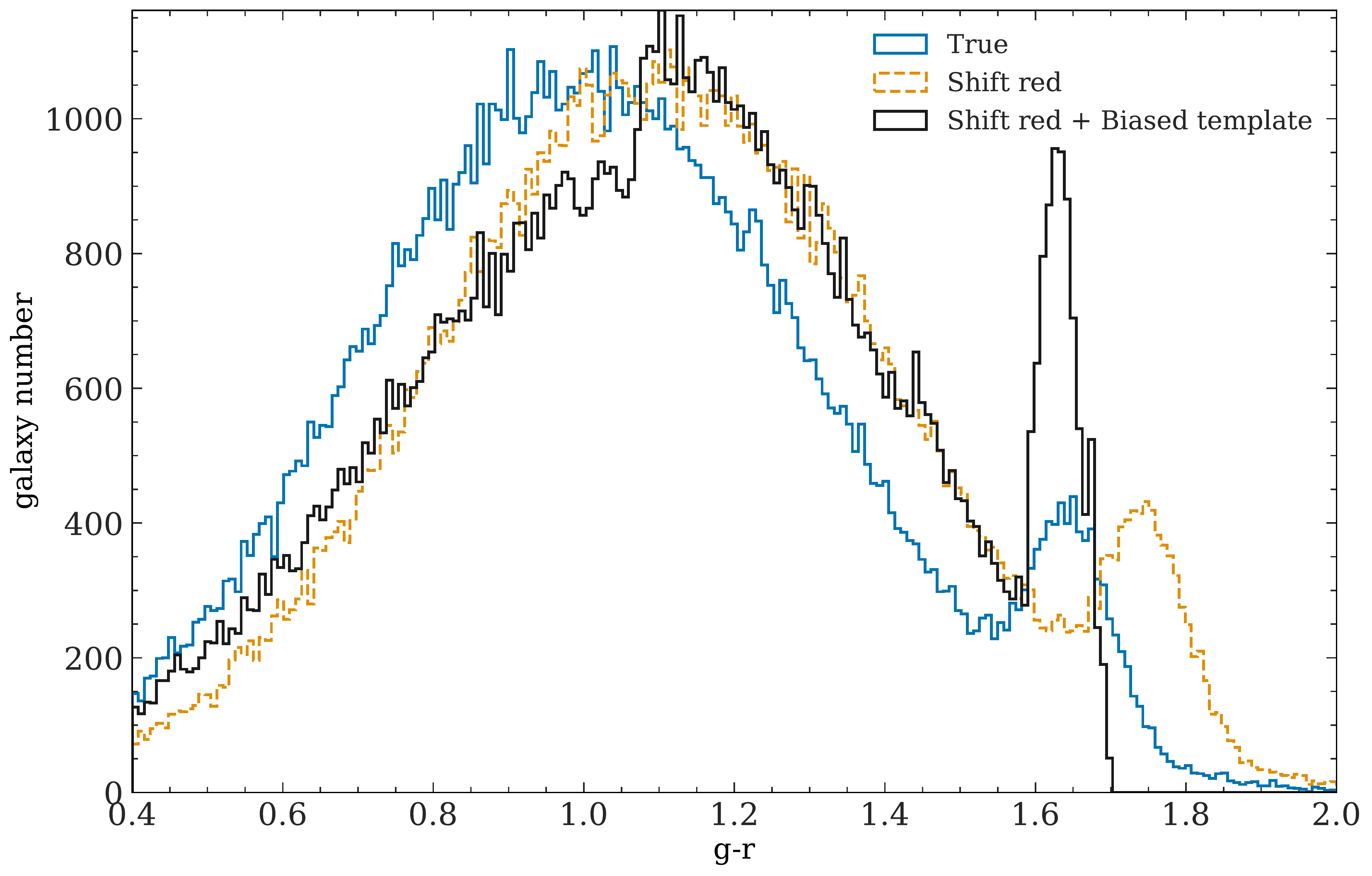}
    \caption{Color distributions of mock samples with the blue histogram representing the true color distribution. The red histogram corresponds to the samples when colors are shifted by $0.1$ magnitude. The black histogram shows the case when we summarize the color using SED templates. To generate the black histogram, we apply the biases shown in the right panel of figure \ref{fig:kcorr} to the galaxies in the red histogram. } 
    \label{fig:app:simcolor}
\end{figure*}

\section{Color template bias simulation}
\label{app:colorbiassim}
This section employs a simple simulation to offer valuable insights into the potential impact of the color biases presented in figure \ref{fig:kcorr} on galaxy color distribution. We accomplish this by generating mock galaxy samples through a combination of two Gaussian distributions. Specifically, we draw 100k blue galaxies with a mean $g-r$ color of $1$ and a standard deviation of $0.3$, followed by drawing 5k red galaxies with a mean $g-r$ of $1.65$ and a standard deviation of $0.05$. The resulting galaxy color distribution is presented as the blue histogram in figure \ref{fig:app:simcolor}, designed to mirror the cosmos samples' color distribution at $z=0.43-0.63$, as depicted in figure 5 of \cite{JoeBuzzard}. Next, we shift the color up by $0.1$ magnitude, a typical value when comparing the $g-r$ color distribution between Buzzard and DES-Y3 data. The resulting color distribution is illustrated as the red histogram in figure \ref{fig:app:simcolor}. Subsequently, we incorporate the color biases displayed in the right panel of figure \ref{fig:kcorr}, resulting in a galaxy sample with the color distribution depicted as the black histogram in figure \ref{fig:app:simcolor}. Strikingly, the black histogram reproduces two key features in figure 5 of \cite{JoeBuzzard}. First, the red-sequence galaxies corresponding to the black histogram have a similar mean color to those in the unaltered galaxy samples. Second, the color distribution of red-sequence galaxies is significantly more peaked than those of unaltered samples. This simple yet compelling simulation further supports our hypothesis regarding the color discrepancies' origin between Buzzard and data.

\section{Differences between Cardinal and Buzzard v2.0}
\label{app:improvement}
This section summarizes the main improvements from Buzzard v2.0 to Cardinal. We divide these improvements into four categories, summarized below.
\begin{enumerate}
    \item Subhalo abundance matching:
    \begin{enumerate}
    \item We consider a luminosity-dependent scatter.
    \item Orphan subhalos are included, whose abundances are controlled by a specially-designed orphan model.
    \item The group–galaxy cross-correlation function is used in addition to the galaxy-galaxy correlation function to constrain the model. 
    \end{enumerate}
    \item Environmentally dependent galaxy color model:
    \begin{enumerate}
    \item We consider a luminosity-dependent correlation between galaxy color and distances to nearby massive halos at fixed galaxy luminosity. 
    \item When using distances to nearby massive halos as a local environment proxy, we normalize them by the halo size. 
    \item We apply a nonlinear transform on distances to nearby massive halos to break the degeneracy of large-scale color-dependent clustering and galaxy color gradient around massive halos. 
    \item We introduce an additional parameter that controls the color of low-mass centrals. 
    \end{enumerate}
    \item Observational effects:
     \begin{enumerate}
     \item We increase the fidelity of faint galaxy photometric uncertainties by swapping the Gaussian distribution used to generate noise of galaxy fluxes to a Poisson distribution.
    \item We empirically correct artifacts in the raytracing shears due to limited resolution. 
     \end{enumerate}
    \item Afterburner:
    \begin{enumerate}
    \item We demonstrate the limitations of kcorrect SED templates in describing red galaxy colors.
    \item To mitigate this problem, we develop a new algorithm that employs the conditional abundance matching technique and a photometric dataset to correct galaxy colors empirically. 
    \item We develop a new algorithm for matching the red sequence between mocks and data that takes into account the slope of the color--magnitude relations. 
    \end{enumerate}
\end{enumerate}
\bibliographystyle{mn2e_2author_arxiv_amp.bst}
\bibliography{sample.bib} %

\end{document}